\documentclass[twocolumn,times,tighten]{aastex701}

\hypersetup{linkcolor=blue,citecolor=blue,filecolor=cyan,urlcolor=blue}

\def\smallplottwo#1#2{\centering \leavevmode
\includegraphics[width=89.5mm]{#1}
 \hfil \includegraphics[width=89.5mm]{#2} }

\def\smallplotthree#1#2#3{\centering \leavevmode
\includegraphics[width=88.8mm]{#1} \includegraphics[width=88.8mm]{#2}
\includegraphics[width=88.8mm]{#3}}

\def\smallplotfour#1#2#3#4{\centering \leavevmode
\includegraphics[width=89.5mm]{#1} \includegraphics[width=89.5mm]{#2}
\hfil \includegraphics[width=89.5mm]{#3} \includegraphics[width=89.5mm]{#4}}

\def\smallplotsix#1#2#3#4#5#6{\centering \leavevmode
\includegraphics[width=89.5mm]{#1} \includegraphics[width=89.5mm]{#2}
\hfil \includegraphics[width=89.5mm]{#3} \includegraphics[width=89.5mm]{#4}
\hfil \includegraphics[width=89.5mm]{#5} \includegraphics[width=89.5mm]{#6}}

\def\smallploteight#1#2#3#4#5#6#7#8{\centering \leavevmode
\includegraphics[width=89.5mm]{#1} \includegraphics[width=89.5mm]{#2}
\hfil \includegraphics[width=89.5mm]{#3} \includegraphics[width=89.5mm]{#4}
\hfil \includegraphics[width=89.5mm]{#5} \includegraphics[width=89.5mm]{#6}
\hfil \includegraphics[width=89.5mm]{#7} \includegraphics[width=89.5mm]{#8}}

\graphicspath{{./}{figures/}}

\shorttitle{Properties of C-rich AGB Stars in the LMC and the Milky Way}
\shortauthors{Suh} 

\begin{document}

\title{Properties of Carbon-rich Asymptotic Giant Branch Stars in the LMC and the Milky Way}

\correspondingauthor{Kyung-Won Suh}
\email{kwsuh@chungbuk.ac.kr}

\author[orcid=0000-0001-9104-9763,sname='Suh']{Kyung-Won Suh}
\affiliation{Department of Astronomy and Space Science, Chungbuk National
University, Cheongju, 28644, Republic of Korea}
\email[show]{kwsuh@chungbuk.ac.kr}

\begin{abstract}
We present a comparative study of carbon-rich asymptotic giant branch (CAGB)
stars in the Large Magellanic Cloud (LMC; 7347 stars) and the Milky Way (7163
stars) using infrared color–magnitude diagrams, spectral energy distributions
(SEDs), two-color diagrams, and variability data. Observed SEDs are compared
with theoretical models to characterize the central stars and their
circumstellar dust envelopes and to estimate distances. For the LMC, a set of
best-fitting CAGB models is derived by fitting observed SEDs with radiative
transfer models, utilizing the galaxy’s well-established distance. For Galactic
CAGB stars, where Gaia DR3 parallaxes are uncertain, we estimate distances by
fitting observed SEDs with the CAGB models validated against LMC stars, and for
Mira variables, from the period–magnitude relation calibrated with LMC Miras. A
comparison of these approaches demonstrates that the SED-based distances are
both reliable and practical for a large sample of Galactic CAGB stars. We find
that CAGB stars in both galaxies show broadly similar infrared properties,
although the LMC sample lacks stars with extremely thick dust envelopes.
\end{abstract}



\section{Introduction} \label{sec:intro}

The asymptotic giant branch (AGB) phase consists of two stages: the early AGB
(E-AGB) and the thermally pulsing AGB (TP-AGB) (e.g., \citealt{iben1983}). E-AGB
stars are generally oxygen-rich (OAGB), whereas carbon-rich AGB (CAGB) stars are
thought to evolve from intermediate-mass OAGB stars ($\approx$ 2 to 4.5
$M_{\odot}$ at solar metallicity, with the lower limit decreasing and the upper
limit increasing at lower metallicity) once thermal pulses begin (e.g.,
\citealt{karakas2014}). More massive OAGB stars enter the TP-AGB phase without
altering their surface chemistry. AGB stars are long-period variables (LPVs),
characterized by large-amplitude pulsations and substantial mass-loss rates,
which play a crucial role in driving envelope ejection and late stellar evolution
(e.g., \citealt{suh2014}).

Although most carbon stars belong to the CAGB class, the category also
encompasses other types. Classical (intrinsic, or type C-N) carbon stars are
generally CAGB stars, whereas non-classical, or extrinsic, carbon stars are not
on the AGB. The latter group includes CH stars, Barium stars, dwarf carbon stars,
early R-type stars, and J-type stars (e.g., \citealt{green2013};
\citealt{suh2024}).

IR observations from Infrared Astronomical Satellite (IRAS), Midcourse Space
Experiment (MSX), Infrared Space Observatory (ISO), AKARI, Two-Micron All-Sky
Survey (2MASS), Wide-field Infrared Survey Explorer (WISE), and Spitzer have been
instrumental in identifying CAGB stars and investigating their properties.

Gaia Data Release 3 (DR3) has yielded precise optical measurements for over one
billion stars (\citealt{rimoldini2023}). The distance determinations
(\citealt{bailer-jones2021}) and extinction estimates (e.g.,
\citealt{lallement2022}) derived from Gaia DR3 constitute essential inputs for
constraining the absolute magnitudes of different stellar populations in the
Milky Way.

\begin{table*}
\centering
\caption{Sample of CAGB stars in the LMC and the Milky Way\label{tab:tab1}}
\begin{tabular}{lllll}
\hline \hline
Group  &Reference &Total Number & Selected\\
\hline
LMC-CAGB\_SAGE & \citet{riebel2012} & 9293 & 7308$^a$  \\
LMC-CAGB\_SAGE-S$^b$ & \citet{sloan2016}; \citet{jones2017} & - & 151   \\
LMC-CAGB\_KDMK & \citet{kontizas2001} & 7760$^c$ & 24+15$^d$\\
LMC-CAGB & combined$^e$  & -  &  7347   \\
\hline
MW-CAGB\_IC & \citet{suh2024} & 4909  & 4909 \\
MW-CAGB\_IC-S & \citet{suh2024} & -  & 91 \\
MW-CAGB\_NI & \citet{suh2024} & 2254  & 2254 \\
MW-CAGB & combined$^e$  & -  &  7163   \\
\hline
\end{tabular}
\begin{flushleft}
$^a$See Section~\ref{sec:magb}.
$^b$Subgroup of LMC-CAGB\_SAGE; objects identified from the Spitzer IRS.
$^c$3933 objects are duplicates of LMC-CAGB\_SAGE objects. A large fraction of this sample could be extrinsic carbon stars.
$^d$Miras from OGLE-III and candidates for new Miras identified from WISE data (\citealt{suh2025}).
$^e$It denotes the merged set of all unique objects in the catalogs listed above.
\end{flushleft}
\end{table*}

For Galactic CAGB stars, distance estimates based on Gaia DR3 parallaxes remain
uncertain (e.g., \citealt{suh2024}). In this study, we use the well-established
distance to the LMC to validate our models and methodology. Distances to Galactic
CAGB stars are then derived by fitting their observed SEDs with a set of
theoretical CAGB models calibrated against LMC CAGB stars at the known distance.
For Galactic CAGB stars that are Mira variables, distances can also be estimated
from the period–magnitude relation (PMR) established for LMC Miras.

We analyze the properties of CAGB stars in the LMC and the Milky Way.
Section~\ref{sec:sample} presents catalogs of known CAGB stars in both galaxies.
Section~\ref{sec:models} describes radiative transfer models for circumstellar
dust shells and the derivation of model parameters, constrained by LMC CAGB
stars. In Section~\ref{sec:seds}, observed spectral energy distributions (SEDs)
are compared with model predictions to characterize the central stars and their
dust envelopes and to estimate theoretical distances. Section~\ref{sec:irpro}
compares observed infrared two-color diagrams (2CDs) and color–magnitude diagrams
(CMDs) with theoretical models. Section~\ref{sec:pul} examines the
period–magnitude relation (PMR) of Mira-type CAGB stars. Section~\ref{sec:sum}
summarizes the main results.

\section{Sample Stars\label{sec:sample}}

We use catalogs of CAGB stars in the LMC and the Milky Way from the available
literature (see Table~\ref{tab:tab1}). For all sample stars, we cross-identified
counterparts from Gaia DR3, 2MASS, IRAS, MSX, ISO, AKARI,  WISE, Spitzer, SIMBAD,
and the American Association of Variable Star Observers (AAVSO; international
variable star index; version 2025 June 29; \citealt{watson2025}). All objects in
the sample are identified by their unique SIMBAD identifiers, and their Gaia
counterparts have been confirmed through SIMBAD.

\subsection{CAGB stars in the LMC\label{sec:magb}}

The Spitzer Space Telescope Legacy program Surveying the Agents of Galaxy
Evolution (SAGE; \citealt{meixner2006}) identified a large population of CAGB
stars in the LMC. Using SAGE data, \citet{riebel2012} compiled a catalog of 7293
CAGB objects. In addition, 151 CAGB stars in the LMC were identified from Spitzer
Infrared Spectrograph (IRS; 5.2–38 $\mu$m) observations (\citealt{sloan2016};
\citealt{jones2017}). Cross-matching these 151 LMC-CAGB\_SAGE-S objects with the
catalog of \citet{riebel2012} yields a total of 7308 CAGB stars in the LMC
(LMC-CAGB\_SAGE objects; see Table~\ref{tab:tab1}). We note that all
LMC-CAGB\_SAGE-S sources are already included in the LMC-CAGB\_SAGE sample.

The Optical Gravitational Lensing Experiment III (OGLE-III; \citealt{sus09})
reported 91,995 LPVs in the LMC, including 1663 classified as Mira variables.

\citet{kontizas2001} compiled a catalog of 7760 carbon stars in the LMC, a large
fraction of which may be extrinsic carbon stars. Among these, 3933 are duplicates
of LMC-CAGB\_SAGE objects. Of the remaining sources, 39 are likely additional
CAGB stars, 24 are identified as Miras from OGLE-III, and 15 are strong
candidates for new Miras based on recently analyzed WISE light curves
(\citealt{suh2025}).

When we combine the two samples, there are 7347 CAGB stars in the LMC (LMC-CAGB
objects; see Table~\ref{tab:tab1}). There are 1184 Mira variables from OGLE-III
in the list of CAGB stars in the LMC (LMC-CAGB objects).

To calculate the absolute magnitudes for LMC-CAGB stars, we adopt a distance of
49.6 kpc to the LMC, a commonly used approximation based on the precise
measurement of 49.59 $\pm$ 0.54 kpc by \citet{pietrzynski2019}.

\subsection{CAGB stars in the Milky Way\label{sec:mw}}

\citet{suh2024} compiled a catalog of CAGB stars in the Milky Way, classified
into two groups: those with IRAS counterparts (MW-CAGB\_IC), identified from the
IRAS source catalog and generally corresponding to brighter or more isolated
stars, and those without IRAS counterparts (MW-CAGB\_NI), identified from the
AllWISE or Gaia DR3 catalogs and typically representing fainter stars or objects
in crowded regions. Most MW-CAGB\_IC sources also have entries in the AllWISE or
Gaia DR3 catalogs. In this study, we adopt the catalog of Galactic CAGB stars
from \citet{suh2024} (MW-CAGB objects; see Table~\ref{tab:tab1}). The
MW-CAGB\_IC-S subset denotes those MW-CAGB\_IC objects for which infrared
spectral data are available (see Section~\ref{sec:seds}).

\begin{table}
\setlength{\tabcolsep}{4pt}
\scriptsize
\caption{Visual and IR bands\label{tab:tab2}}
\centering
\begin{tabular}{lllll}
\hline \hline
Band &$\lambda_{ref}$ ($\mu$m)	&ZMF (Jy) &Telescope &Reference$^a$\\
\hline
Bp[0.5]  &0.511	&	3552	&	Gaia DR3	& \citet{rimoldini2023}\\
G[0.6]  &0.622	&	3229	&	Gaia DR3    & \citet{rimoldini2023}	\\
Rp[0.8] &0.777	&	2555	&	Gaia DR3    & \citet{rimoldini2023}	\\
J[1.2]  &1.235	&	1594	&	2MASS	& \citet{cohen2003}\\
H[1.7]  &1.662	&	1024	&	2MASS	& \citet{cohen2003}\\
K[2.2]  &2.159	&	666.7	&	2MASS	& \citet{cohen2003}\\
W1[3.4]	&3.35	&	306.681	&	WISE    & \citet{jarrett2011}	\\
W2[4.6]	&4.60	&	170.663	&	WISE    & \citet{jarrett2011}	\\
S3[5.8] &5.731	&	115.0   &	Spitzer & A	\\
S4[8.0] &7.872  &	64.9	&	Spitzer & A	\\
MA[8.3]  &8.28	&	58.49	&	MSX    & \citet{egan2003}	\\
AK[9]   &9	    &	56.262	&	AKARI    & \citet{murakami2007}	\\
IR[12]	&12     &  28.3 	&	IRAS     & \citet{beichman1988}	\\
W3[12]$^b$	&12	    &  28.3 	&	WISE & \citet{jarrett2011}\\
MC[12.1]  &12.13	&	26.51	&	MSX    &  \citet{egan2003}       	\\
MD[14.7]  &14.65	&	18.29	&	MSX    & \citet{egan2003}	\\
AK[18]   &18	&	12.001	&	AKARI    & \citet{murakami2007}	\\
ME[21.3]  &21.34	&	8.8  	&	MSX    & \citet{egan2003}	\\
W4[22]	&22.08	&	8.284	&	WISE    & \citet{jarrett2011}	\\
S5[24]  &23.68  &	7.17	&	Spitzer & B	\\
IR[25]	&25 & 6.73 	&	IRAS      & \citet{beichman1988}	\\
IR[60]	&60 & 1.19 	&	IRAS      & \citet{beichman1988}	\\
AK[65]  &65	& 0.965	&	AKARI    & \citet{murakami2007}	\\
AK[90]  &90	& 0.6276	&	AKARI    & \citet{murakami2007}	\\
IR[100]	&100 & 0.43 	&	IRAS      & \citet{beichman1988}	\\
AK[140]  &140	&0.1859	&	AKARI    & \citet{murakami2007}	\\
AK[160]  &160	&0.1487	&	AKARI    & \citet{murakami2007}	\\
\hline
\end{tabular}
\begin{flushleft}
$^a$A: \url{https://irsa.ipac.caltech.edu/data/SPITZER/docs/irac/iracinstrumenthandbook},
B: \url{https://irsa.ipac.caltech.edu/data/SPITZER/docs/mips/mipsinstrumenthandbook}.
$^b$For W3[12], $\lambda_{ref}$ and ZMF values are adopted from \citet{suh2020}.
\end{flushleft}
\end{table}

\subsection{Observational Data\label{sec:photdata}}

Table~\ref{tab:tab2} presents the visual and IR bands used in this study for
constructing SEDs, CMDs and 2CDs. The table includes the reference wavelength
($\lambda_{ref}$) and zero-magnitude flux (ZMF) for each band, which are
essential for deriving theoretical models (see Section~\ref{sec:models}) and
building SEDs from observational data. For a comprehensive discussion of IR
photometric data and ZMF values, refer to Section 2.1 of \citet{suh2021}.

For all sample stars, we cross-matched counterparts from Gaia DR3, 2MASS, IRAS,
 MSX, ISO, AKARI, WISE, Spitzer, SIMBAD, OGLE-III, and AAVSO by selecting the nearest
sources within the angular resolution (beam size) of each telescope (see
\citealt{suh2021}; \citealt{suh2024}). All objects are identified by their unique
SIMBAD identifiers, and their Gaia counterparts were verified through SIMBAD.

IRAS, AKARI, and MSX data are available for most Galactic CAGB stars, but only
for a small subset of those in the LMC. In contrast, Spitzer data cover the
majority of LMC-CAGB stars through the SAGE program, but only a limited fraction
of Galactic CAGB stars through the Galactic Legacy Infrared Mid-Plane Survey
Extraordinaire (GLIMPSE; \citealt{churchwell2009}), which does not include
observations in the S5[24] band. By comparison, WISE and 2MASS data are available
for most CAGB stars in both the LMC and the Milky Way.

This study incorporates only high-quality observational data across all
wavelength bands for photometric analysis (quality better than 1 for IRAS, MSX,
and AKARI data; quality better than U for WISE data).

\subsection{Distances and extinction for Galactic CAGB stars}\label{sec:ext}

Recent parallax distance data from \citet{bailer-jones2021} (r\_med\_geo: the
median of the geometric distance) derived from Gaia DR3, are crucial for
determining the distances of many classes of stars in our Galaxy. Gaia DR3
provides readily available distances for a significant fraction of CAGB stars.

Gaia detectors operating in the visual bands cannot reliably detect CAGB stars
with thick dust envelopes, often resulting in missing data for these objects. In
addition, the Gaia counterparts provided by SIMBAD may be inaccurate for some
objects, as it is often difficult to identify reliable counterparts using
available photometric data and theoretical models, especially in crowded regions.
Furthermore, CAGB stars frequently exhibit non-symmetric motions in their outer
envelopes (e.g., \citealt{kt2012}), which can lead to inaccurate distance
measurements in Gaia DR3. Therefore, distances derived from comparative analyses
between observed and theoretical model SEDs may provide alternative or more
reliable estimates (see Section~\ref{sec:seds}).

Once the distance and position of each object were established, we obtained
visual-band extinction values ($A_V$) from the database of \citet{lallement2022}.
Extinction values for Gaia and 2MASS bands were then derived using the
optical-to-MIR extinction law described by \citet{wang2019}.

The derived distance and extinction data are employed to create a range of CMDs
and IR 2CDs. In this study, extinction corrections are applied to the colors and
magnitudes for the IR CMDs and 2CDs constructed from 2MASS band data (see
Table~\ref{tab:tab2}).

\section{Theoretical Dust Shell Models\label{sec:models}}

To calculate theoretical model SEDs for CAGB stars, we use radiative transfer
models for spherically symmetric dust shells surrounding central stars. The
radiative transfer code RADMC-3D
(\url{https://github.com/dullemond/radmc3d-2.0}), following the approaches of
\citet{sk2013} and \citet{suh2024}, is applied. Detailed descriptions of the
theoretical models and their limitations can be found in \citet{suh2020}.

A continuous power-law dust density distribution ($\rho \propto r^{-2}$) is
assumed, with a dust formation temperature ($T_c$) of 1000 K. The inner radius of
the dust shell corresponds to $T_c$, while the outer radius is defined by a dust
temperature of 30 K. The dust optical depth ($\tau_{10}$) is referenced at a
wavelength of 10 $\mu$m.

For the dust opacity, the optical constants of amorphous carbon (AMC) and SiC
grains are taken from \citet{suh2000} and \citet{pegouri1988}, respectively. We
adopt two distinct dust opacity models: pure AMC and a simple mixture of AMC and
SiC (20\% by mass). Spherical dust grains with a uniform radius of 0.1 $\mu$m are
assumed.

We adopt a stellar blackbody temperature ($T_*$) between 2000 and 3000 K and a
stellar luminosity ($L_*$) ranging from $5\times10^{3}$ to $1\times10^{4}$
$L_{\odot}$.

\begin{table}
\caption{37 models for typical CAGB stars$^a$\label{tab:tab3}}
\centering
\begin{tabular}{llllll}
\hline \hline
Model$^b$	&	Dust$^c$	& $\tau_{10}$	& $T_*$ (K)	& $L_*$$^d$	\\
\hline
CA1  	&	AMC       	&	0.0001	&	3000	&	5.65	\\
     	&	AMC       	&	0.001	&	3000	&	5.65	\\
     	&	AMC       	&	0.005	&	3000	&	5.65	\\
     	&	AMC+SiC   	&	0.005	&	3000	&	5.65	\\
     	&	AMC       	&	0.007	&	2700	&	6.5	\\
     	&	AMC+SiC   	&	0.007	&	2700	&	6.5	\\
CA2  	&	AMC       	&	0.01	&	2700	&	6.5	\\
     	&	AMC+SiC   	&	0.01	&	2700	&	6.5	\\
     	&	AMC       	&	0.02	&	2700	&	6.5	\\
     	&	AMC+SiC   	&	0.02	&	2700	&	6.5	\\
     	&	AMC       	&	0.03	&	2700	&	6.5	\\
     	&	AMC+SiC   	&	0.03	&	2700	&	6.5	\\
     	&	AMC       	&	0.05	&	2500	&	6.5	\\
     	&	AMC+SiC   	&	0.05	&	2500	&	6.5	\\
     	&	AMC       	&	0.07	&	2500	&	6.5	\\
     	&	AMC+SiC   	&	0.07	&	2500	&	6.5	\\
CA3  	&	AMC       	&	0.1	&	2500		&	6.7	\\
     	&	AMC+SiC   	&	0.1	&	2500		&	6.7	\\
     	&	AMC       	&	0.2	&	2500		&	6.7	\\
     	&	AMC+SiC   	&	0.2	&	2500		&	6.7	\\
     	&	AMC       	&	0.3	&	2500		&	6.7	\\
	    &	AMC+SiC	    &	0.3	&	2500	   	&	6.7	\\
     	&	AMC       	&	0.4	&	2500		&	6.7	\\
	    &	AMC+SiC	    &	0.4 &	2500	   	&	6.7	\\
CA4  	&	AMC       	&	0.5	&	2300	&	8	\\
     	&	AMC+SiC   	&	0.5	&	2300	&	8	\\
     	&	AMC       	&	0.6	&	2300	&	8	\\
     	&	AMC       	&	0.7	&	2300	&	8	\\
     	&	AMC       	&	0.8	&	2300	&	8	\\
     	&	AMC       	&	0.9	&	2300	&	8	\\
CA5  	&	AMC       	&	1	&	2200	&	9	\\
     	&	AMC       	&	1.2	&	2200	&	9	\\
     	&	AMC       	&	1.5	&	2200    &	9	\\
CA6  	&	AMC       	&	2	&	2000	&	10	\\
     	&	AMC       	&	2.5	&	2000	&	10	\\
     	&	AMC       	&	3	&	2000	&	10	\\
CA7  	&	AMC       	&	4	&	2000	&	10	\\
\hline
\end{tabular}
\begin{flushleft}
\scriptsize
$^a$See Section~\ref{sec:models} for details. For all models, $T_c$ = 1000 K.
$^b$Seven representative CAGB models were used to produce CMDs and 2CDs.
$^c$AMC: pure AMC; AMC+SIC: a simple mixture of AMC and SiC (20\% by mass)
$^d$Stellar luminosity in $10^{3} L_{\odot}$.
Based on this luminosity, the scaled logarithmic distances can be sampled using a Gaussian Monte Carlo spread (GS)
in $\log_{10} d$ ($\sigma = 0.0798$~dex) derived from LMC-CAGB stars (see Figure~\ref{f1}),
corresponding to a total luminosity variation of approximately $2.1\times$ from minimum to maximum (see Section~\ref{sec:lum-lcagb}).
\end{flushleft}
\end{table}

\subsection{Modeling LMC-CAGB stars}\label{sec:lum-lcagb}

The absolute luminosity of CAGB stars in the Milky Way remains uncertain due to
large errors in distance estimates. In contrast, the LMC, with its
well-determined distance, provides an ideal setting for investigating the
intrinsic luminosity of CAGB stars.

For the full sample of 7347 LMC-CAGB stars (Table~\ref{tab:tab1}), we tested an
extensive grid of radiative transfer models covering a broad range of parameter
combinations to identify the most suitable set of CAGB models. The observed SEDs
were fitted using a least-squares optimization method (see
Sections~\ref{sec:seds} and \ref{sec:sedlmc}).

\begin{figure}
\centering
\smallplotthree{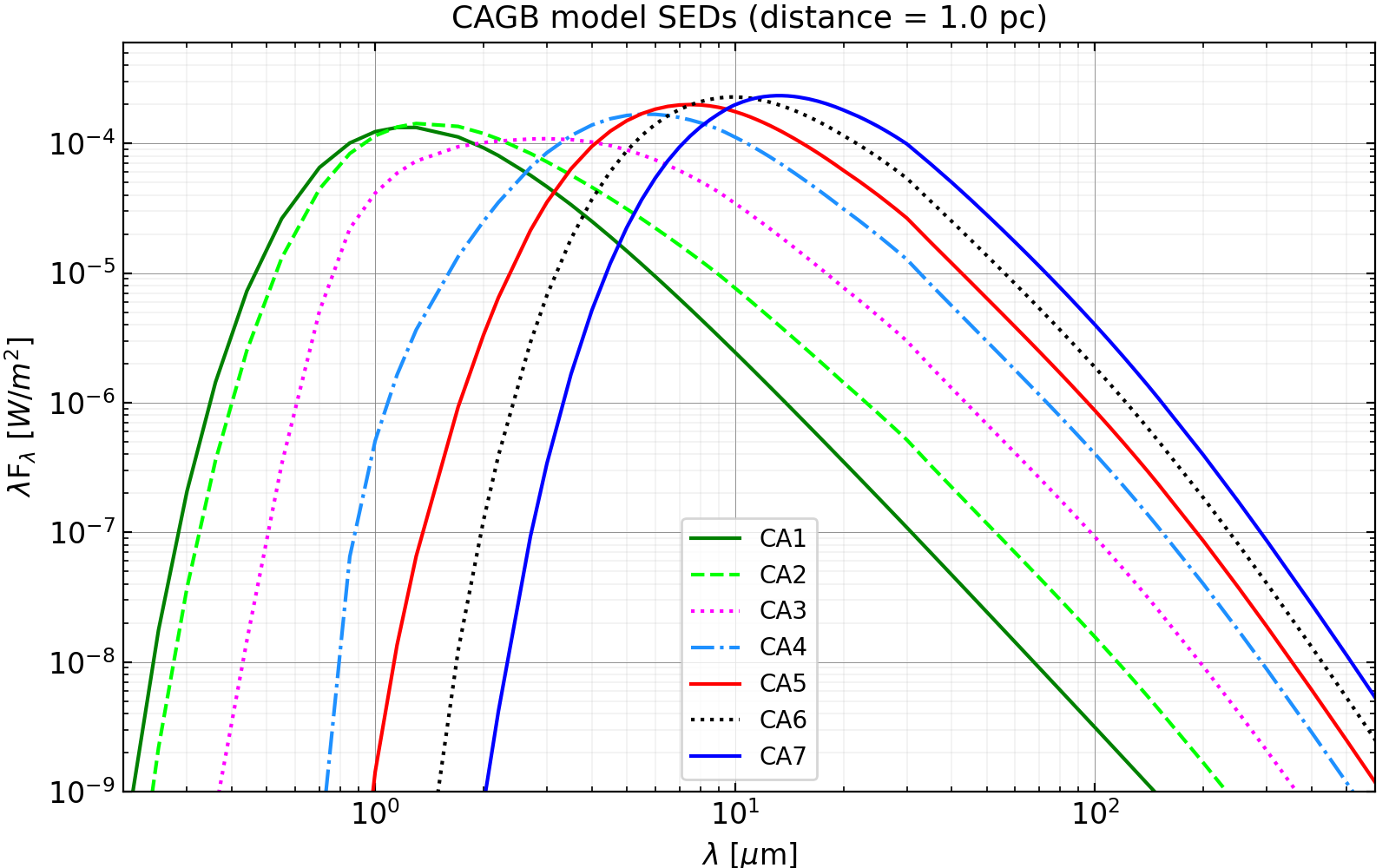}{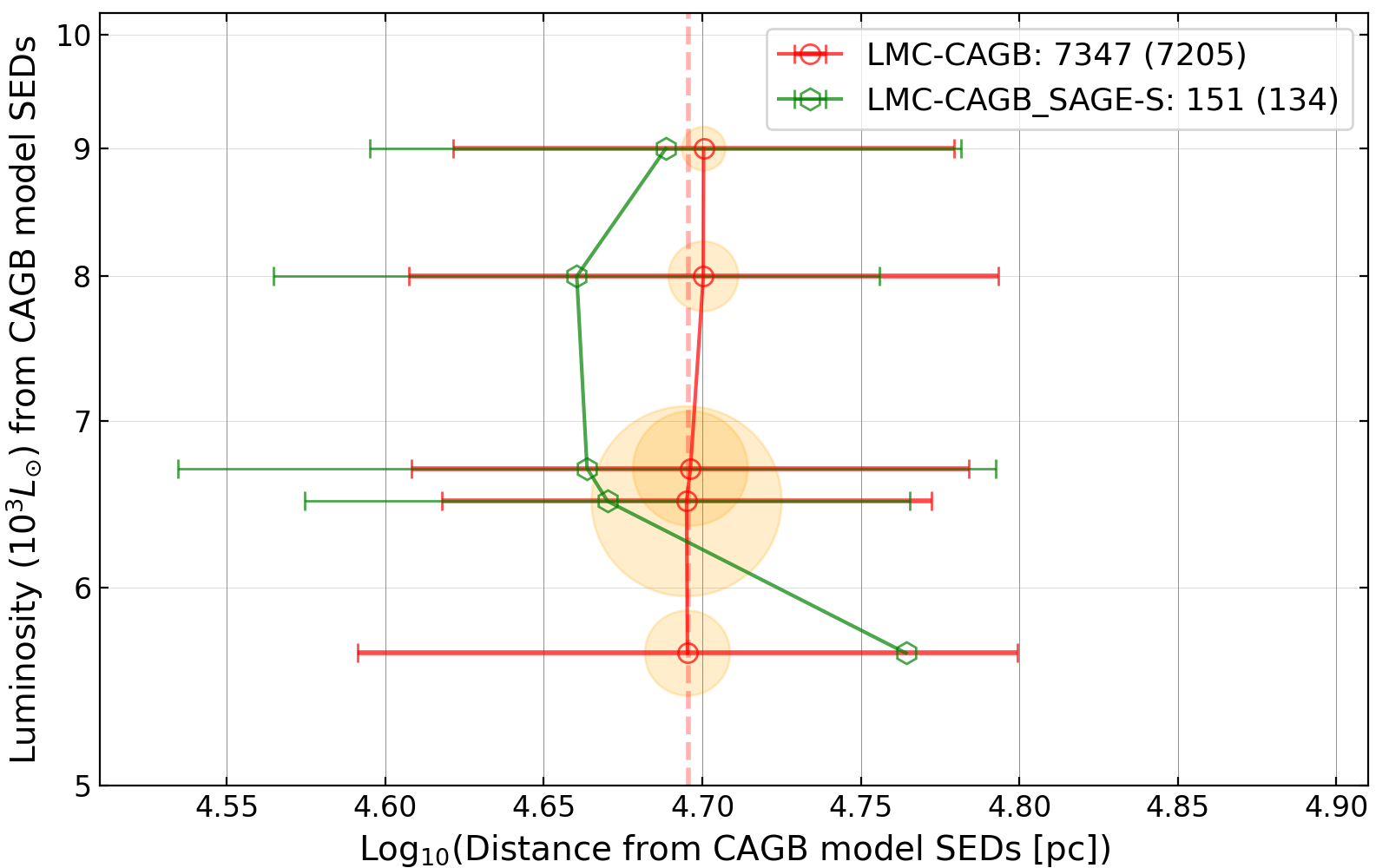}{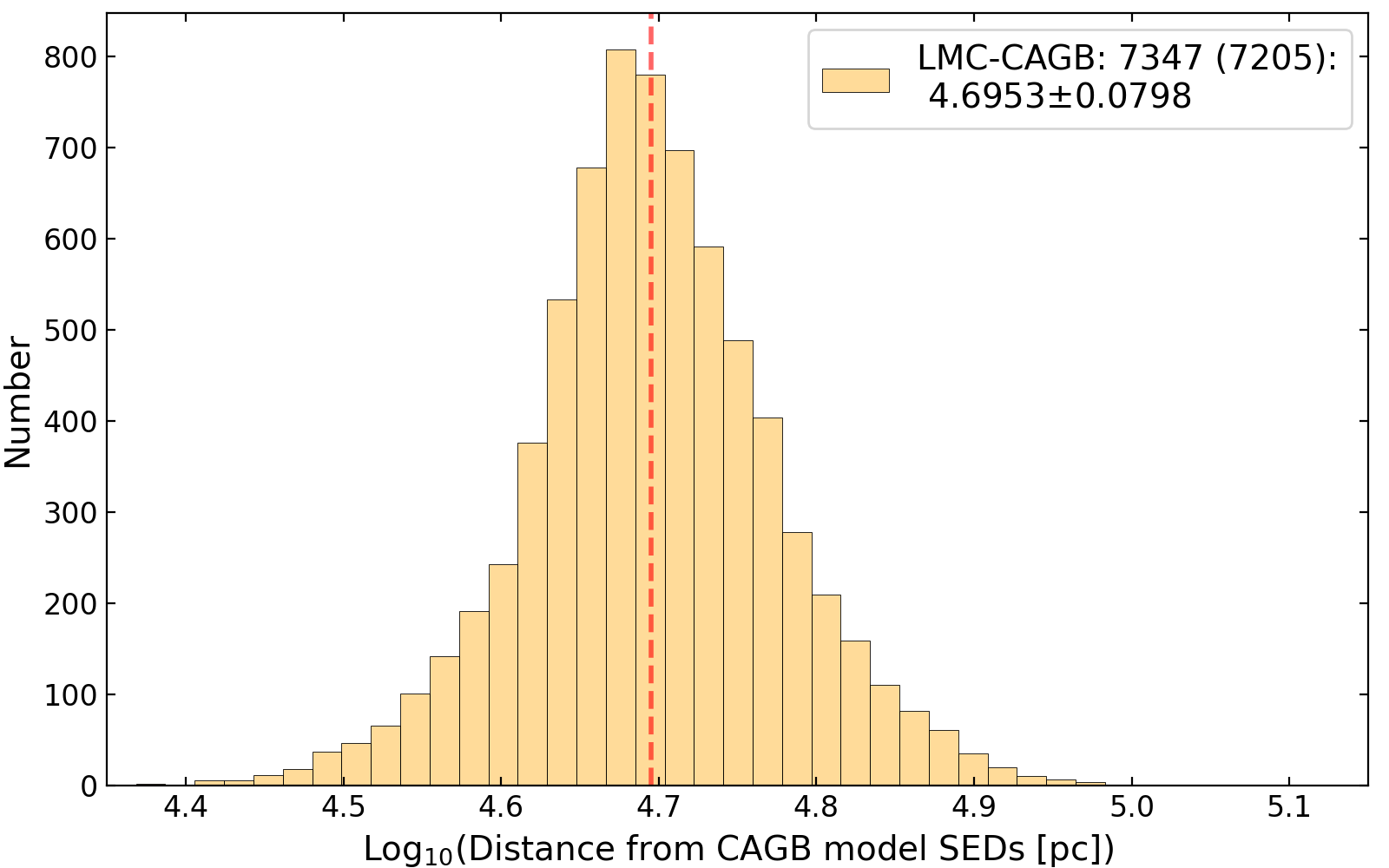}
\caption{The top panel shows CAGB model SEDs for seven representative models (see Table~\ref{tab:tab3}).
For LMC-CAGB stars, the figure shows the distance distribution (middle) and
the relations between dust optical depth and luminosity (bottom)
derived from CAGB model SEDs (Table~\ref{tab:tab3}).
See Section~\ref{sec:lum-lcagb} for details.
} \label{f1}
\end{figure}

\begin{figure*}
\centering
\smallplotfour{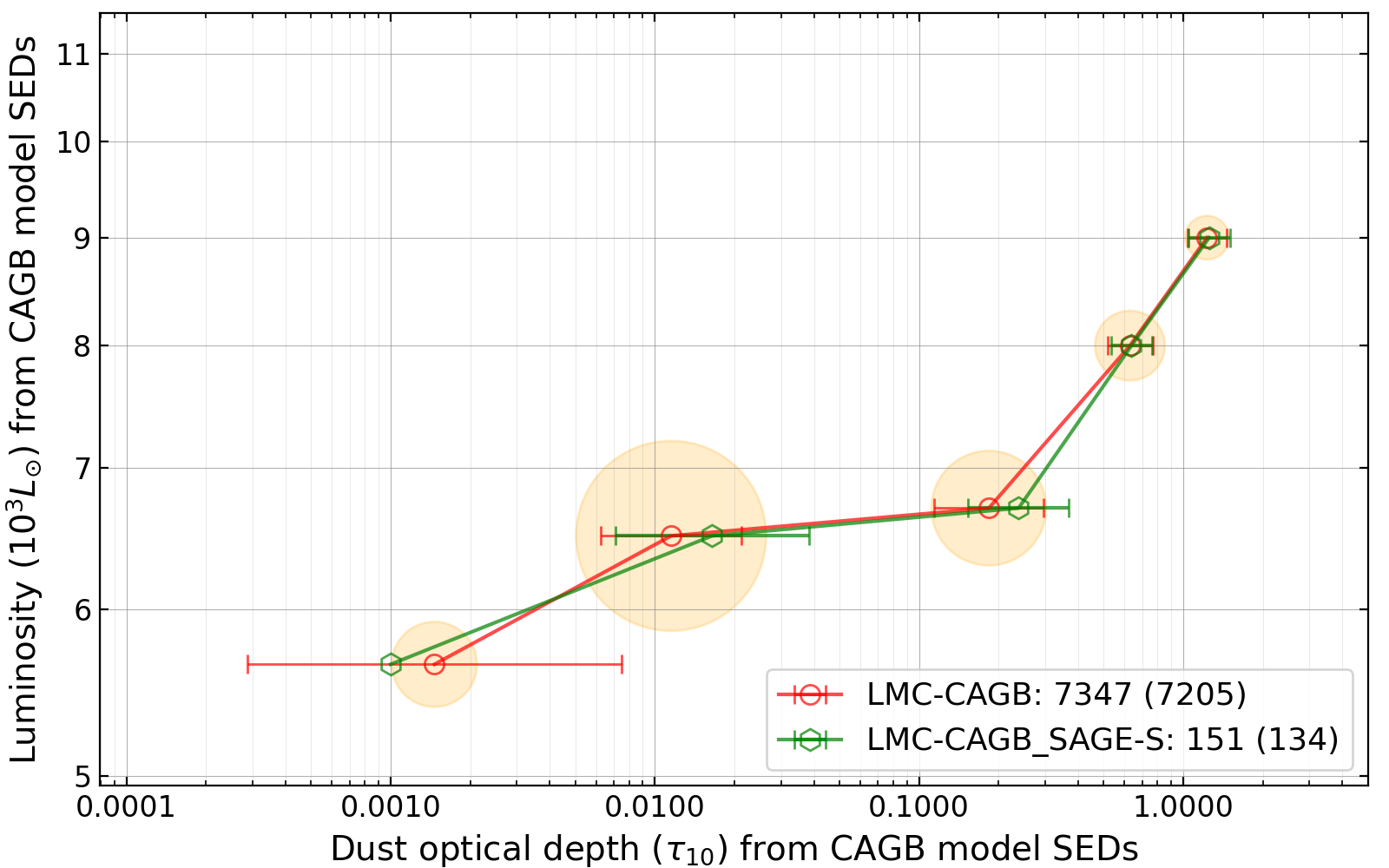}{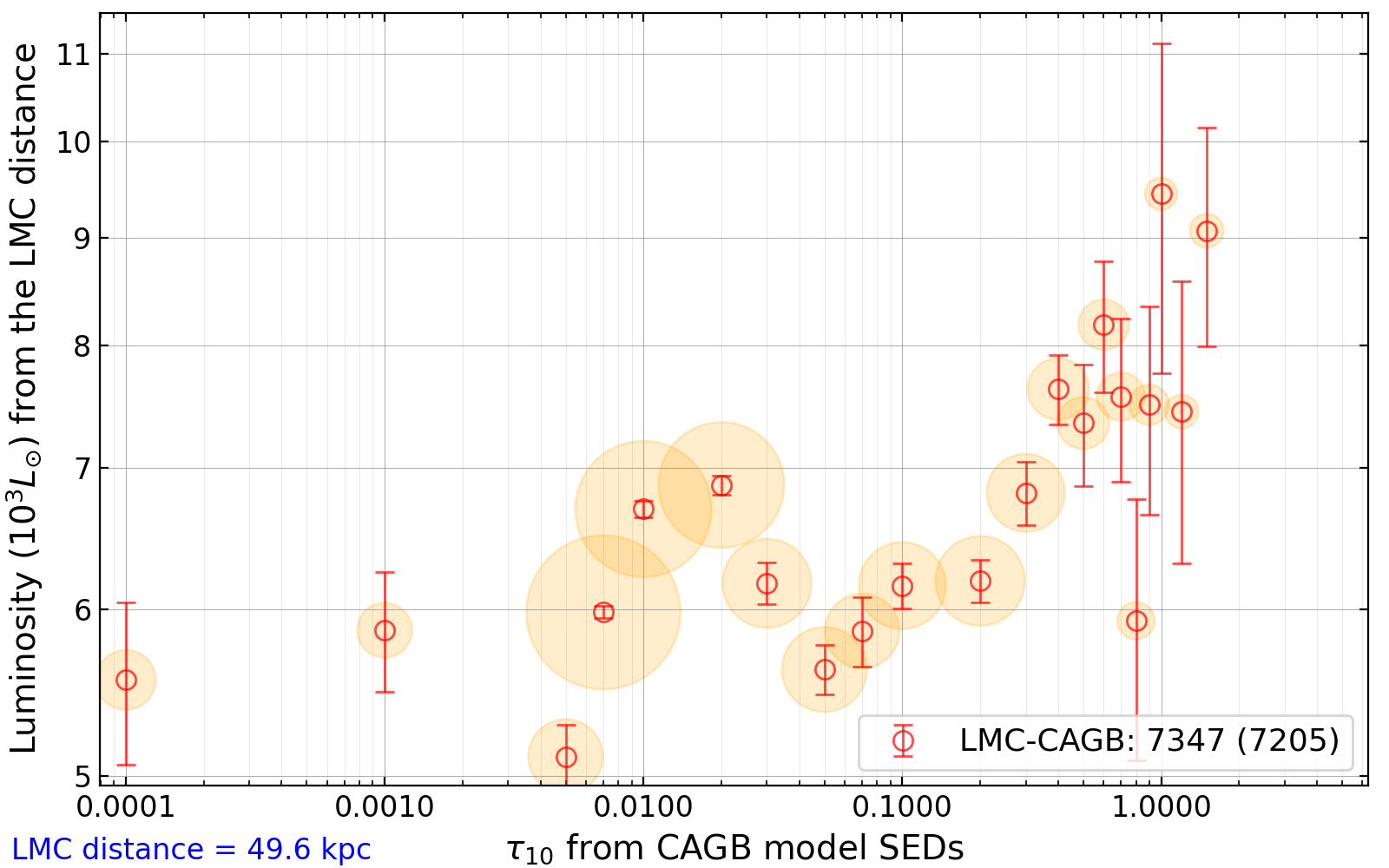}{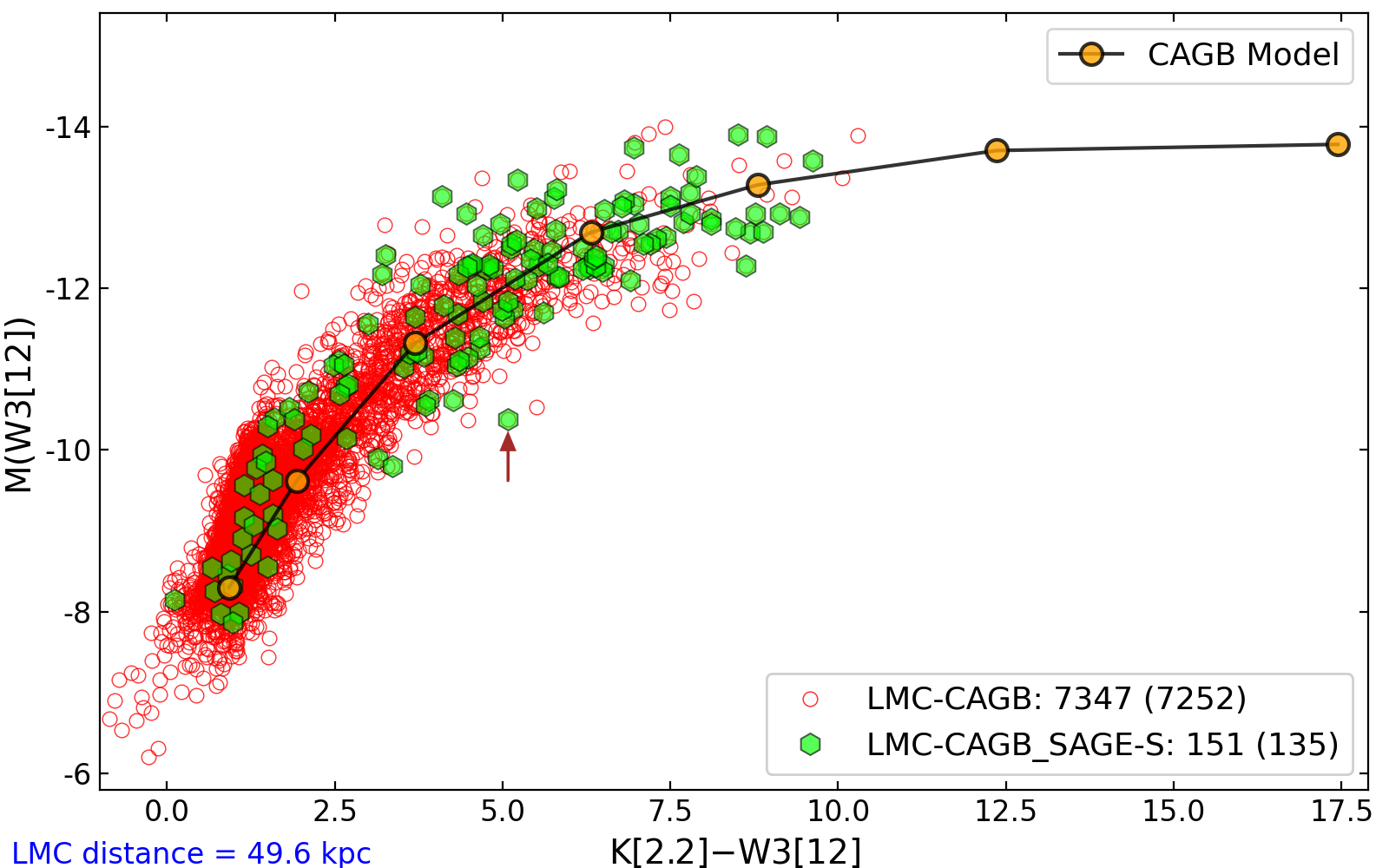}{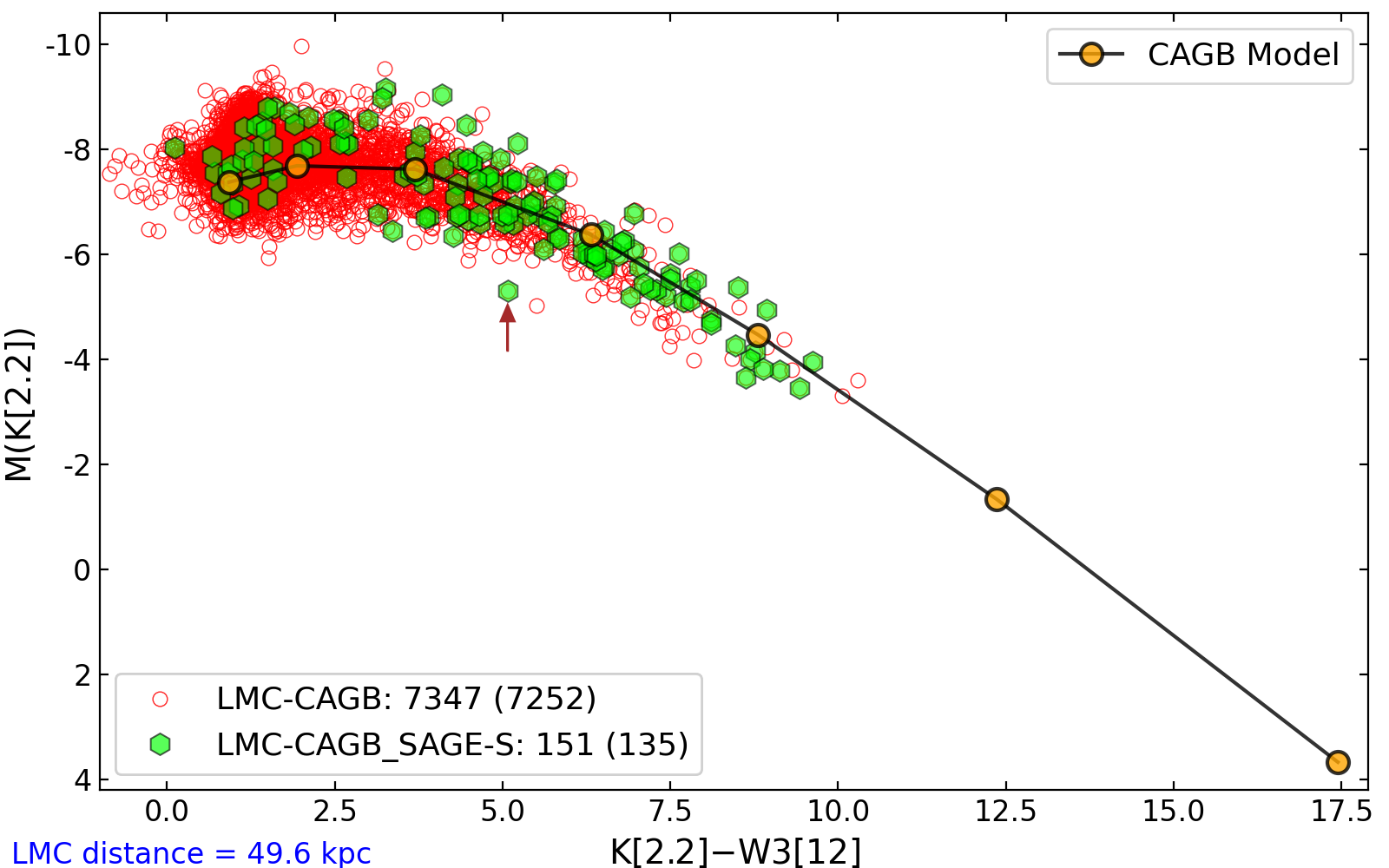}
\caption{
For LMC-CAGB stars, the top panels show the relations between optical depth and luminosity
derived using two different methods based on CAGB models.
In the top panels, the areas of the filled circular markers are proportional to the square root of the number of objects at each point.
The bottom panels present IR CMDs for LMC-CAGB stars, with the brown arrow indicating an object
likely to be an extrinsic carbon star (see Section~\ref{sec:sedlmc}).
For CAGB models (AMC, $T_c$ = 1000 K): $\tau_{10}$ = 0.0001, 0.01, 0.1, 0.5, 1, 2, and 4 from left to right (see Table~\ref{tab:tab3}).
The total number of objects in each subgroup is given, with the values in parentheses
denoting the number of plotted sources for which data are available.
See Section~\ref{sec:lum-lcagb} for details.}
\label{f2}
\end{figure*}

We began the procedure with the initial set of CAGB models for Galactic CAGB
stars derived by \citet{suh2024} and applied them to the LMC-CAGB stars, refining
and adjusting the model parameters. The sample was divided into five groups
according to the dust optical depth ($\tau_{10}$), and for each group we
determined the stellar temperature, spanning 2000–3000~K, that best reproduced
the observed SED shapes of the LMC-CAGB stars. This process yielded five
temperature groups: 3000, 2700, 2500, 2300, and 2200~K. Using the same five
$\tau_{10}$ groups, we then computed SED-based scaling distance distributions for
a range of stellar luminosities from $5\times10^{3}$ to
$1\times10^{4}$~$L_{\odot}$.

We adopted five luminosity groups (5.65, 6.5, 6.7, 8, and 9
$\times10^{3}$~$L_{\odot}$) that minimized the difference between the derived
scaling distances and the known LMC distance as the best-fitting values. Each
luminosity group corresponds to the same temperature group, except for the subset
with $\tau_{10}$ between 0.05 and 0.07 in the 2500 K temperature group, for which
the adopted luminosity is 6.5 $\times10^{3}$~$L_{\odot}$ instead of 6.7
$\times10^{3}$~$L_{\odot}$. In addition, we prepared a sixth group representing
Galactic CAGB stars with very large dust optical depths ($\tau_{10} > 1.5$). See
Table~\ref{tab:tab3} and the middle panel of Figure~\ref{f1}.

The LMC-CAGB stars lack objects with very large dust optical depths ($\tau_{10}
> 1.5$), whereas many such objects are present among Galactic CAGB stars (e.g.,
\citealt{suh2024}), for which a stellar temperature of 2000 K reproduces the
observed SEDs well. For these heavily obscured objects, we extended the trend
derived for lower optical depths in the LMC to a sixth group with a stellar
temperature of 2000 K and a luminosity of $1.0\times10^{4}$~$L_{\odot}$ (see
Table~\ref{tab:tab3}).

Table~\ref{tab:tab3} lists the final set of the 37 CAGB models that best
reproduced the observed SED shapes of LMC-CAGB stars and minimized the
differences between the scaling and true distances. The models are arranged in
order of increasing $\tau_{10}$ and include the corresponding dust opacity and
stellar parameters ($T_*$ and $L_*$). The top panel of Figure~\ref{f1} presents
SEDs for seven representative models.

The middle panel of Figure~\ref{f1} shows the relation between distance and
luminosity, both derived from the CAGB model SEDs (Table~\ref{tab:tab3}). Across
the five luminosity groups, the derived distances are consistently in good
agreement with the true distance of the LMC. In this diagram, the areas of the
filled circular markers are proportional to the square root of the number of
objects at each point.

The best-fitting CAGB model (from Table~\ref{tab:tab3}) and the corresponding
scaling distance were derived for each LMC-CAGB star. Reliable results were
obtained for 7205 of the 7347 LMC-CAGB stars (see Section~\ref{sec:seds}). The
bottom panel of Figure~\ref{f1} shows the logarithmic distribution of the derived
distances for the 7205 LMC-CAGB stars. The median distance (49.6 kpc) agrees well
with the established value for the LMC, and the small dispersion in $\log_{10} d$
(0.0798 dex) demonstrates the consistency and reliability of the CAGB models
(Table~\ref{tab:tab3}).

The upper panels of Figure~\ref{f2} show the relations between optical depth and
luminosity for the LMC-CAGB stars, derived using two different methods based on
the CAGB model SEDs. The upper-left panel shows the relation obtained using the
luminosities derived by the method described in the previous paragraphs (see
Table~\ref{tab:tab3}; Figure~\ref{f1}). In contrast, the upper-right panel
presents an alternative approach in which the luminosity was determined from the
derived distance scaled to the true LMC distance of 49.6 kpc. This method reveals
the detailed luminosity distribution for each $\tau_{10}$, which exhibits larger
scatter. When the data from the five luminosity groups based on $\tau_{10}$ (as
listed in Table~\ref{tab:tab3}) are combined, both methods yield similar results.
In these diagrams, the areas of the filled circular markers are proportional to
the square root of the number of objects at each point.

The bottom panels of Figure~\ref{f2} present WISE–2MASS CMDs for LMC-CAGB stars,
plotting the absolute magnitudes M(W3[12]) and M(K[2.2]) against the
K[2.2]$-$W3[12] color. The observed absolute magnitudes, computed assuming a
distance of 49.6 kpc, agree reasonably well with the CAGB models.

We find that M(W3[12]) traces the absolute luminosity reasonably well and can
serve as a useful proxy for the luminosity function. In contrast, M(K[2.2])
exhibits fainter magnitudes for brighter CAGB stars with larger dust optical
depths, because circumstellar dust absorbs shorter-wavelength light and re-emits
it at longer wavelengths. Therefore, absolute magnitudes in NIR bands or at
shorter wavelengths cannot be used as reliable proxies for the luminosity
function for CAGB stars.

The dispersion in $\log_{10} d$ (0.0798 dex; corresponding to roughly 20 \% in
distance; bottom panel of Figure~\ref{f1}) for LMC-CAGB stars likely arises from
three factors: (1) positional variations within the LMC, whose radius is about 10
\% of its mean distance, implying a $\sim$5 \% dispersion in distance; (2) the
fact that a given $\tau_{10}$ value can correspond to multiple luminosities, as
stars with different initial masses may exhibit different luminosities; and (3)
intrinsic luminosity variations caused by the pulsations of CAGB stars. The
observed dispersion in $\log_{10} d$ can therefore be regarded as representing
the overall uncertainty in luminosity, since the contribution of the first factor
is relatively minor. The measured dispersion in $\log_{10} d$ corresponds to a
luminosity dispersion of 0.160 dex for the CAGB models (see
Table~\ref{tab:tab3}).

\begin{figure*}
\centering
\smallplotsix{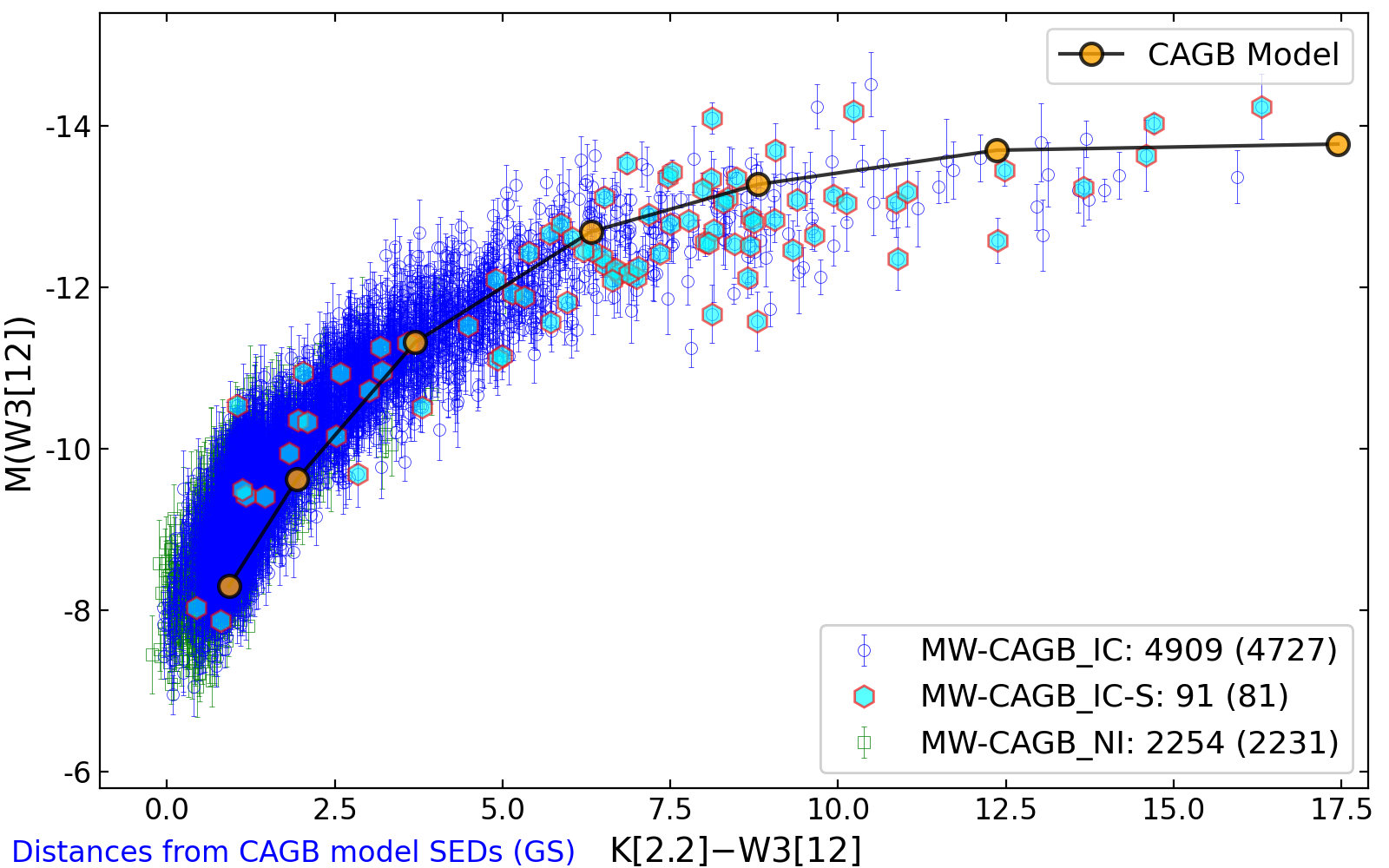}{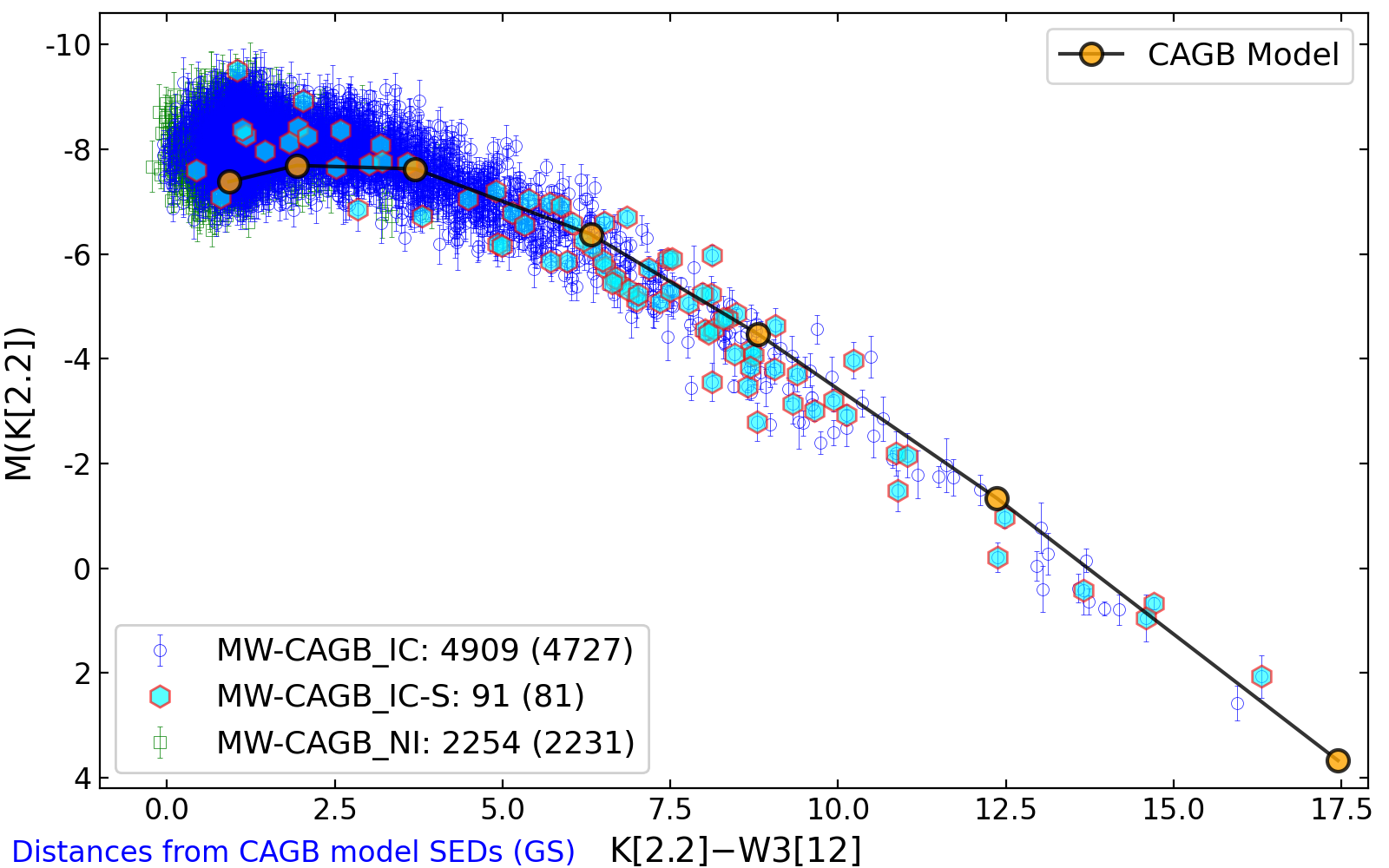}{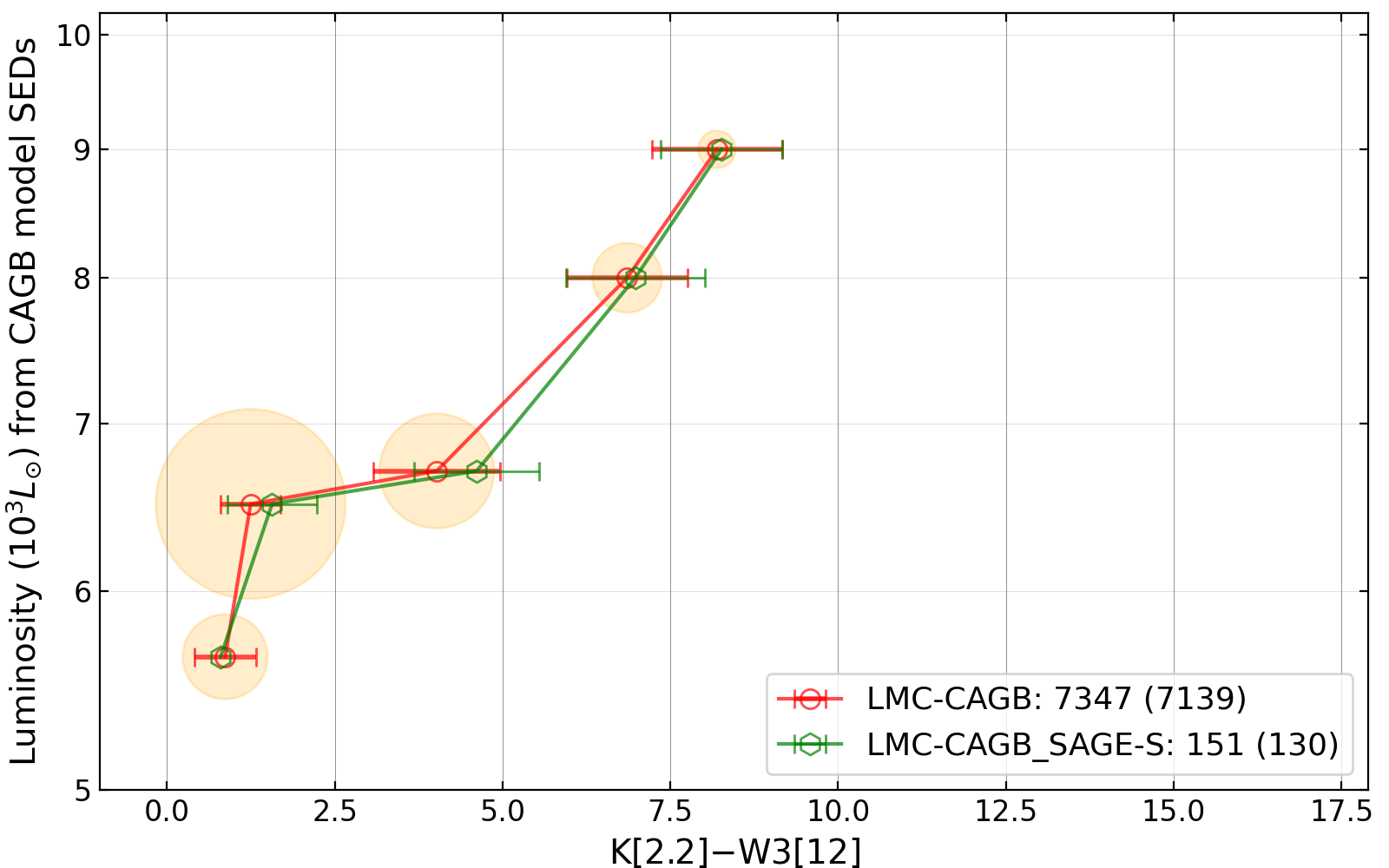}{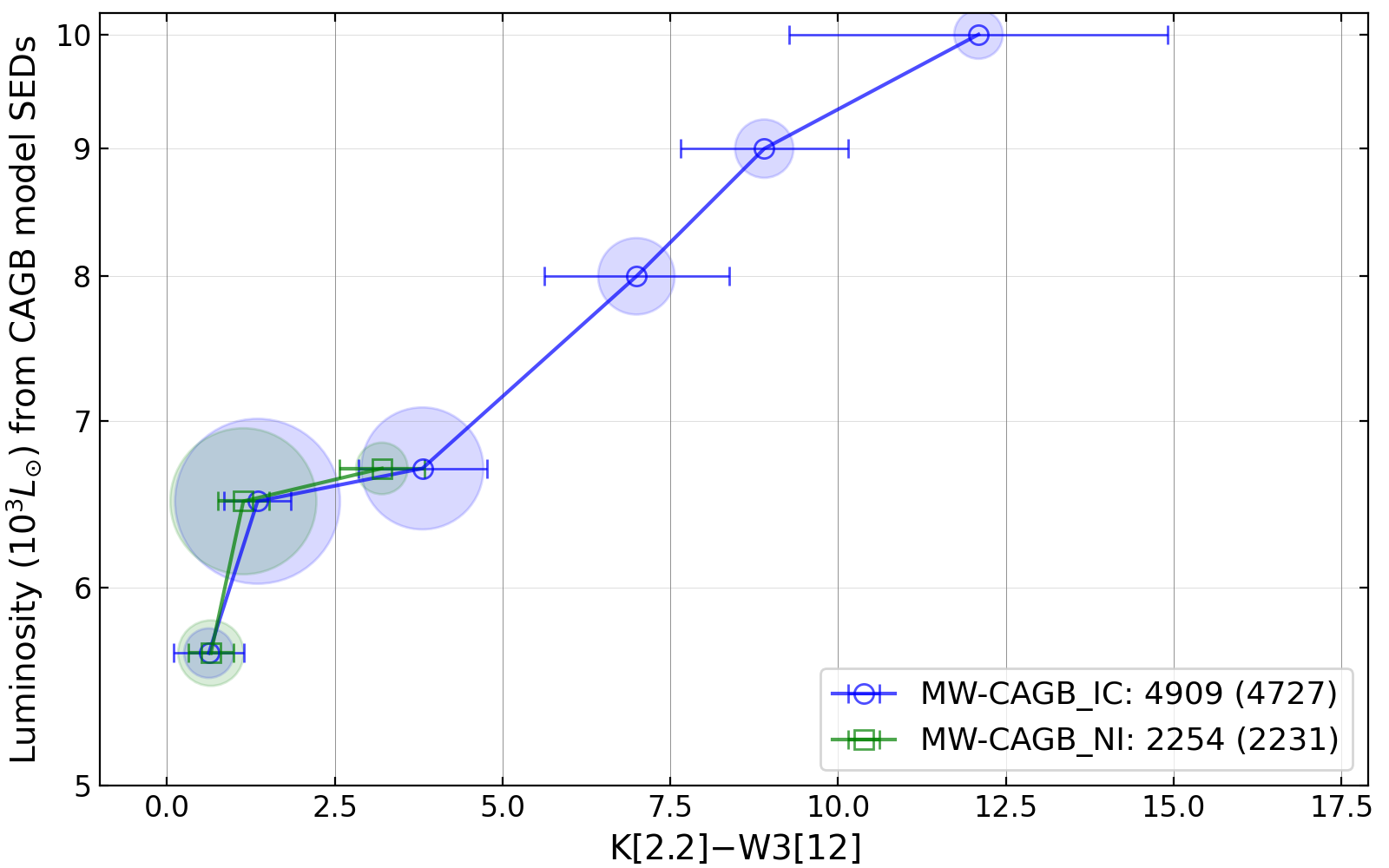}{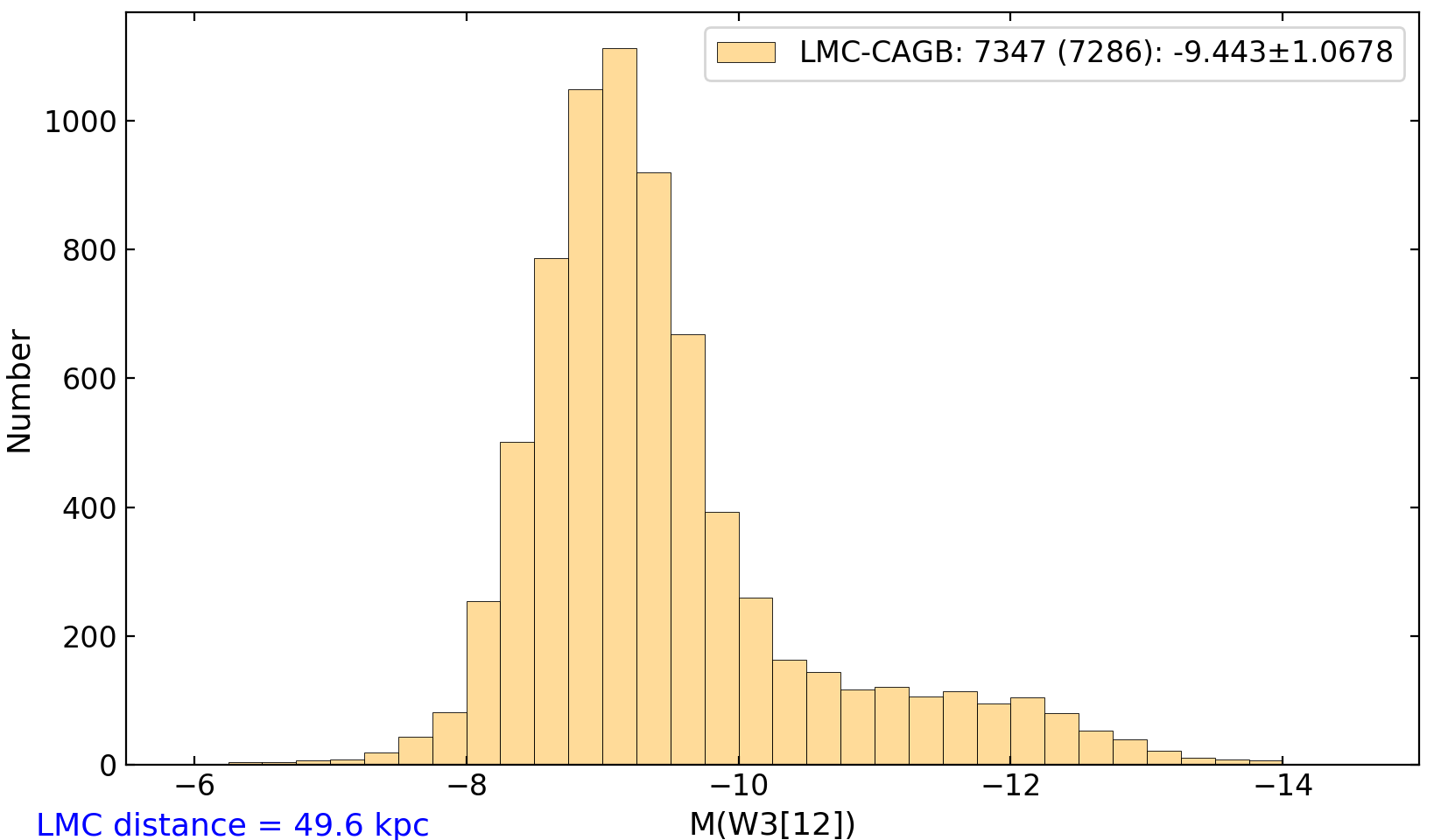}{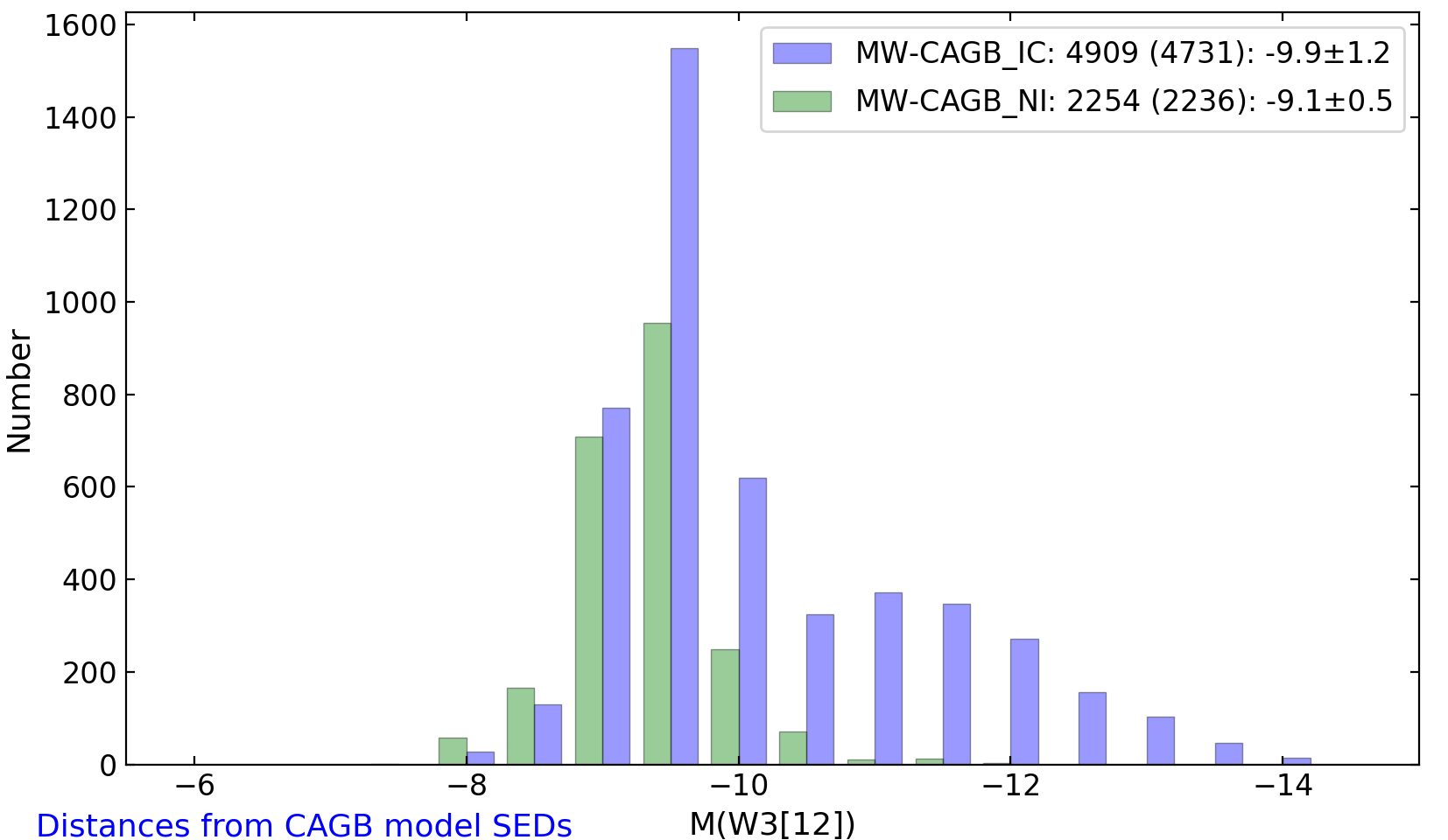}
\caption{The top panels show IR CMDs for Galactic CAGB stars.
For Galactic CAGB stars, distances were derived from CAGB model SEDs, with Gaussian Monte Carlo scatter (GS)
in $\log_{10} d$ ($\sigma$=$\pm$0.0798 dex; obtained from LMC-CAGB stars; see Figure~\ref{f1}).
For CAGB stars in the LMC and the Milky Way, the middle panels show the relations between
optical depth and luminosity, and the bottom panels show the luminosity functions.
In the middle panels, the areas of the filled circular markers are proportional to the square root of the number of objects at each point.
The total number of objects is given, with the values in parentheses
denoting the number of plotted sources for which data are available.
See Section~\ref{sec:lum-gcagb} for details.
} \label{f3}
\end{figure*}

\subsection{Application to Galactic CAGB stars}\label{sec:lum-gcagb}

After confirming the reliability and uncertainties of the CAGB models for
LMC-CAGB stars, we extended the analysis to the Milky Way sample (7163 MW-CAGB
objects; Table~\ref{tab:tab1}), using the 37 CAGB models listed in
Table~\ref{tab:tab3} with observed SEDs (see Sections~\ref{sec:seds} and
\ref{sec:sedmw} for details).

The top panels of Figure~\ref{f3} present the WISE–2MASS CMDs for Galactic CAGB
stars, plotting M(W3[12]) and M(K[2.2]) against the K[2.2]$-$W3[12] color.
Distances were derived by fitting the observed SEDs with the CAGB models (see
Table~\ref{tab:tab3}; Section~\ref{sec:seds}). To account for uncertainties in
the luminosity of the CAGB models for Galactic CAGB stars, we applied Gaussian
Monte Carlo scatter (GS) in distance ($\log_{10} d$) with $\sigma = \pm 0.0798$
dex, consistent with the dispersion measured for LMC-CAGB stars (see the right
panel of Figure~\ref{f1}). CAGB stars in the LMC and the Milky Way exhibit very
similar distributions on the CMDs (see Figures~\ref{f2} and \ref{f3}) and show
good overall agreement with the CAGB models, displaying a comparable degree of
spread.

The middle panels of Figure~\ref{f3} show the relations between the observed
color (K[2.2]$-$W3[12]) and the luminosity derived from the CAGB model SEDs
(Table~\ref{tab:tab3}) for CAGB stars in the LMC and the Milky Way. Objects with
higher derived luminosities exhibit redder observed colors, consistent with the
characteristic behavior of the CAGB models, in which larger dust optical depths
correspond to higher luminosities (see Table~\ref{tab:tab3}). The number
distributions reveal both similarities and differences between the two galaxies.
Although the overall trends are similar, Galactic CAGB stars display a broader
range of colors than the LMC-CAGB stars. In particular, the LMC lacks objects
with very red K[2.2]$-$W3[12] colors corresponding to large dust optical depths
($\tau_{10} > 1.5$) at brighter luminosities.

The bottom panels of Figure \ref{f3} show the W3[12] band luminosity functions
for CAGB stars in both galaxies. They exhibit similar overall shapes, each
resembling a combination of two roughly Gaussian components: a main one with
M(W3[12]) between $-8$ and $-10.5$ mag, and a secondary one comprising luminous
sources with M(W3[12]) between $-10.5$ and $-14$ mag. Compared with the Milky
Way, the LMC contains fewer of these luminous sources (M(W3[12]) $< -10.5$ mag),
which correspond to “extreme carbon stars” representing the late stage of the
CAGB phase (see \citealt{gruendl2008}). Among Galactic CAGB stars, MW-CAGB\_NI
objects, which are generally fainter than MW-CAGB\_IC objects (see
Section~\ref{sec:mw}), are rare at this bright end of the luminosity function.

\subsection{Limitations of the Theoretical Models\label{sec:limit}}

The dust shell models adopted in this study do not account for gas-phase
radiative processes (see Section~\ref{sec:models}). However, CAGB stars exhibit a
variety of gas-phase emission and absorption features in the NIR and MIR bands
arising from circumstellar molecules such as C$_2$, C$_2$H$_2$, HCN, and CN
(e.g., \citealt{LeBertre2005}). Consequently, the discrepancies between the
models and observations are expected to be more pronounced at wavelengths where
gas-phase radiation is significant.

Moreover, the spherical symmetry assumed in the dust shell models neglects the
possibility of nonspherical dust geometries. As a result, CAGB stars with
asymmetric dust envelopes can display markedly different SEDs that diverge
substantially from the predictions of the theoretical models (e.g.,
\citealt{suh2016}).

\section{SEDs and Distances of CAGB stars}\label{sec:seds}

\citet{suh2024} compared observed SEDs with theoretical models for a subset of
Galactic CAGB stars. In this work, we extend that analysis to the full samples of
CAGB stars in both the LMC and the Milky Way (see Table~\ref{tab:tab1}), using
the radiative transfer models described in Section~\ref{sec:models}.

For constructing the observed SEDs, photometric data were collected from Gaia
DR3, 2MASS, IRAS, MSX, AKARI, Spitzer, and AllWISE. For LMC-CAGB stars, Spitzer
IRS (\citealt{jones2017}) spectral data were included when available. For
Galactic CAGB stars, IR spectra from the IRAS Low Resolution Spectrograph (LRS;
\citealt{kwok1997}) and ISO Short Wavelength Spectrometer (SWS;
\citealt{sylvester1999}) were used when accessible.

The observed SEDs of LMC-AGB stars were fitted using a least-squares optimization
method, resulting in the 37 CAGB models (Table~\ref{tab:tab3}; see
Section~\ref{sec:lum-lcagb} and Figure~\ref{f1}) that best reproduced the
observed SED shapes and minimized the differences between the scaling and true
distances of LMC-CAGB stars. These LMC-validated CAGB models were then applied to
the full sample of Galactic CAGB stars.

For the complete samples of CAGB stars in the LMC and the Milky Way, the
best-fitting CAGB model ($\tau_{10}$) and the corresponding scaling distance,
together with its uncertainty, were derived through least-squares optimization
using the set of 37 CAGB models (Table~\ref{tab:tab3}). To ensure the reliability
of the results, only fits with a coefficient of determination ($R^2$) exceeding
0.6 between the observed and model SEDs were retained. Applying this criterion,
reliable best-fitting CAGB models and scaling distances were obtained for 7205 of
the 7347 LMC-CAGB stars (98.1 \%), 4884 of the 4909 MW-CAGB\_IC stars (99.5 \%),
and 2238 of the 2254 MW-CAGB\_NI stars (99.3 \%).

Figure~\ref{f4} compares the observed SEDs with CAGB models (see
Table~\ref{tab:tab3}) for eight representative CAGB stars in the LMC (upper four
panels) and the Milky Way (lower four panels), arranged in order of increasing
dust optical depth ($\tau_{10}$). The comparisons show an overall similarity in
the observed SEDs of CAGB stars in the LMC and the Milky Way.

For each panel in Figure~\ref{f4}, the $\tau_{10}$ value and the corresponding
scaling distance are shown in the bottom-right corner. A suffix “s2” on the
$\tau_{10}$ values indicates that the dust opacity is modeled with a simple
mixture of AMC and SiC (20 \% by mass). The SiC feature at 11.3 $\mu$m is
reasonably well reproduced by the CAGB models. In general, many CAGB stars with
moderate dust optical depths ($\tau_{10}=0.007$–0.5) exhibit prominent SiC
features, whereas those with dust optical depths outside this range tend not to,
consistent with previous studies (e.g., \citealt{suh2000}).

\subsection{SEDs of LMC-CAGB stars}\label{sec:sedlmc}

\begin{figure*}
\centering
\smallploteight{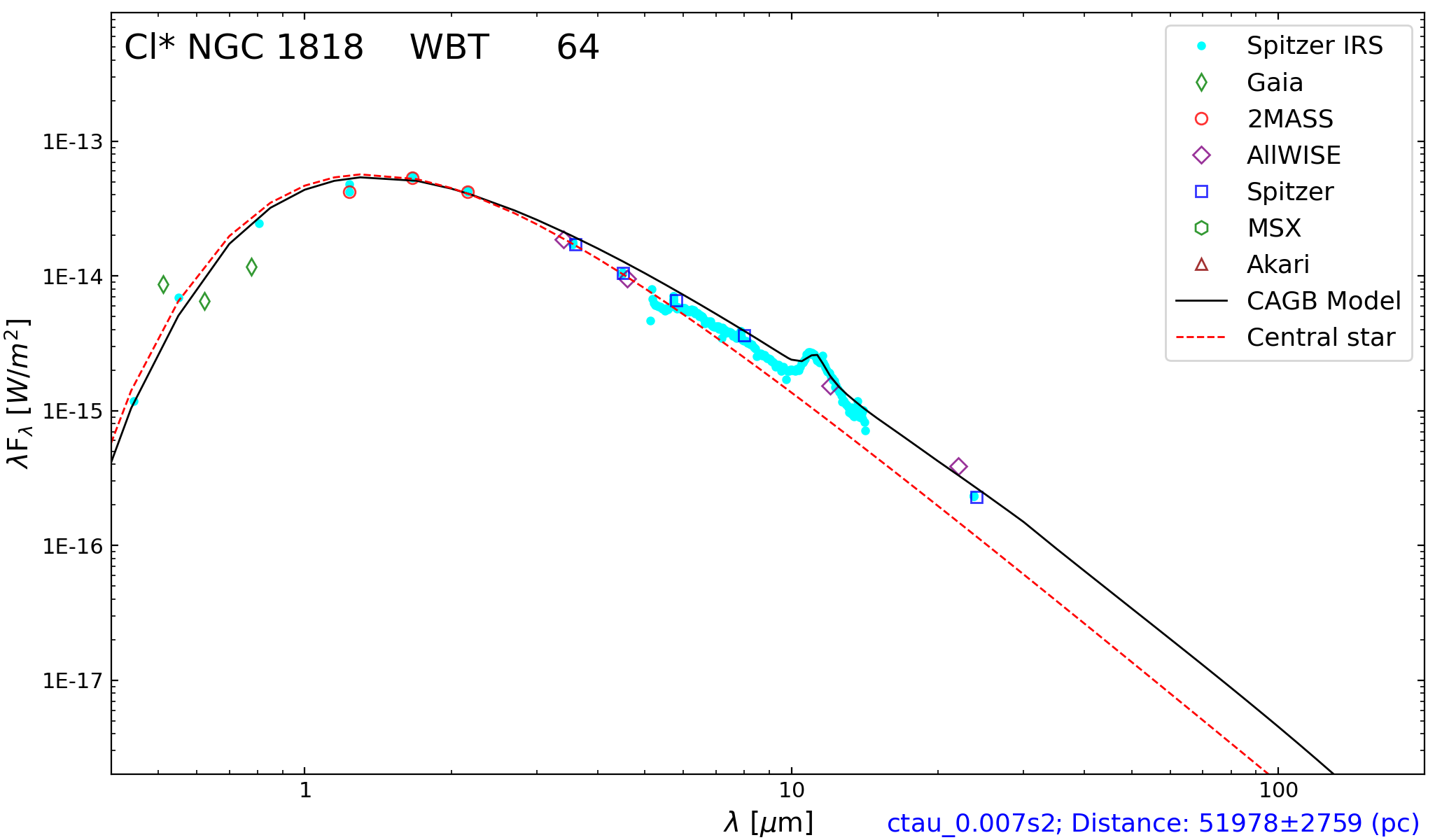}{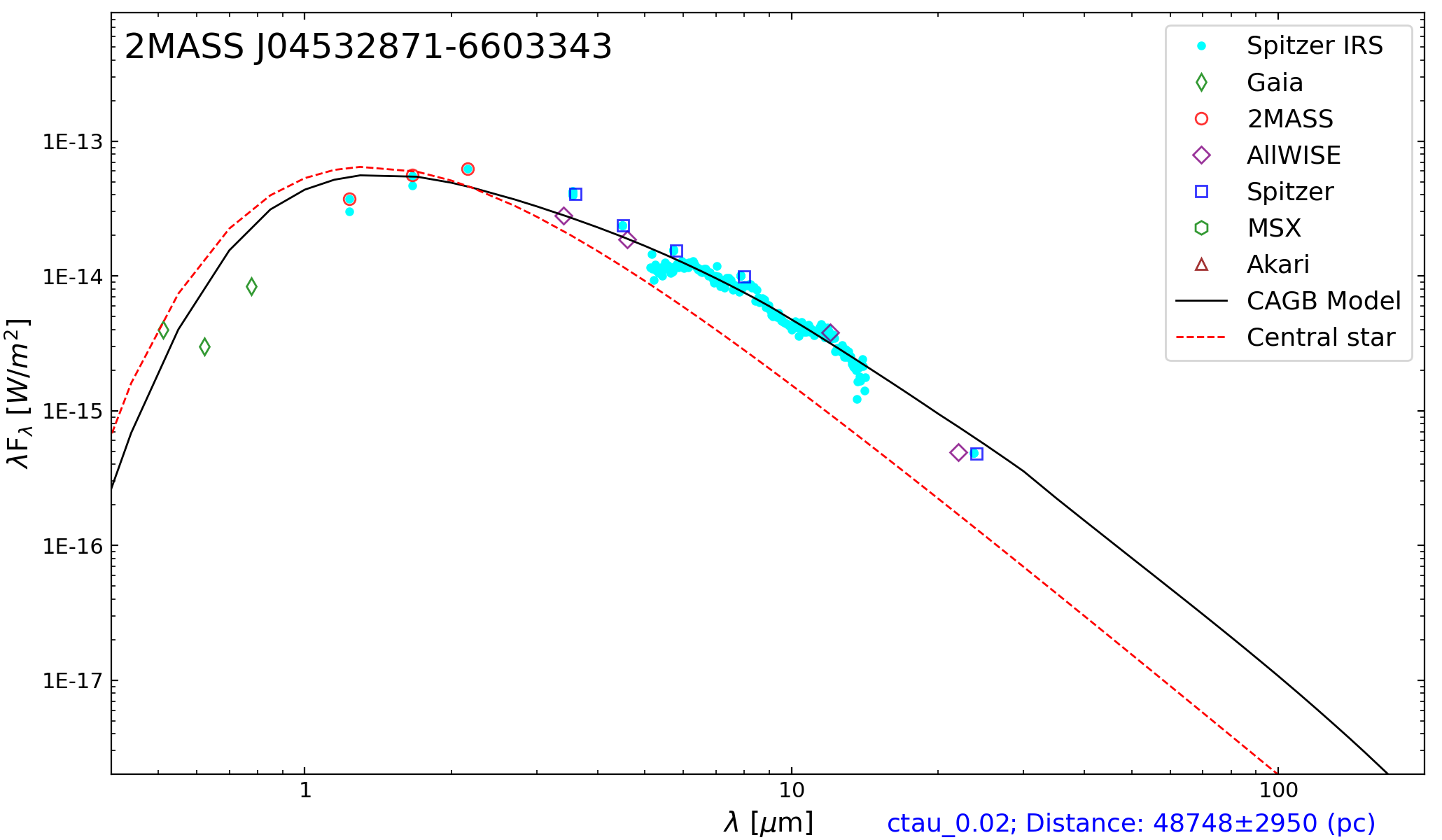}{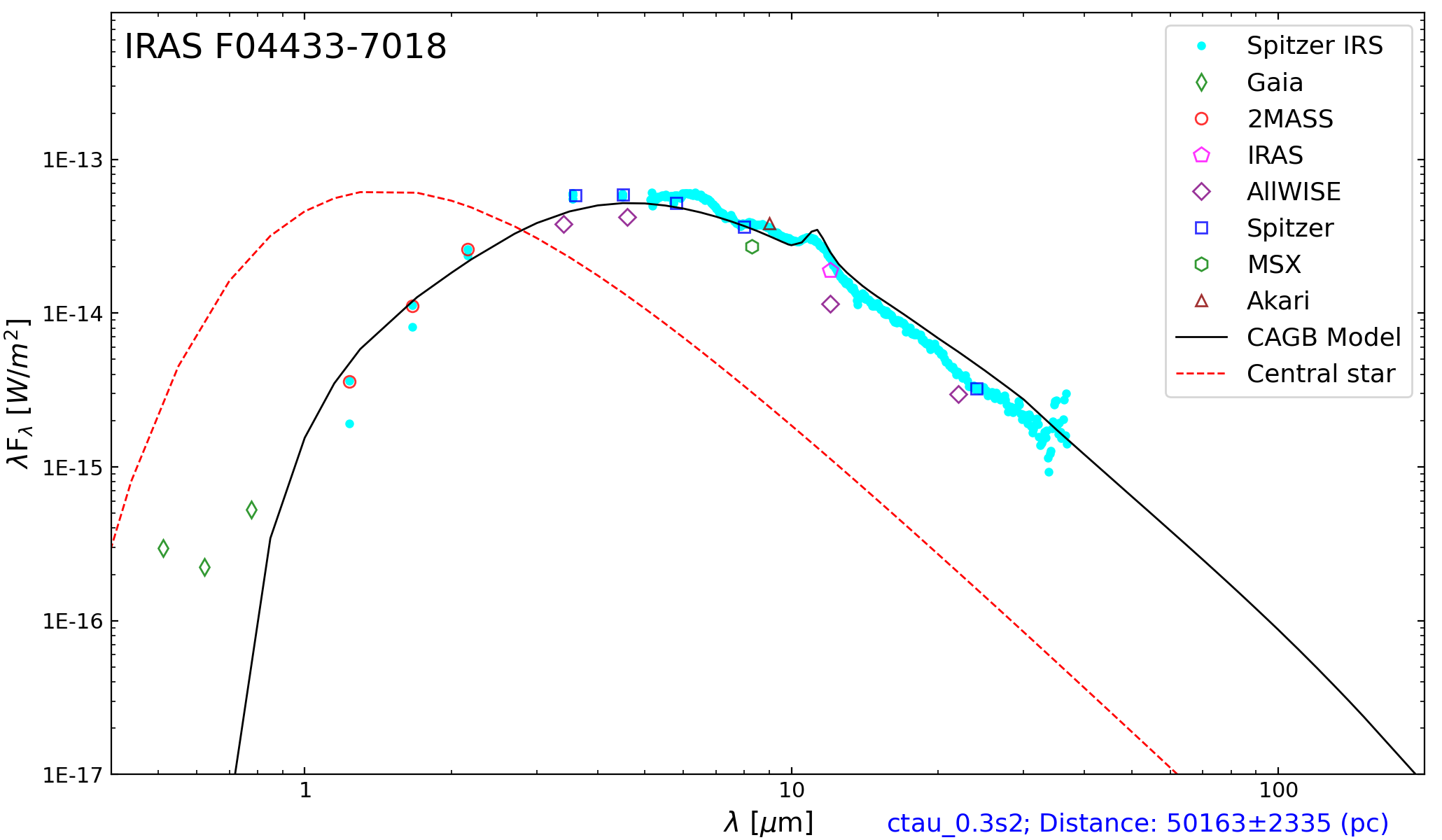}{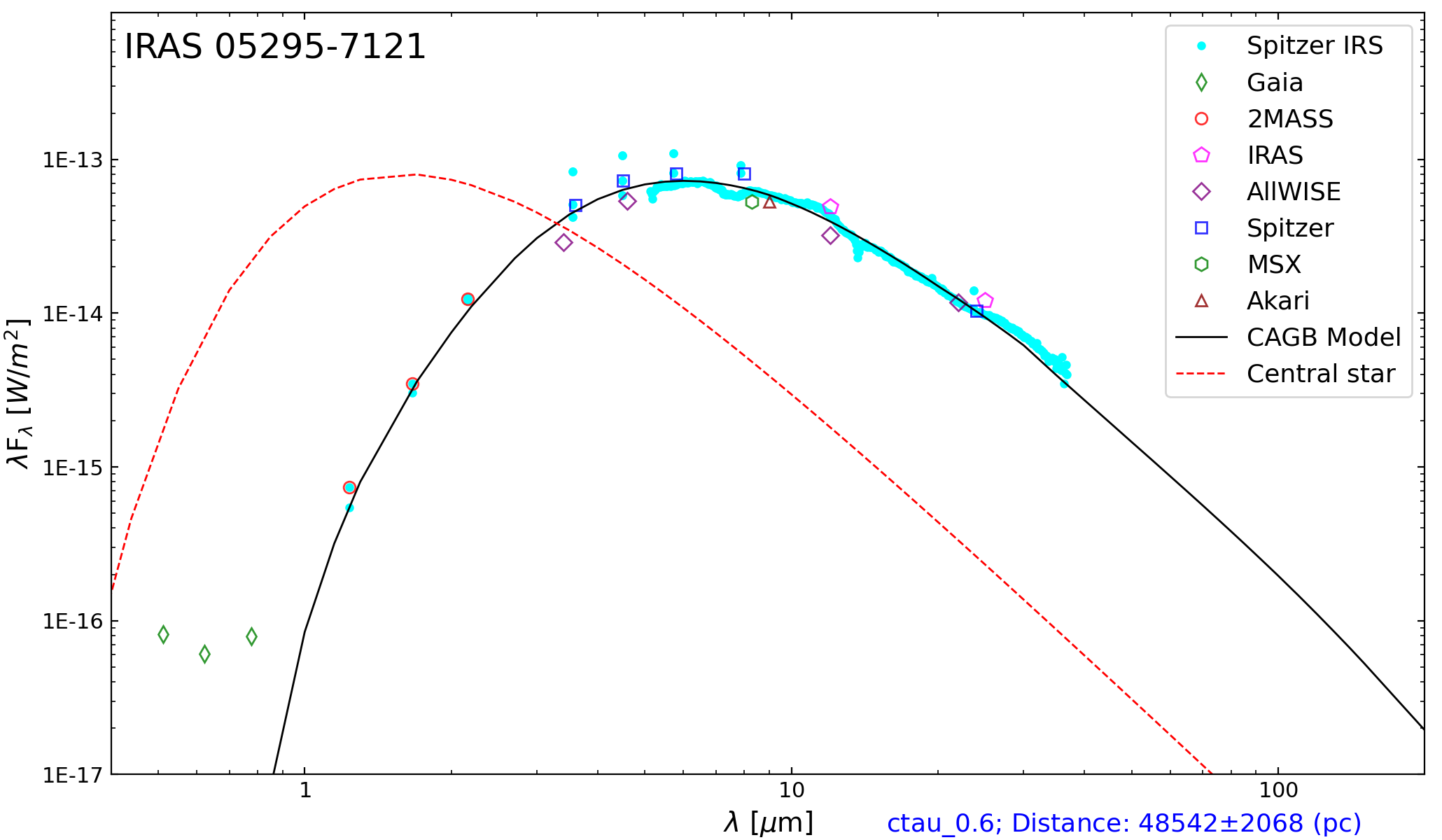}{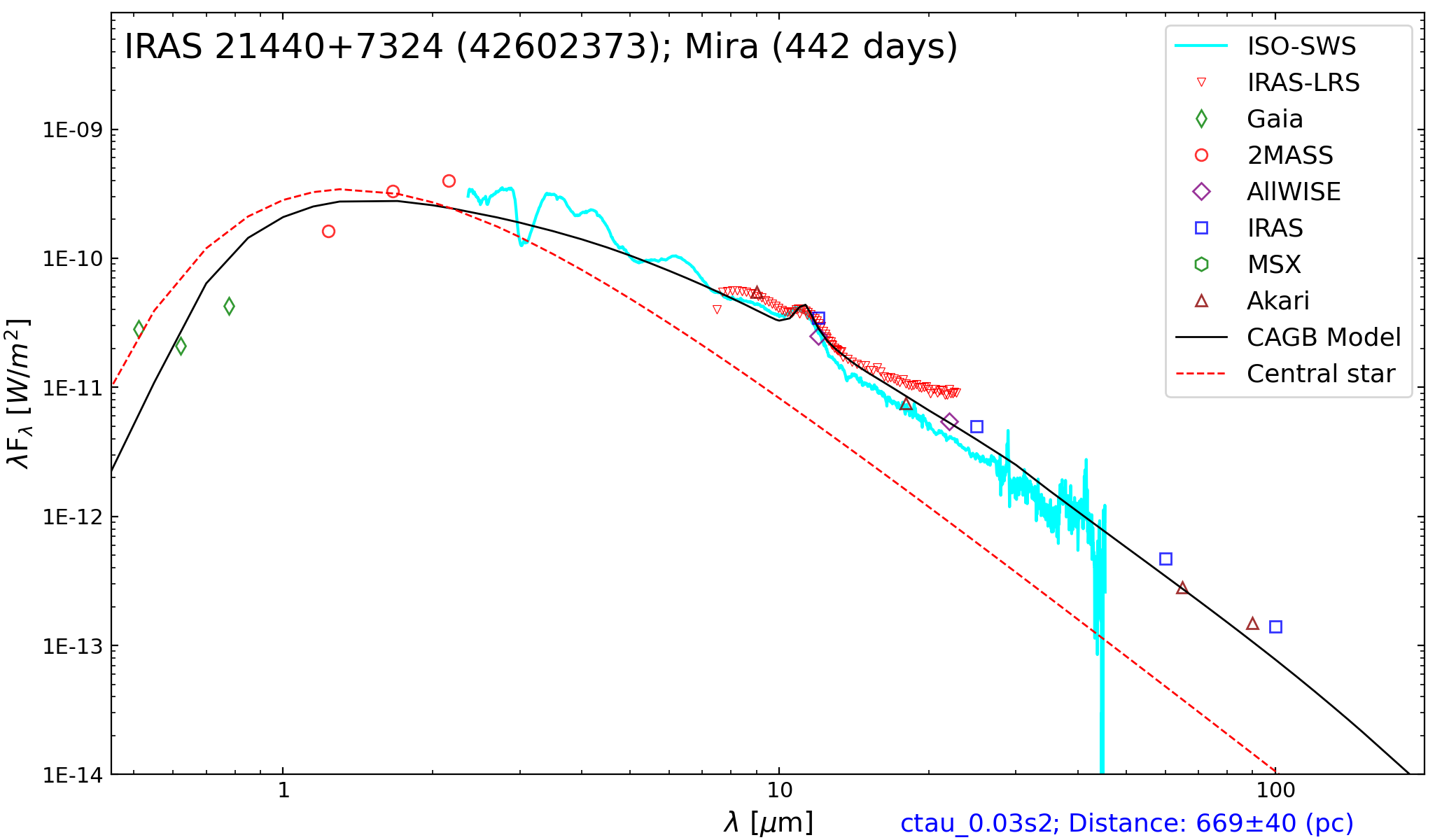}{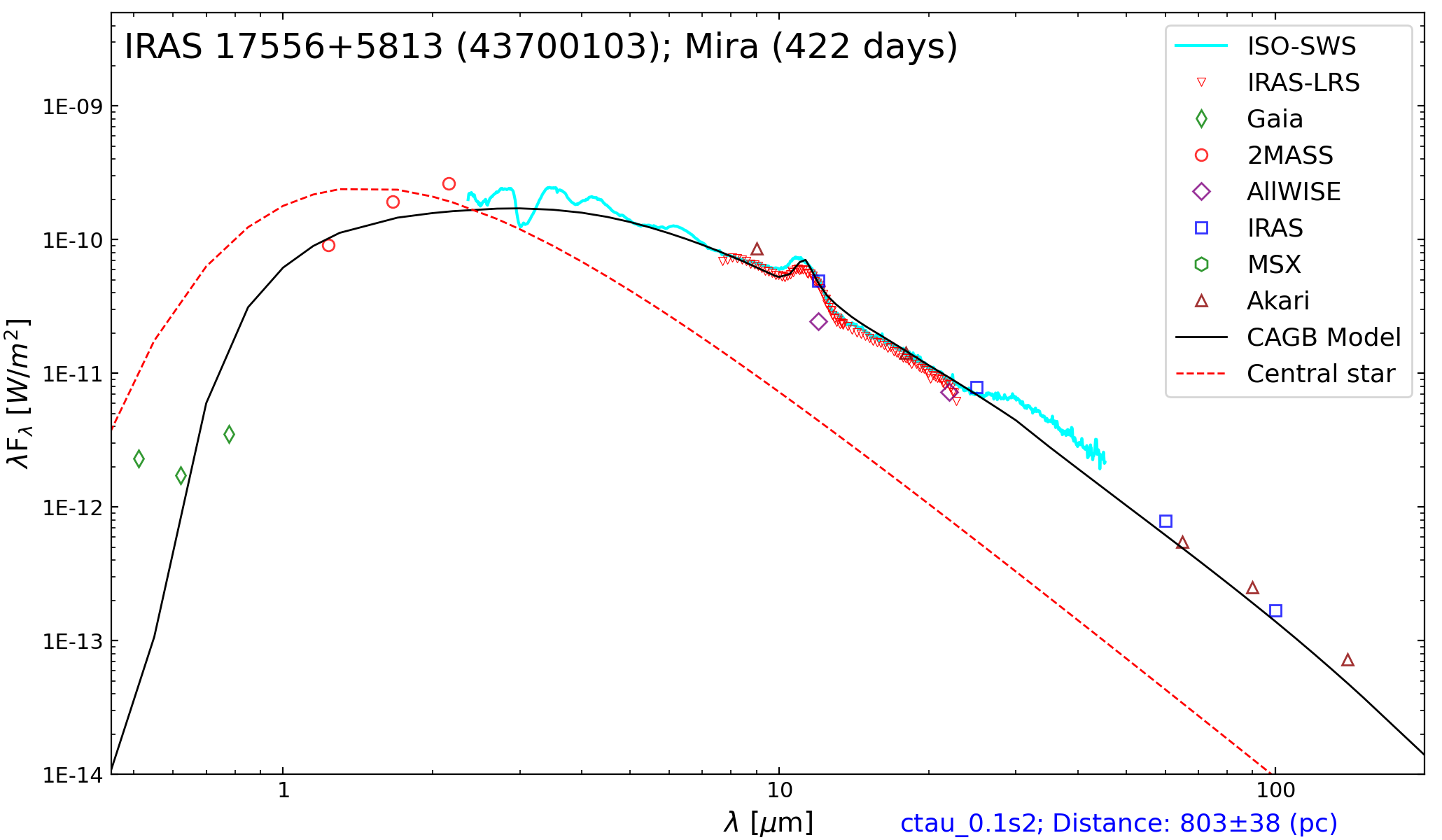}{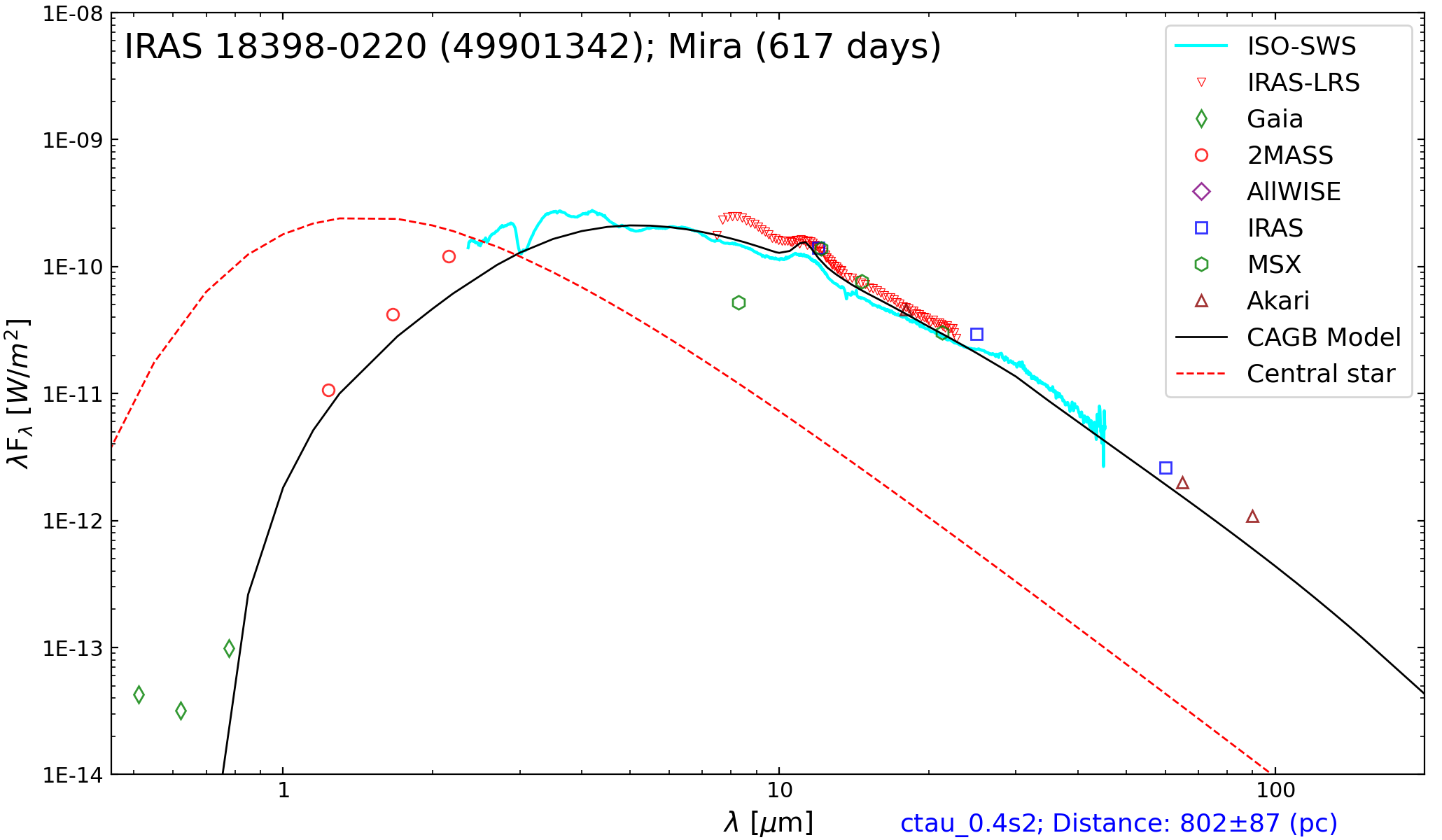}{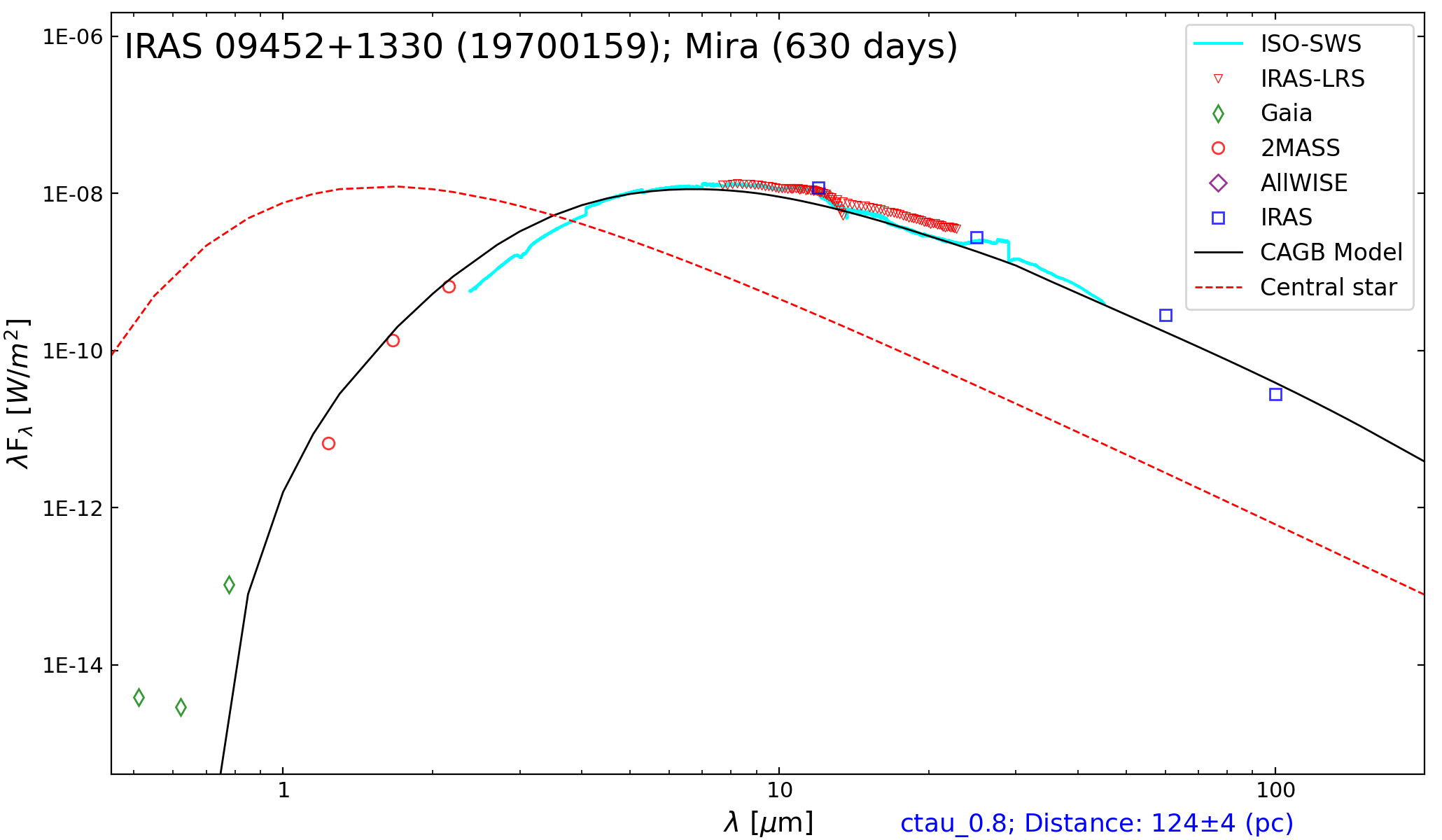}
\caption{
Observed and model SEDs for CAGB stars.
The upper four panels show LMC stars, and the lower four show Milky Way stars.
Each panel indicates the $\tau_{10}$ value and the corresponding scaling distance (bottom right).
A suffix “s2” on $\tau_{10}$ denotes AMC+SiC dust opacity (20 \% SiC).
See Section~\ref{sec:seds} for details.}\label{f4}
\end{figure*}

The upper four panels of Figure~\ref{f4} show the SEDs of four representative
LMC-CAGB stars belonging to the subgroup LMC-CAGB\_SAGE-S, for which infrared
spectral data (Spitzer IRS) are available. The observed SEDs are reasonably well
reproduced by the theoretical model SEDs (Table~\ref{tab:tab3}).

\begin{figure*}
\centering
\smallplottwo{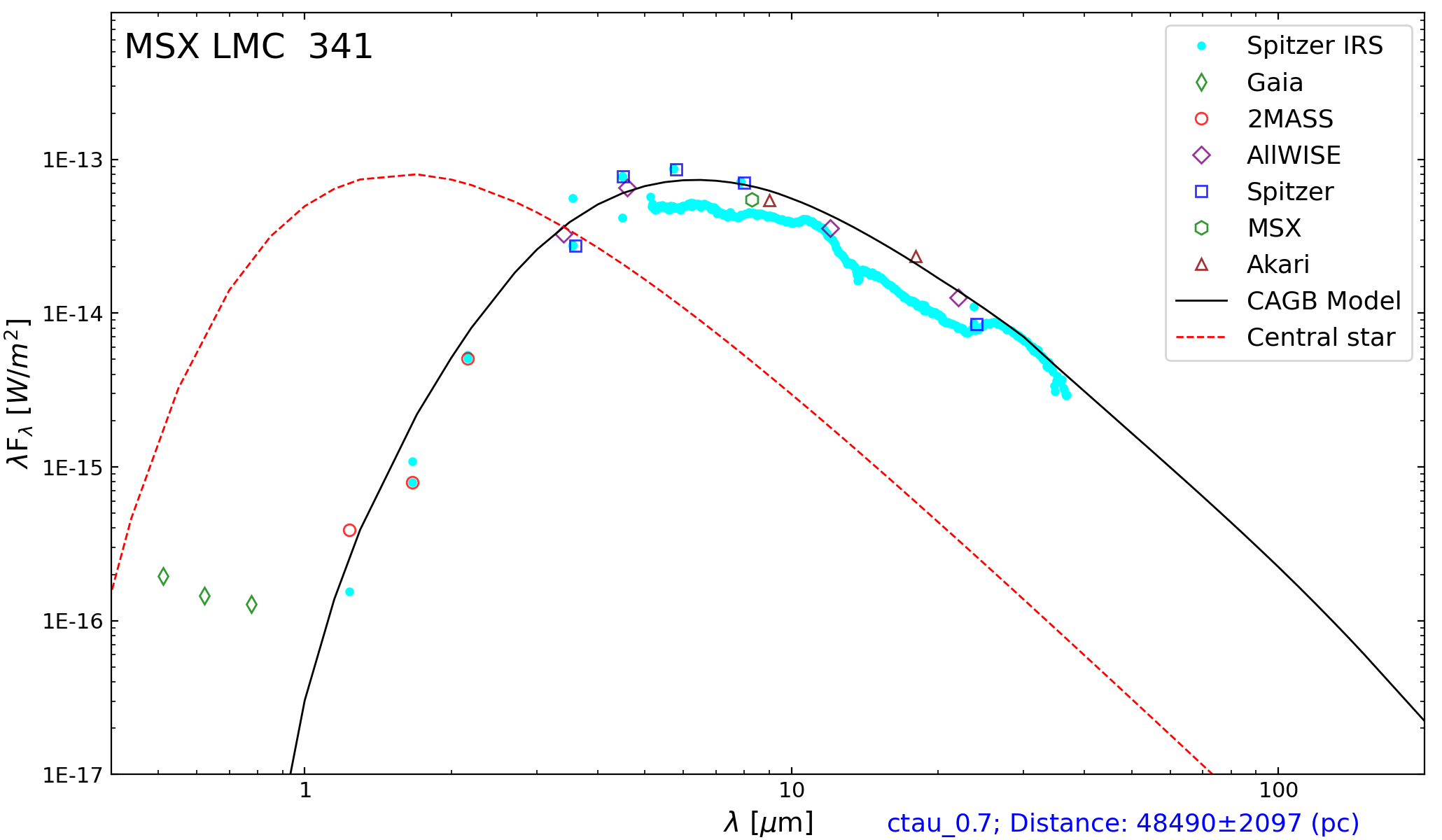}{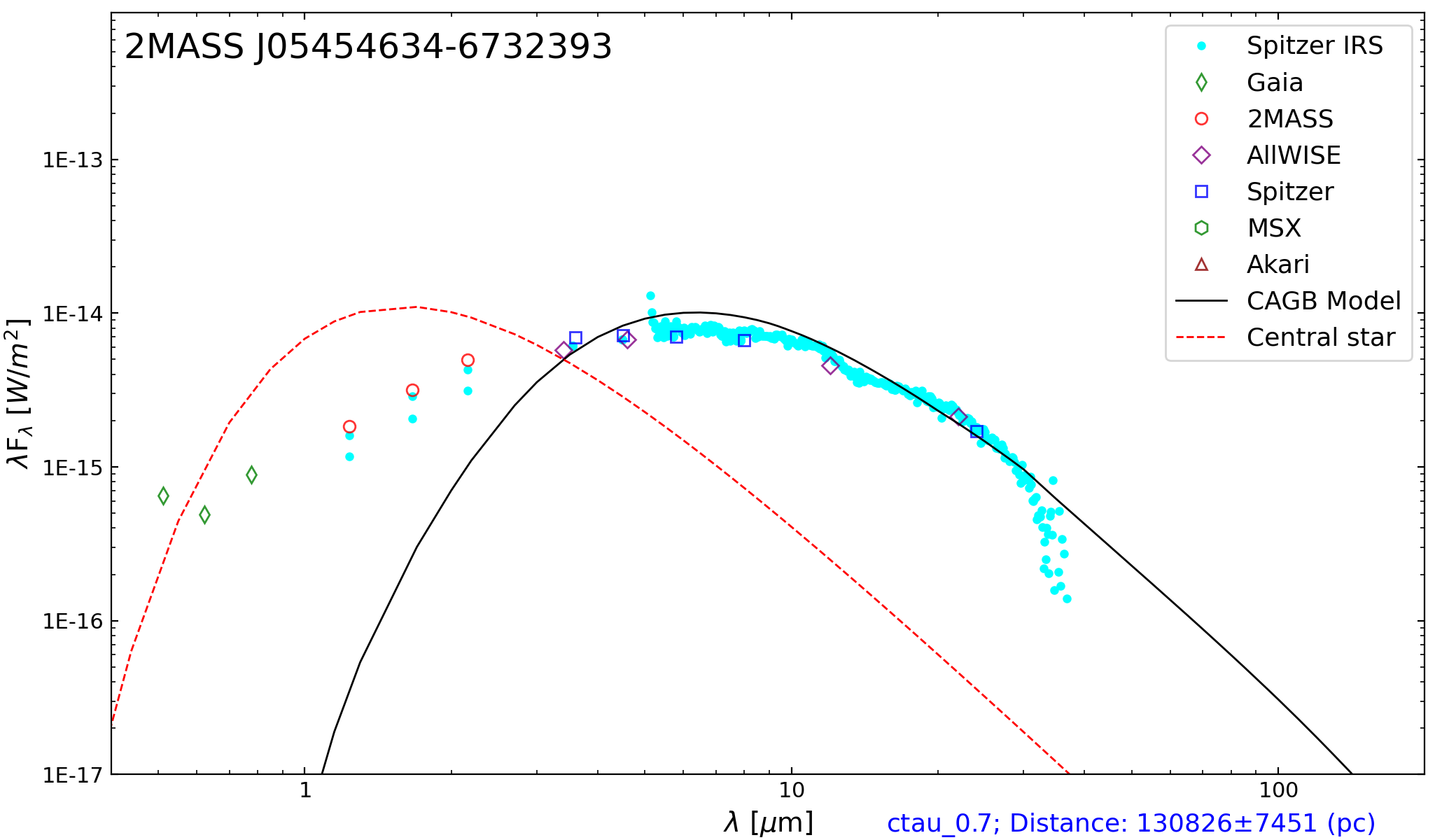}
\caption{
Observed and model SEDs for two LMC-CAGB stars with thick dust envelopes.
The object in the right panel is likely an extrinsic carbon star (see Section~\ref{sec:sedlmc}).
}\label{f5}
\end{figure*}

The two panels in Figures~\ref{f5} compare the SEDs of two stars from the
LMC-CAGB sample. The object in the left panel (MSX LMC 341) appears to be a
typical CAGB star with a thick dust envelope. In contrast, the object in the
right panel (2MASS J05454634$-$6732393) is the only source with a scaled distance
greater than 100 kpc and does not resemble a typical CAGB star. Its SED shape is
unusual, and the best-fitting CAGB model implies an extremely large distance
($\sim$131 kpc). If the object is instead located at the distance of the LMC
(49.6 kpc), its absolute luminosity would be only $\sim$1200 $L_{\odot}$, about 6
times fainter than that of a typical CAGB star. This object is therefore more
likely an extrinsic carbon star, which is thought to be a binary system
consisting of a giant (or dwarf) star and a white dwarf that was once a CAGB star
(see \citealt{suh2024}). The brown arrow in the CMDs of LMC-CAGB stars (see the
bottom panels of Figure~\ref{f2}) marks the location of the object.

\begin{figure*}
\centering
\smallplotfour{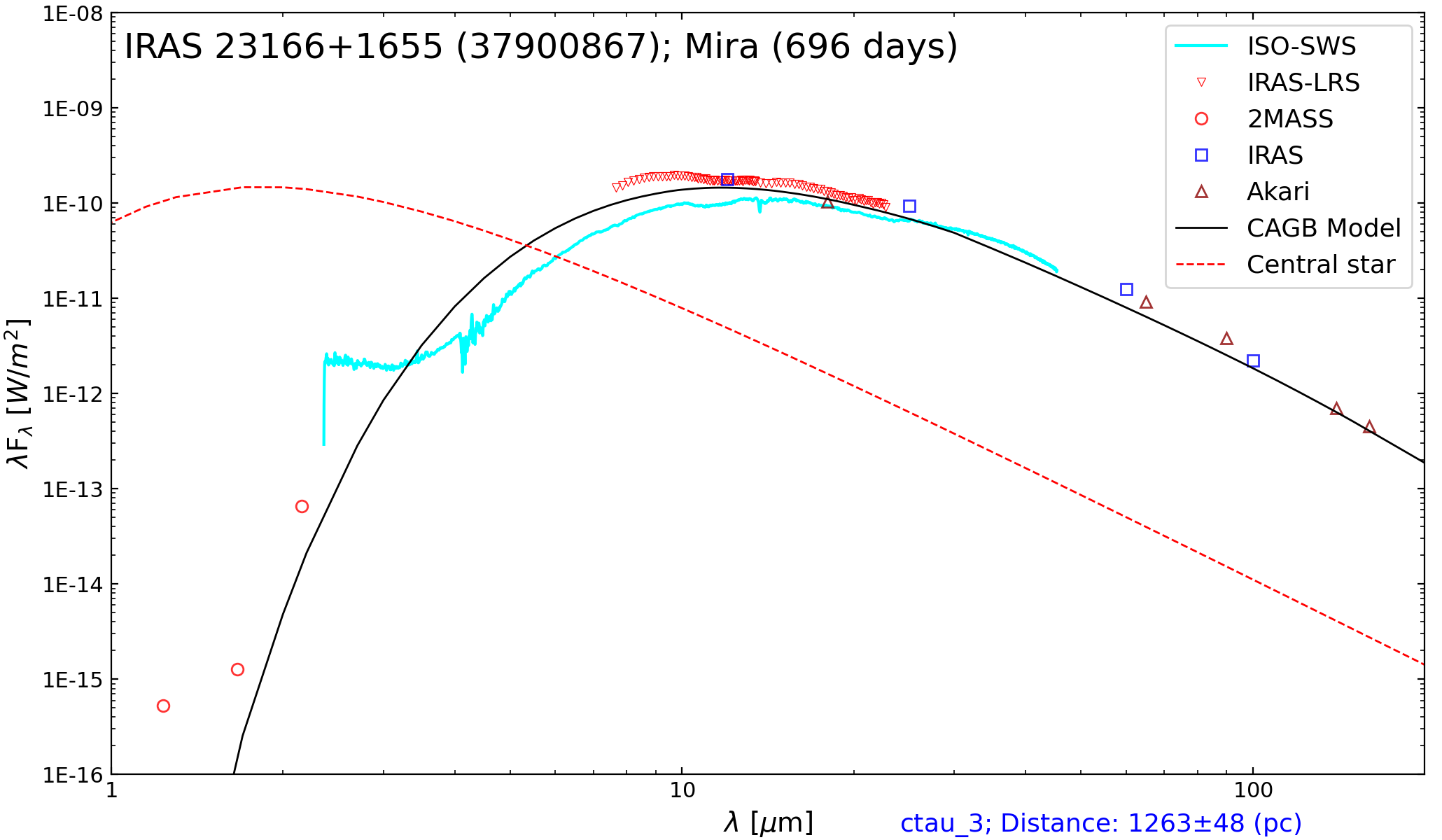}{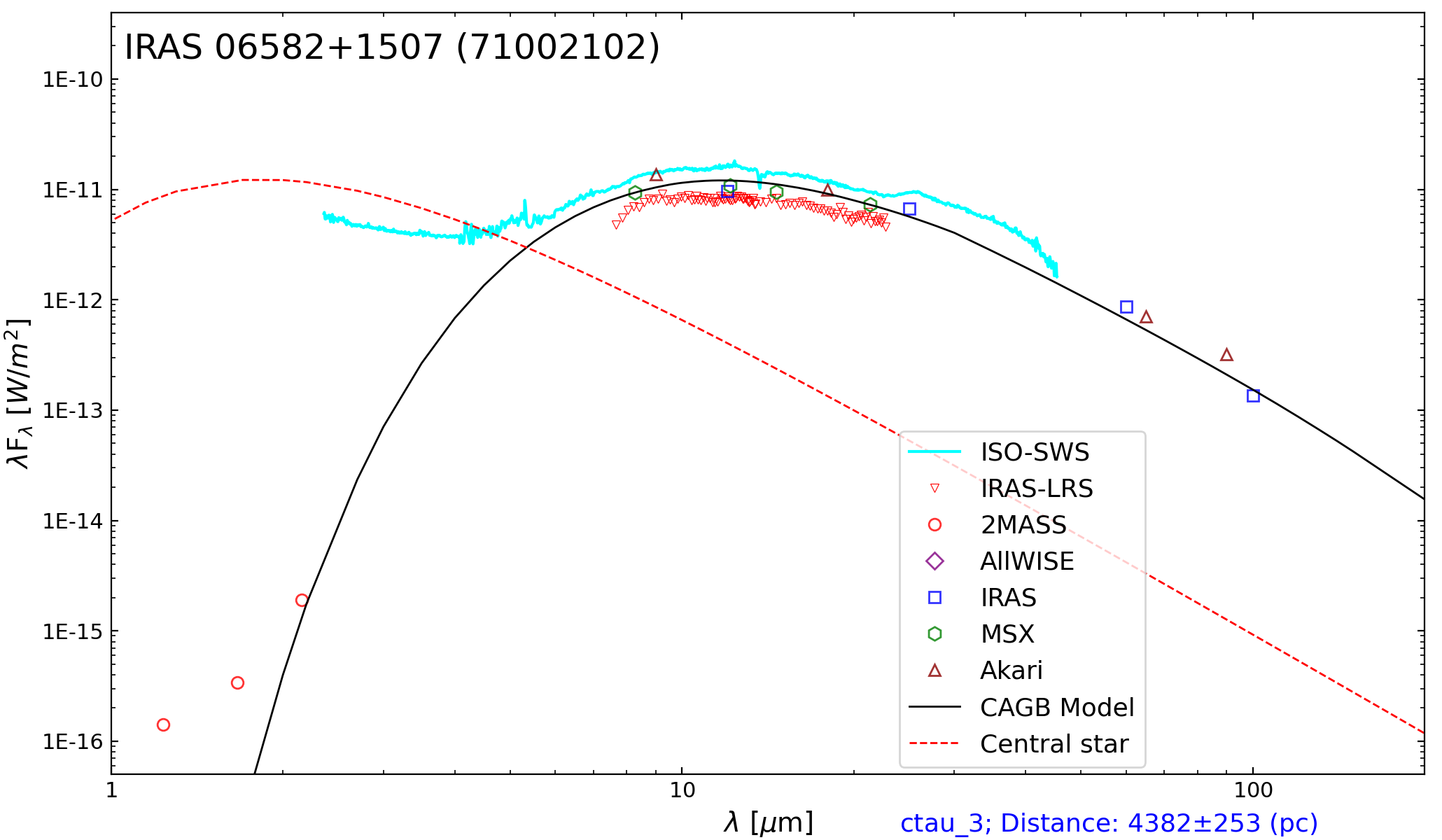}{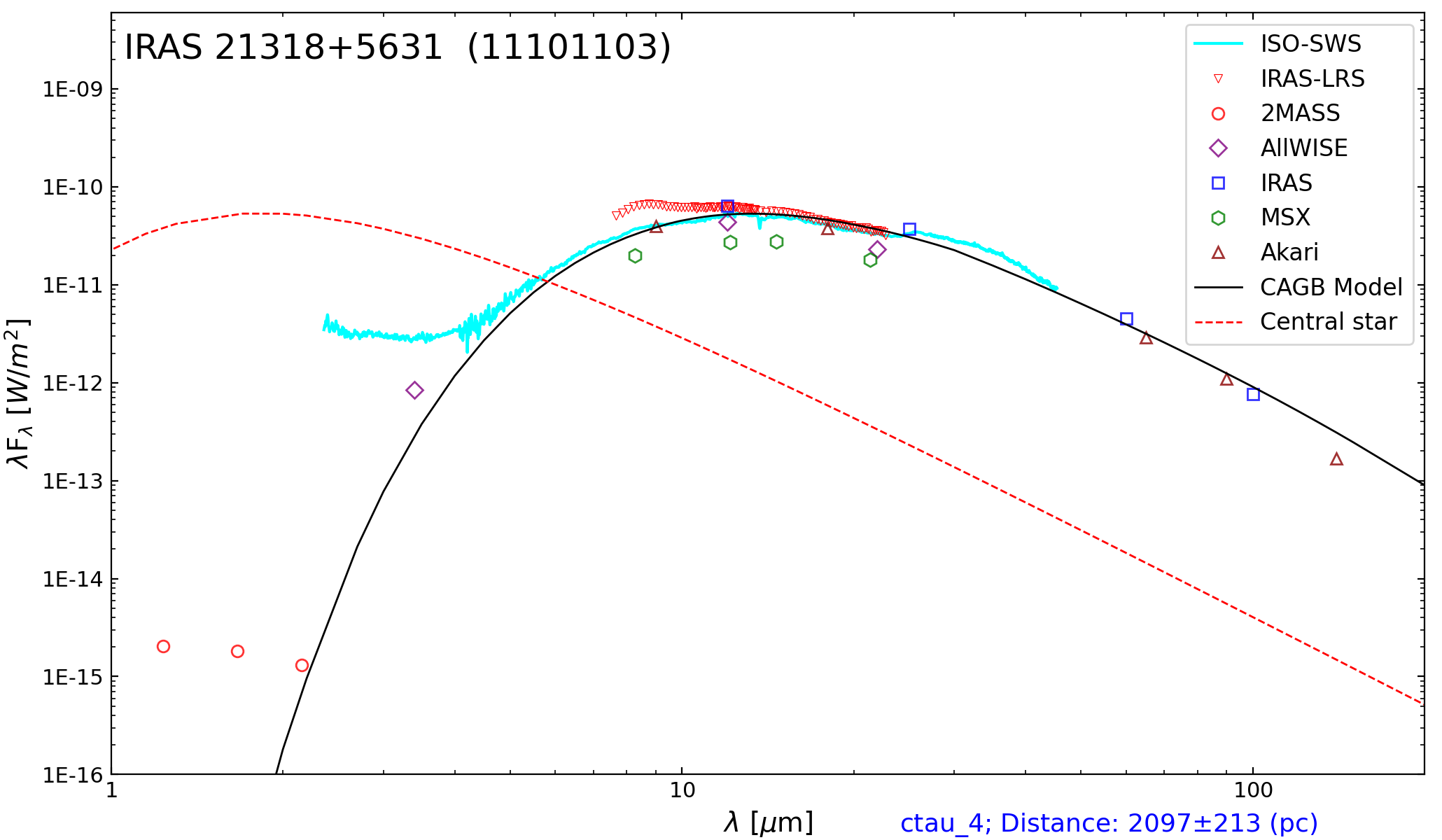}{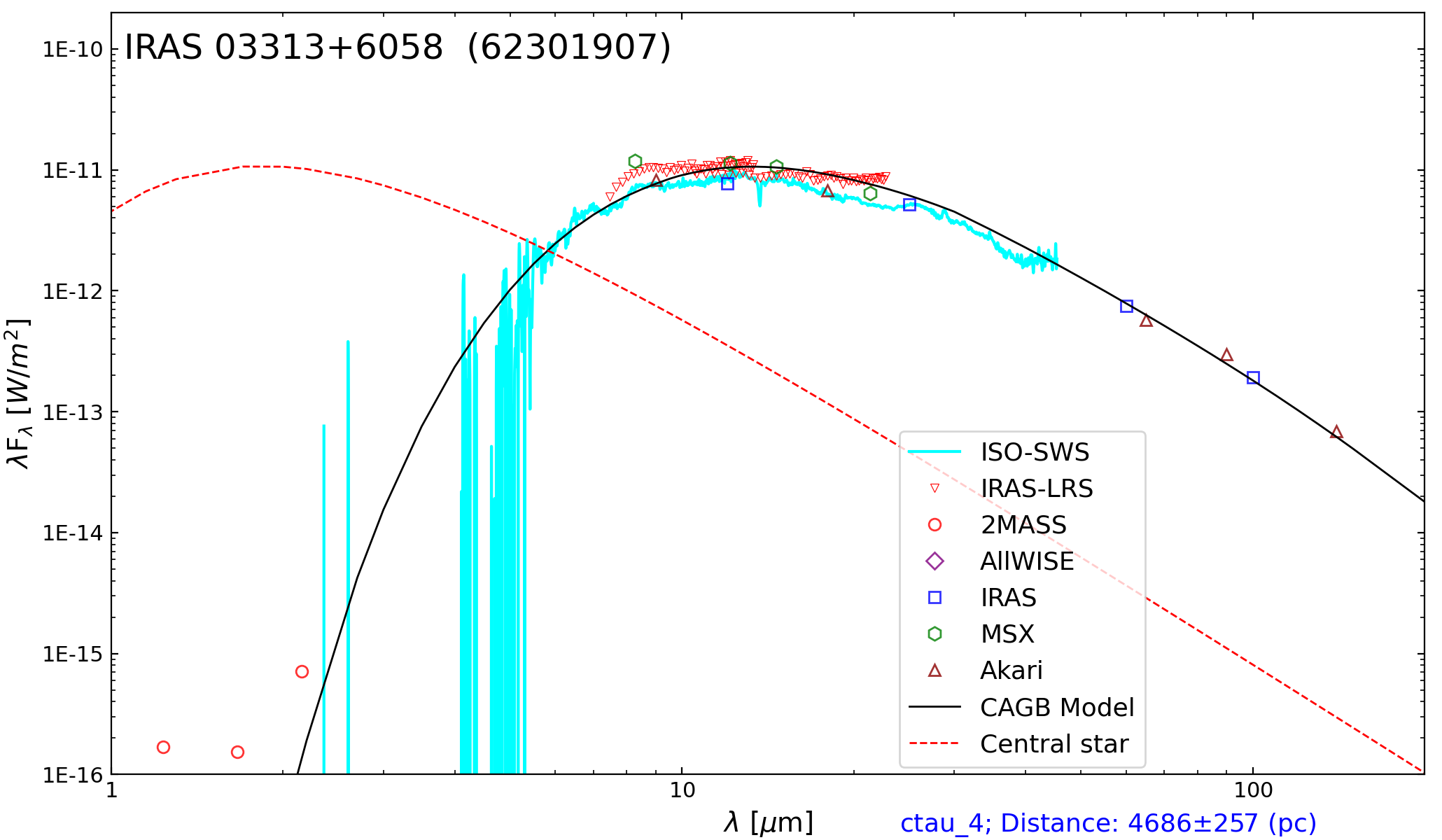}
\caption{Observed and model SEDs for Galactic CAGB stars with very thick dust envelopes ($\tau_{10} > 1.5$).
Such heavily obscured objects are not seen among LMC-CAGB stars (see Section~\ref{sec:sedmw}).} \label{f6}
\end{figure*}

\subsection{SEDs of Galactic CAGB stars}\label{sec:sedmw}

The subgroup of MW-CAGB\_IC objects (see Table~\ref{tab:tab1}) designated as
MW-CAGB\_IC-S consists of 91 well-studied CAGB stars with available infrared
spectral data (IRAS LRS or ISO SWS). The lower four panels of Figures~\ref{f4}
show observed SEDs together with CAGB model SEDs for four representative Galactic
CAGB stars from this subset. The SEDs of Galactic CAGB stars exhibit
characteristics that are closely comparable to those of LMC-CAGB stars.

Figure~\ref{f6} presents observed and model SEDs for four representative stars
with very large dust optical depths ($\tau_{10} > 1.5$), which are not seen among
LMC-CAGB stars.

\subsection{Distances of Galactic CAGB stars}\label{sec:sed-d}

\begin{figure*}
\centering
\smallplotsix{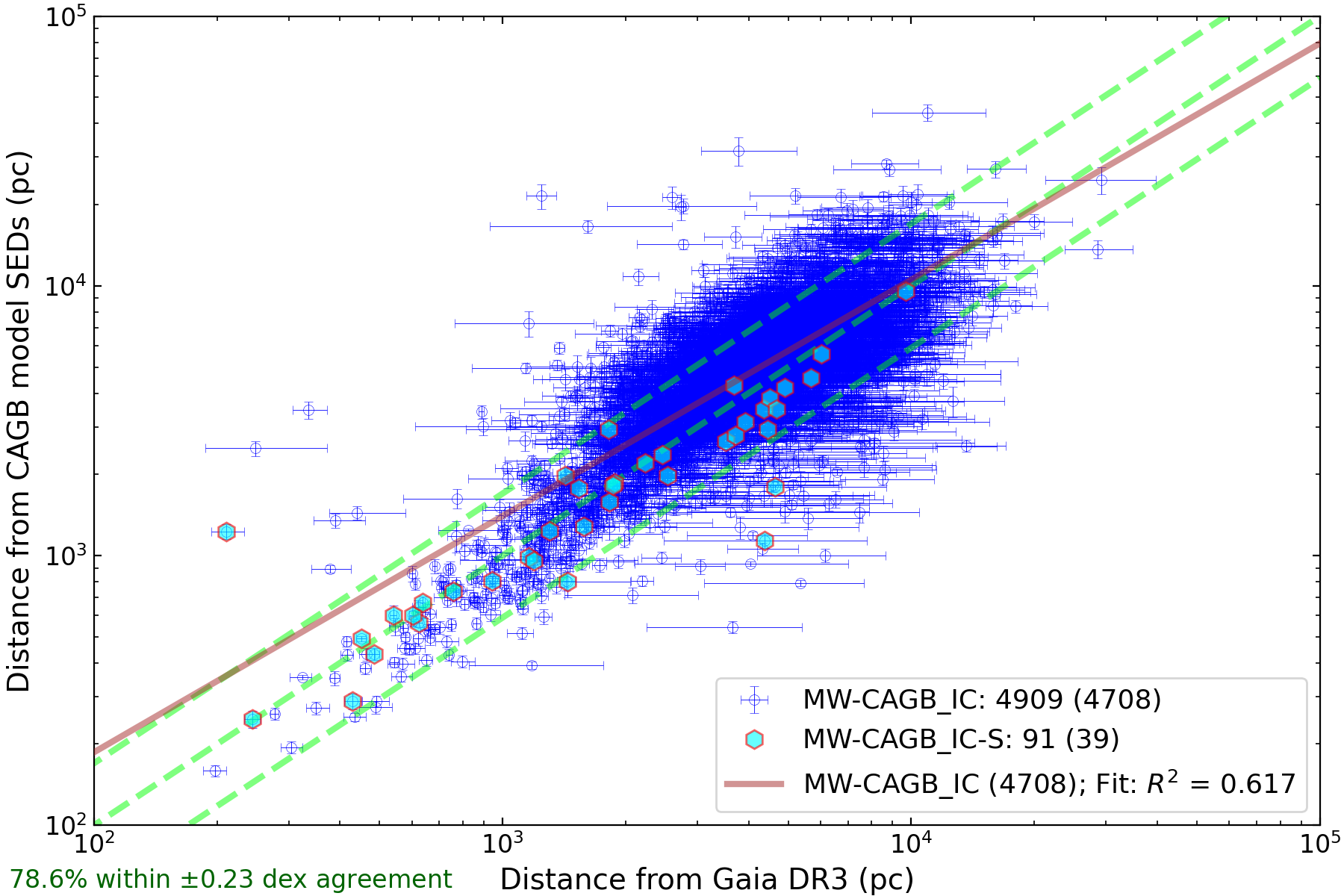}{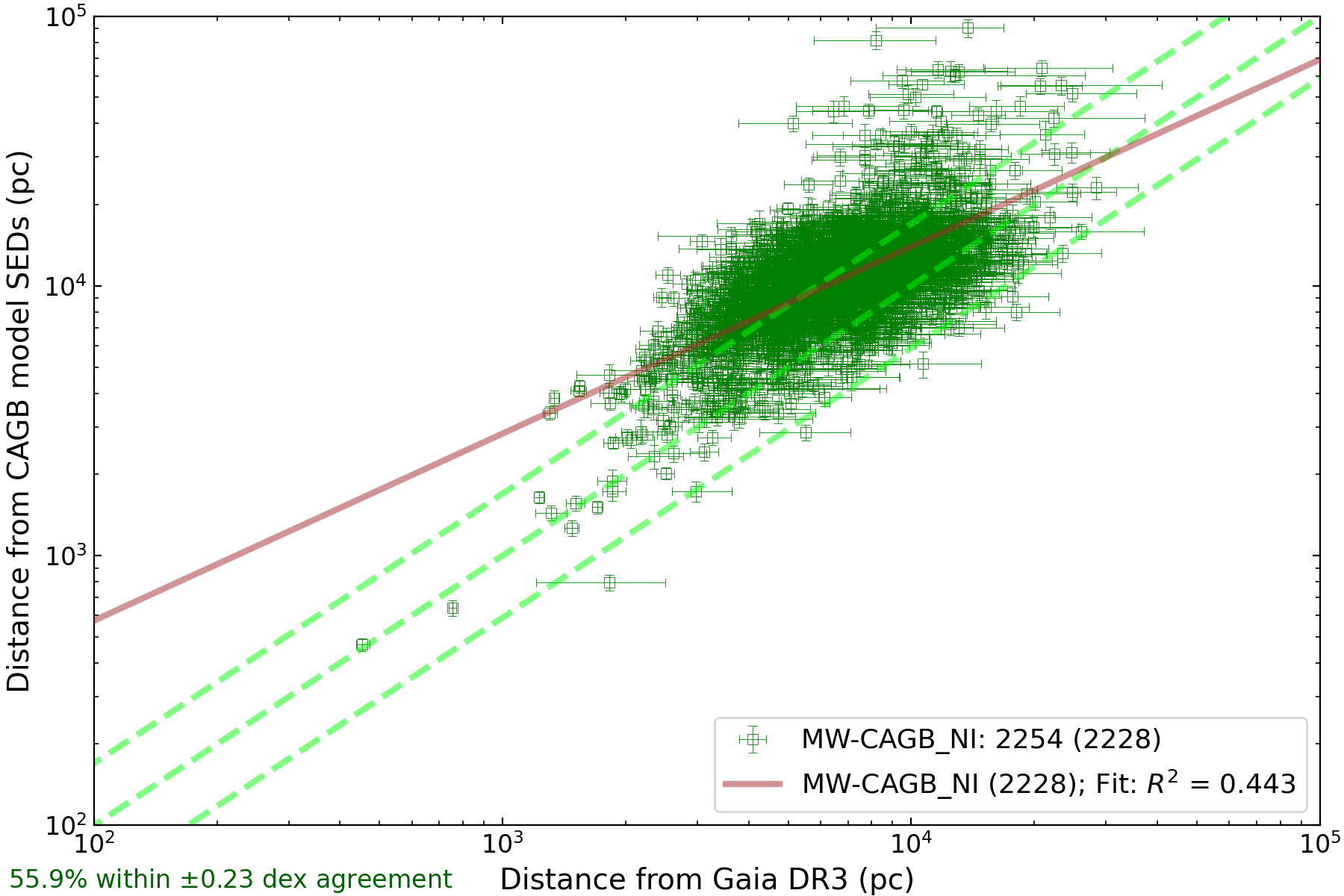}{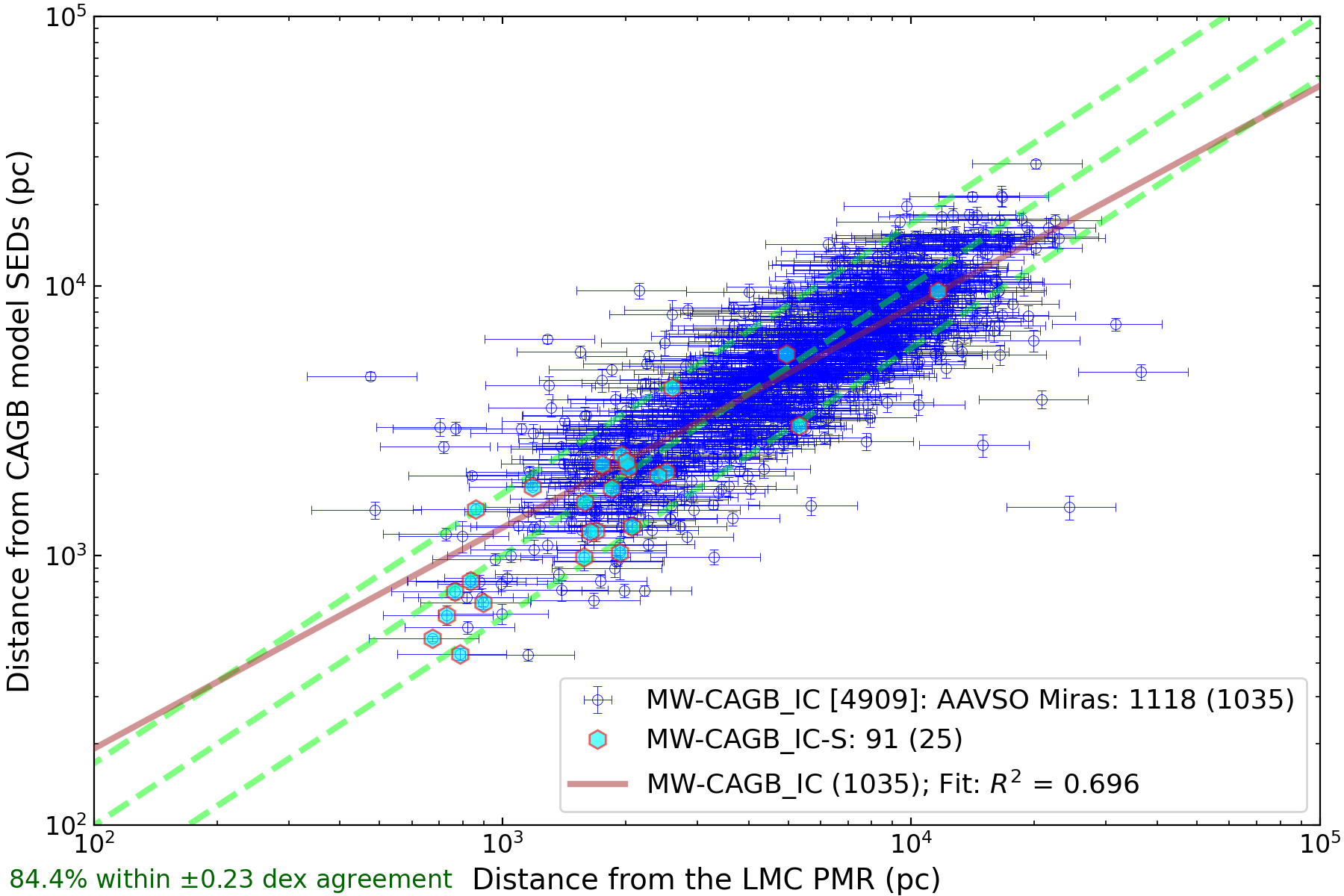}{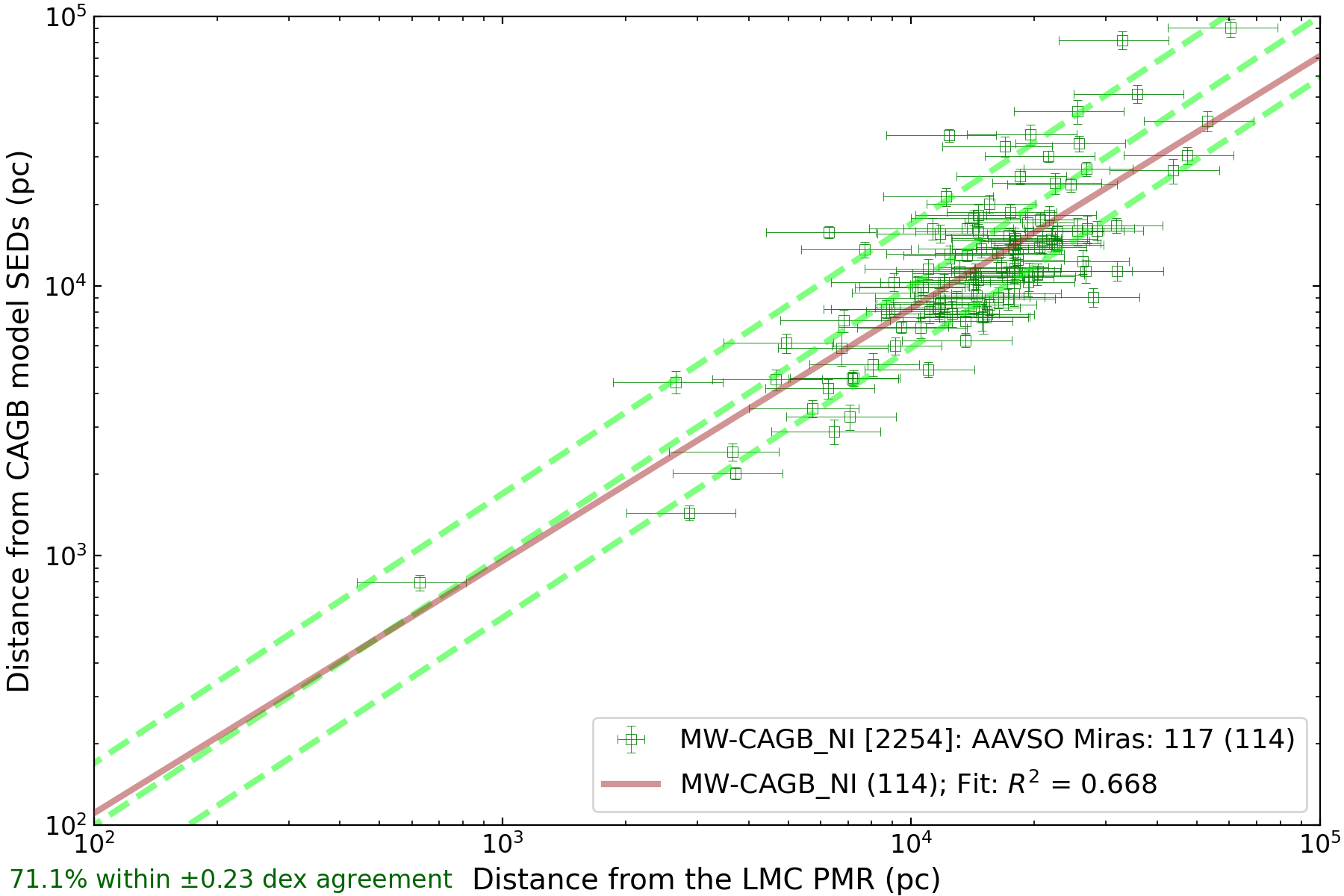}{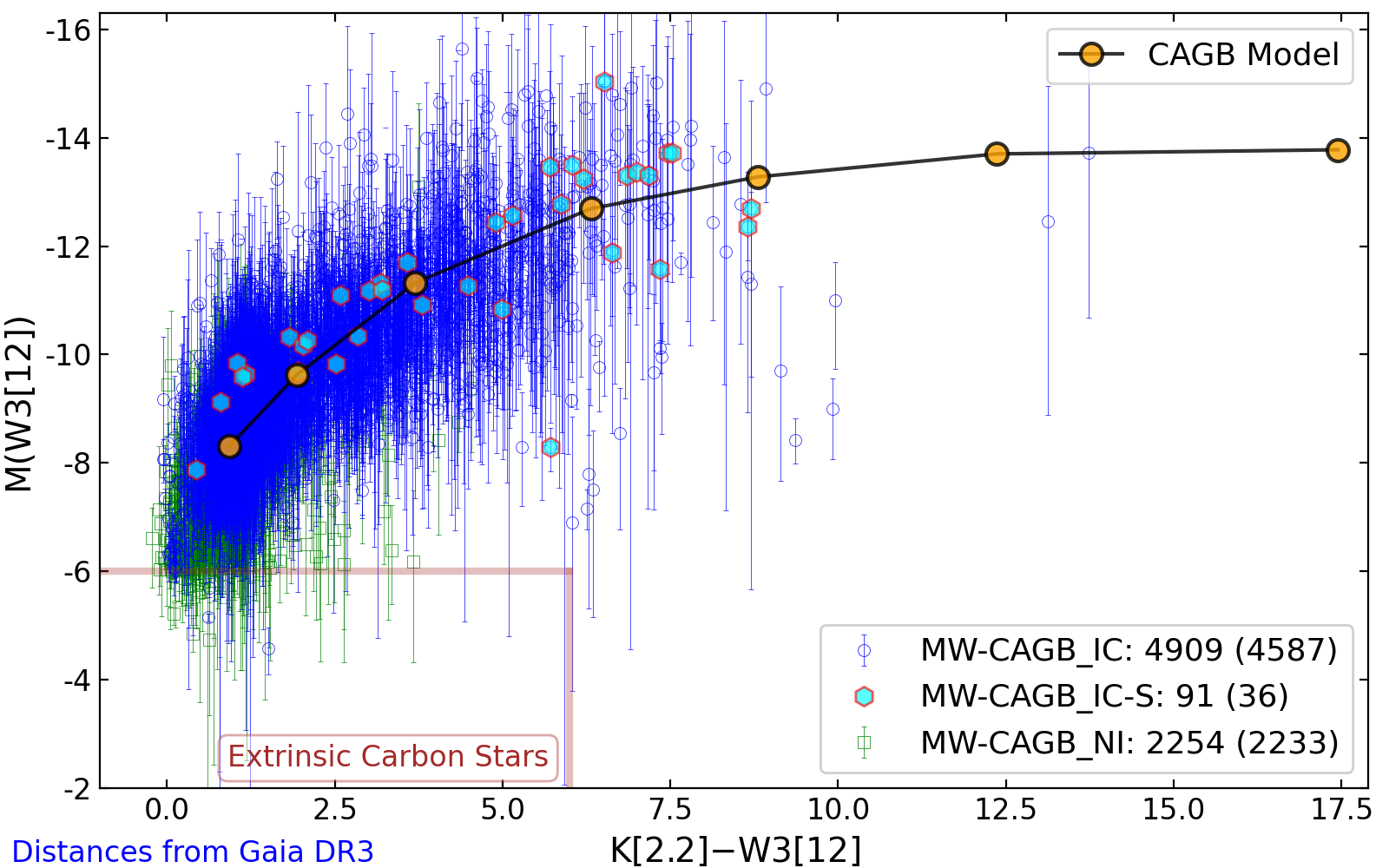}{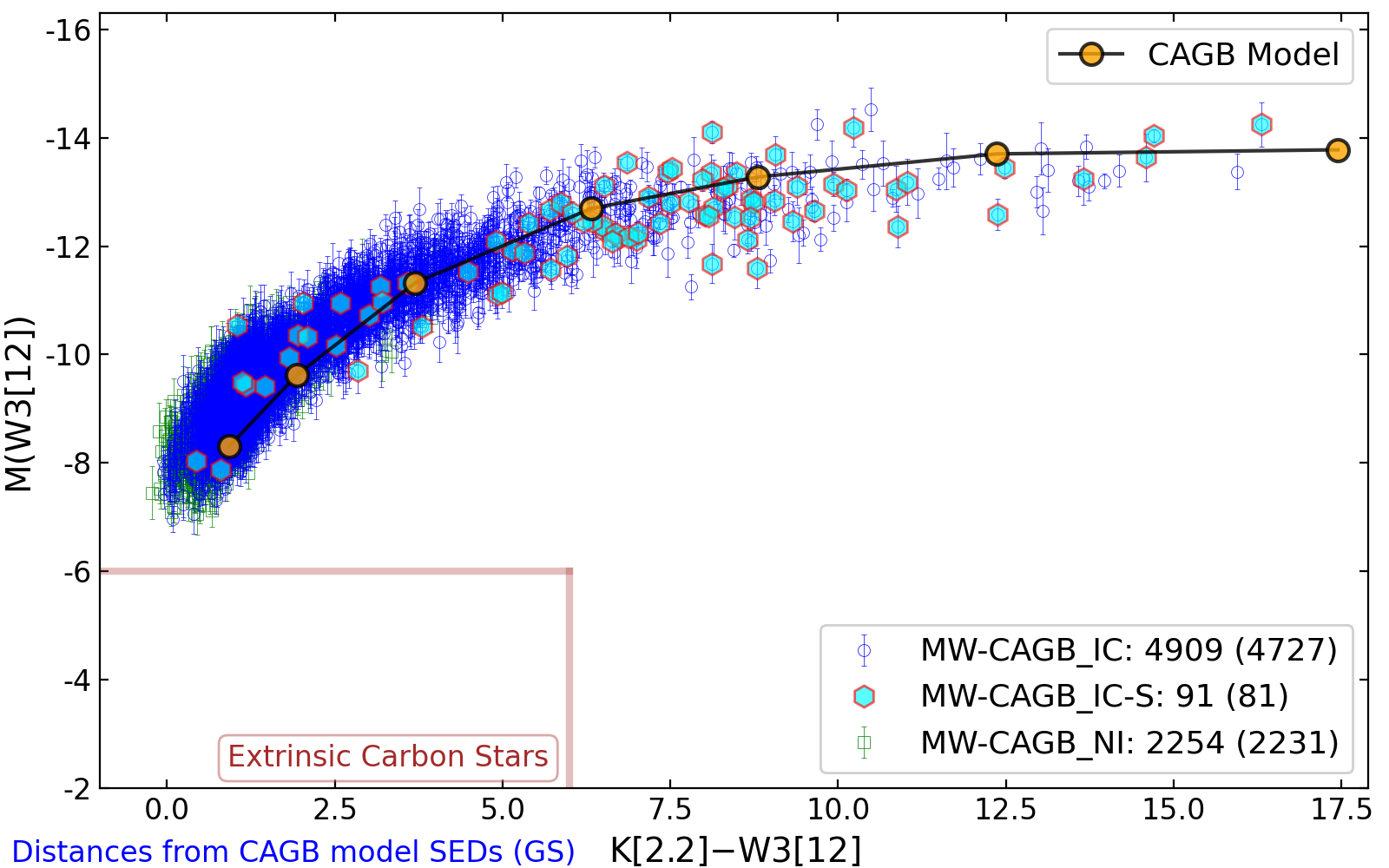}\caption{
For Galactic CAGB stars, the upper 4 panels show comparisons between distances derived from the CAGB model SEDs,
Gaia DR3 parallax measurements, and the PMR obtained from Miras in the LMC.
The bottom panels show IR CMDs for Galactic CAGB stars using distances from Gaia DR3 (left) and CAGB model SEDs (right).
For CAGB models (AMC, $T_c$ = 1000 K): $\tau_{10}$=0.0001, 0.01, 0.1, 0.5, 1, 2, and 4 from left to right (see Table~\ref{tab:tab3}).
The error bars represent the uncertainties in the derived distances.
The total number of objects in each subgroup is given, with the values in parentheses
denoting the number of plotted sources for which data are available.
See Section~\ref{sec:sed-d} for details.}
\label{f7}
\end{figure*}

For Galactic CAGB stars, the top panels of Figure~\ref{f7} compare distances
derived from Gaia DR3 parallax measurements with those obtained from theoretical
CAGB model SEDs, shown separately for MW-CAGB\_IC (left) and MW-CAGB\_NI (right)
stars. Although considerable scatter is present, the two distance estimates are
generally consistent, with MW-CAGB\_IC objects showing better overall agreement.
The upper-left panel also highlights a comparison for the MW-CAGB\_IC-S subgroup,
a subset of CAGB\_IC stars (see Section~\ref{sec:sedmw}).

Of the 4909 MW-CAGB\_IC objects (see the upper-left panel of Figure~\ref{f7}),
78.6\% fall within a $\pm$0.23 dex agreement between the two distance estimates,
as indicated by the lime dashed lines in the panel. This corresponds to distances
agreeing within a factor of 1.7 (approximately -41\% to +70\% difference). When
restricting the sample to the 114 objects with distances from the CAGB model SEDs
within 900 pc, the agreement improves to 84.2\%. Of the 2254 MW-CAGB\_NI objects
(see the upper-right panel of Figure~\ref{f7}), 55.9\% fall within a $\pm$0.23
dex agreement between the two distance estimates.

The middle panels of Figure~\ref{f7} compare distances derived from the
period–magnitude relation (PMR) calibrated with LMC Miras (see
Section~\ref{sec:pul}; Equation~\ref{eq:1}) with those obtained from CAGB model
SEDs for Mira-type CAGB stars. The two methods show strong consistency: among
1035 Mira-type MW-CAGB\_IC stars (middle-left panel), 84.4 \% agree within
$\pm$0.23 dex, while for 114 Mira-type MW-CAGB\_NI stars (middle-right panel),
71.1 \% fall within the same range. Although both approaches have inherent
limitations and uncertainties, SED fitting and the PMR provide the only practical
means of deriving reliable distances from observed photometric and variability
data for large samples of CAGB stars.

The lower-left panel of Figure~\ref{f7} shows the WISE–2MASS CMD for Galactic
CAGB stars, using distances derived from Gaia DR3 parallaxes. The diagram
exhibits considerable scatter, with many absolute magnitudes lying outside the
luminosity-function range for CAGB stars discussed in
Section~\ref{sec:lum-gcagb}, and falling within the region occupied exclusively
by extrinsic carbon stars, as identified by \citet{suh2024}.

In contrast, the lower-right panel of Figure~\ref{f7} presents the WISE–2MASS CMD
for Galactic CAGB stars, using distances derived from theoretical CAGB model
SEDs. The CMD exhibits significantly less scatter, with the resulting absolute
magnitudes largely confined within the expected luminosity function limits. This
indicates that distances derived from model SEDs are, in general, more reliable
than Gaia DR3 parallax-based distances for a large sample of CAGB stars.

Among the three distance estimates for Galactic CAGB stars, model-based SED
distances are found to be both reliable and practical for a large sample of
Galactic CAGB stars. Gaia DR3 parallaxes have inherent limitations for AGB stars,
and PMR-based distances apply only to Mira variables. In contrast, CAGB model SED
distances can be derived whenever observed SEDs of sufficient quality and
wavelength coverage are available.

\begin{figure*}
\centering \smallplotfour{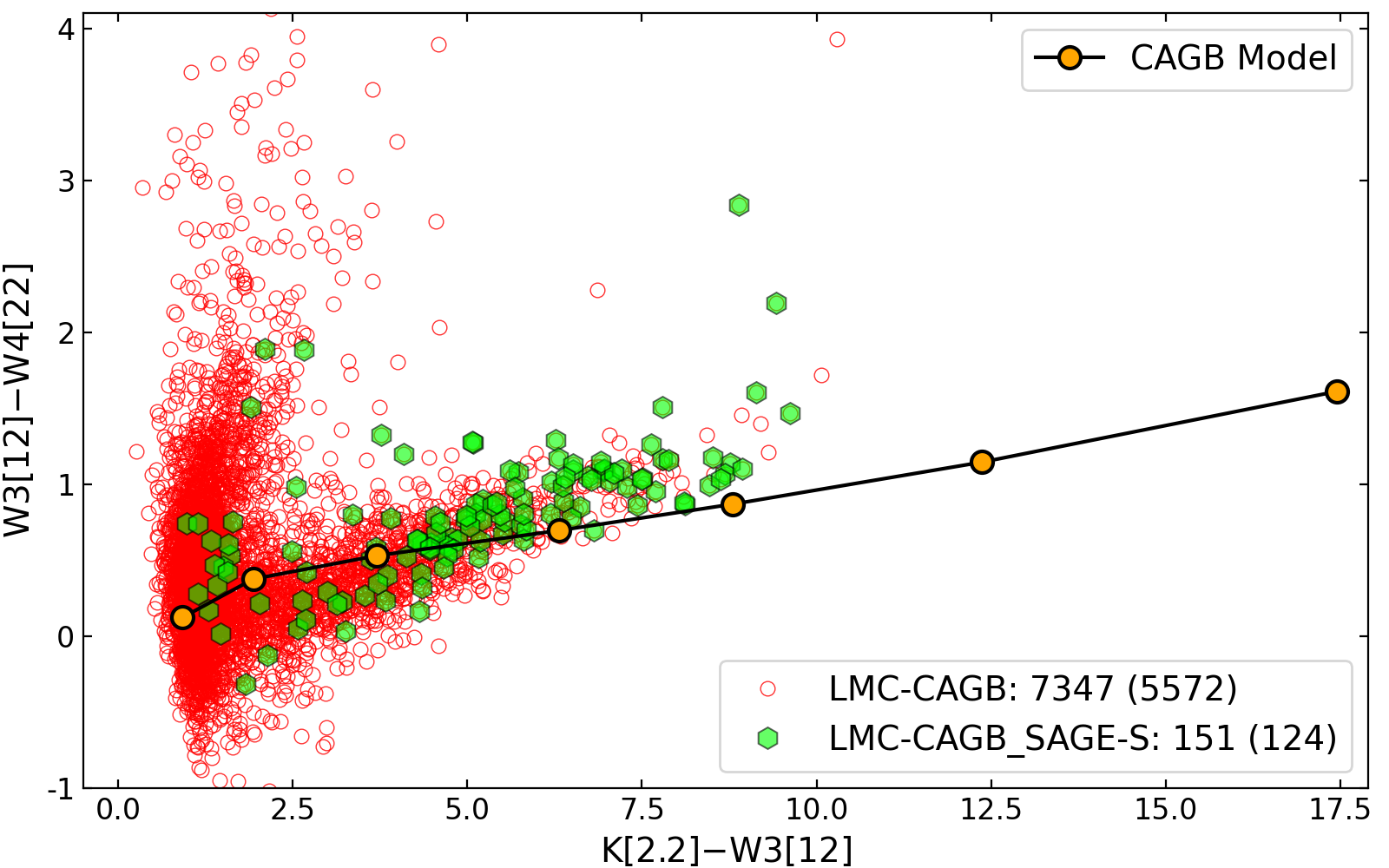}{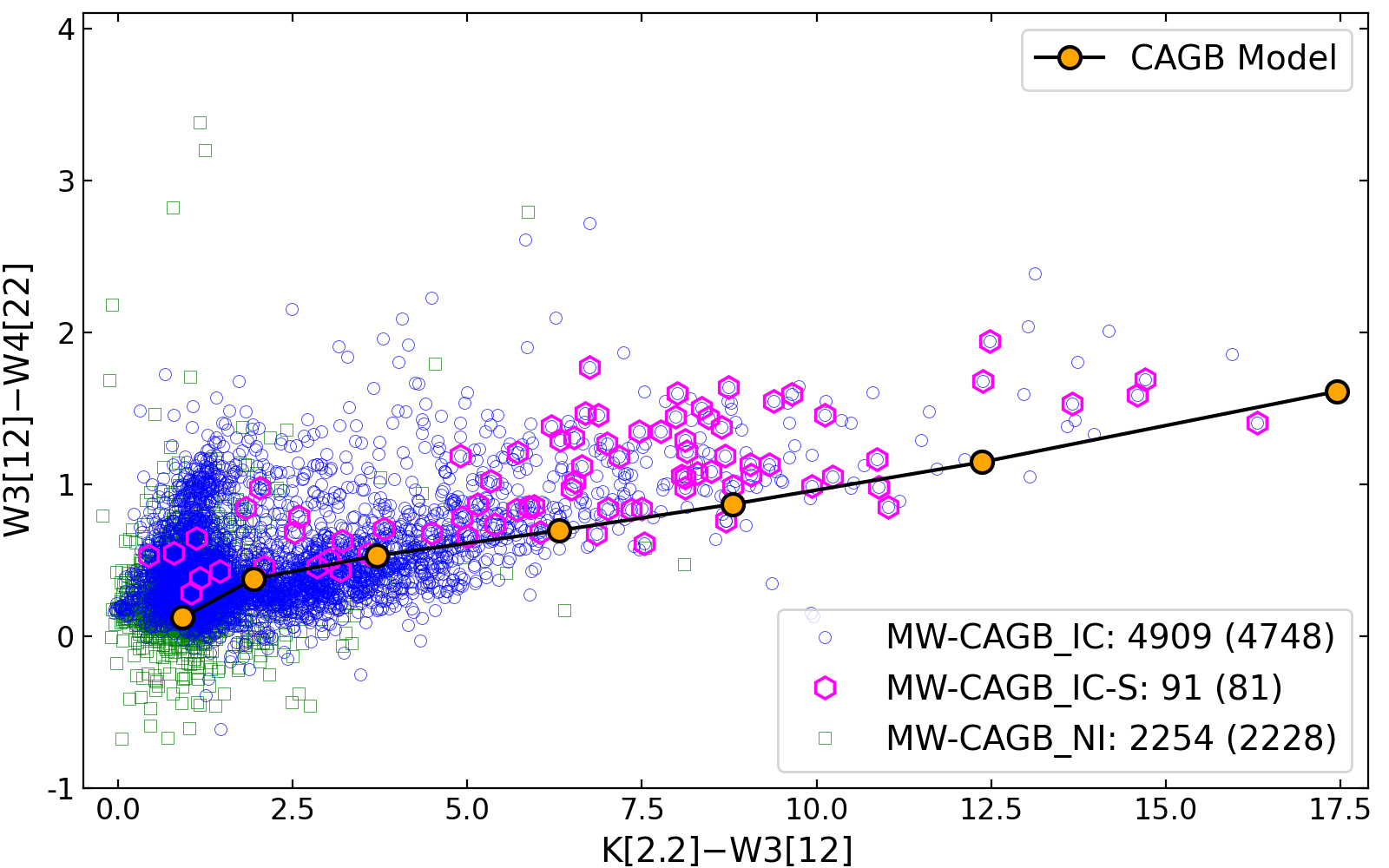}{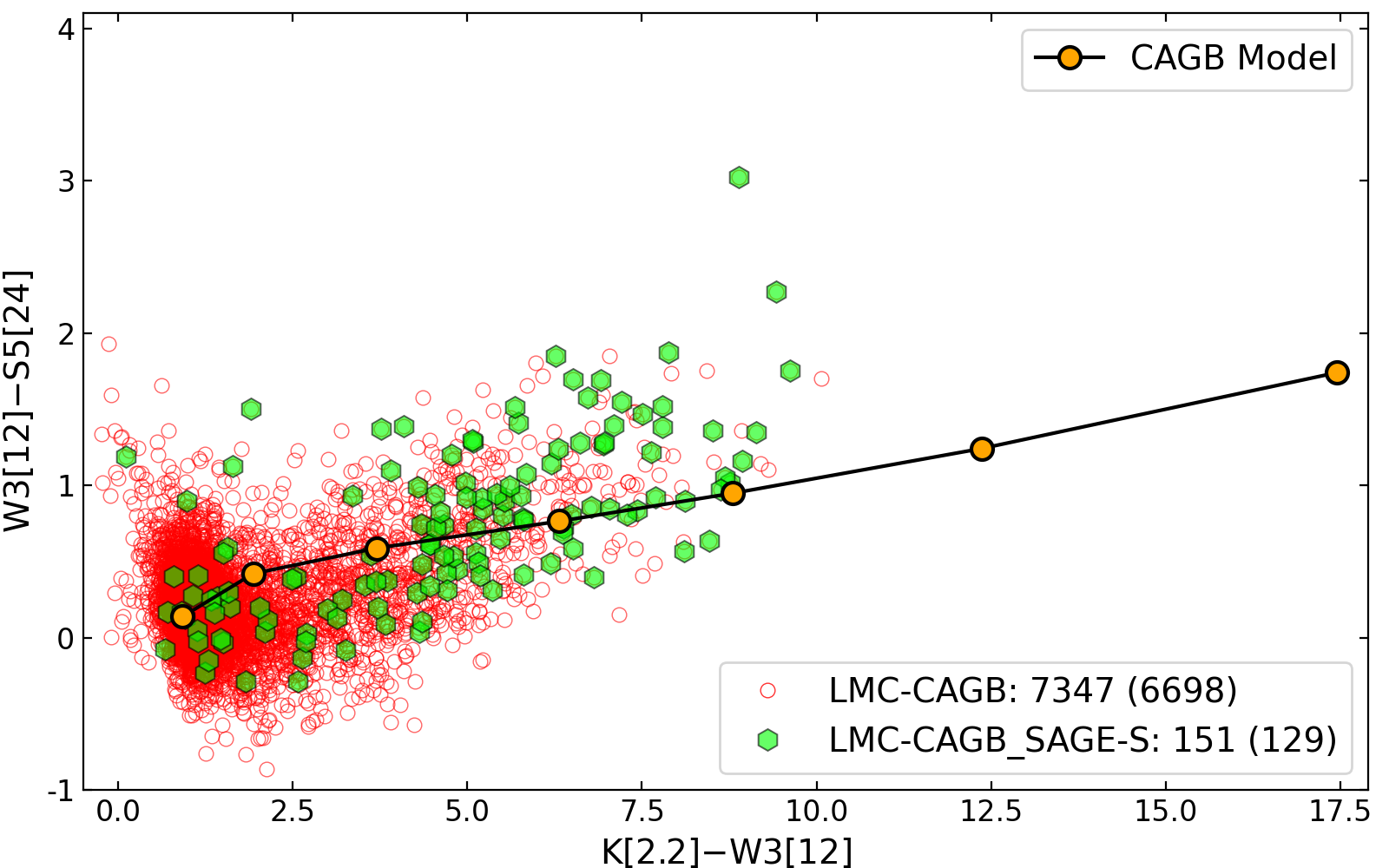}{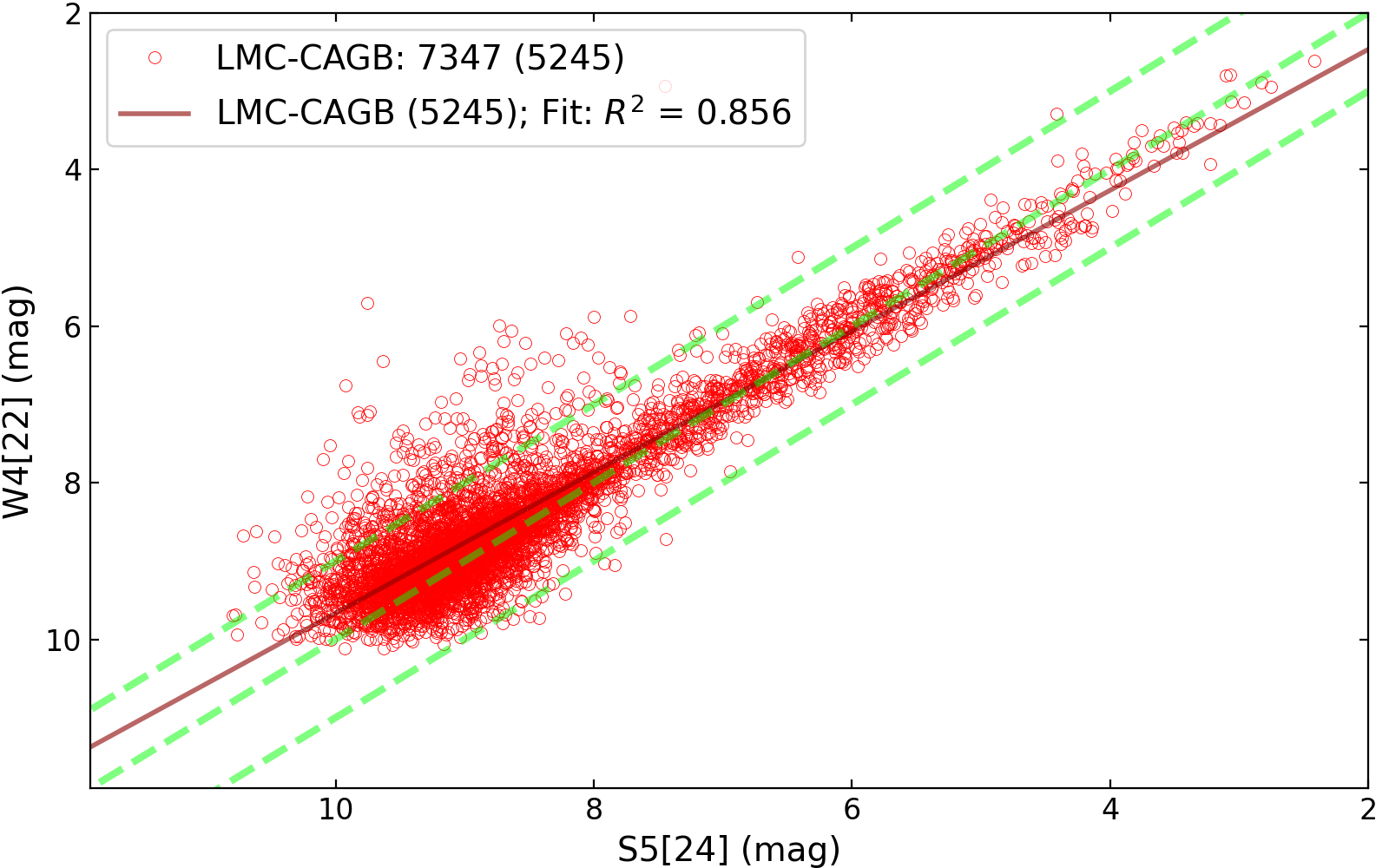}\caption{
The top and lower-left panels show IR 2CDs for CAGB stars in the LMC and the Milky Way,
while the lower-right panel compares fluxes in the W4[22] and S5[24] bands.
For CAGB models (AMC, $T_c$ = 1000 K): $\tau_{10}$=0.0001, 0.01, 0.1, 0.5, 1, 2, and 4 from left to right (see Table~\ref{tab:tab3}).
The total number of objects in each subgroup is given, with the values in parentheses
denoting the number of plotted sources for which data are available.
See Section~\ref{sec:irpro}.} \label{f8}
\end{figure*}

\begin{figure*}
\centering
\smallplottwo{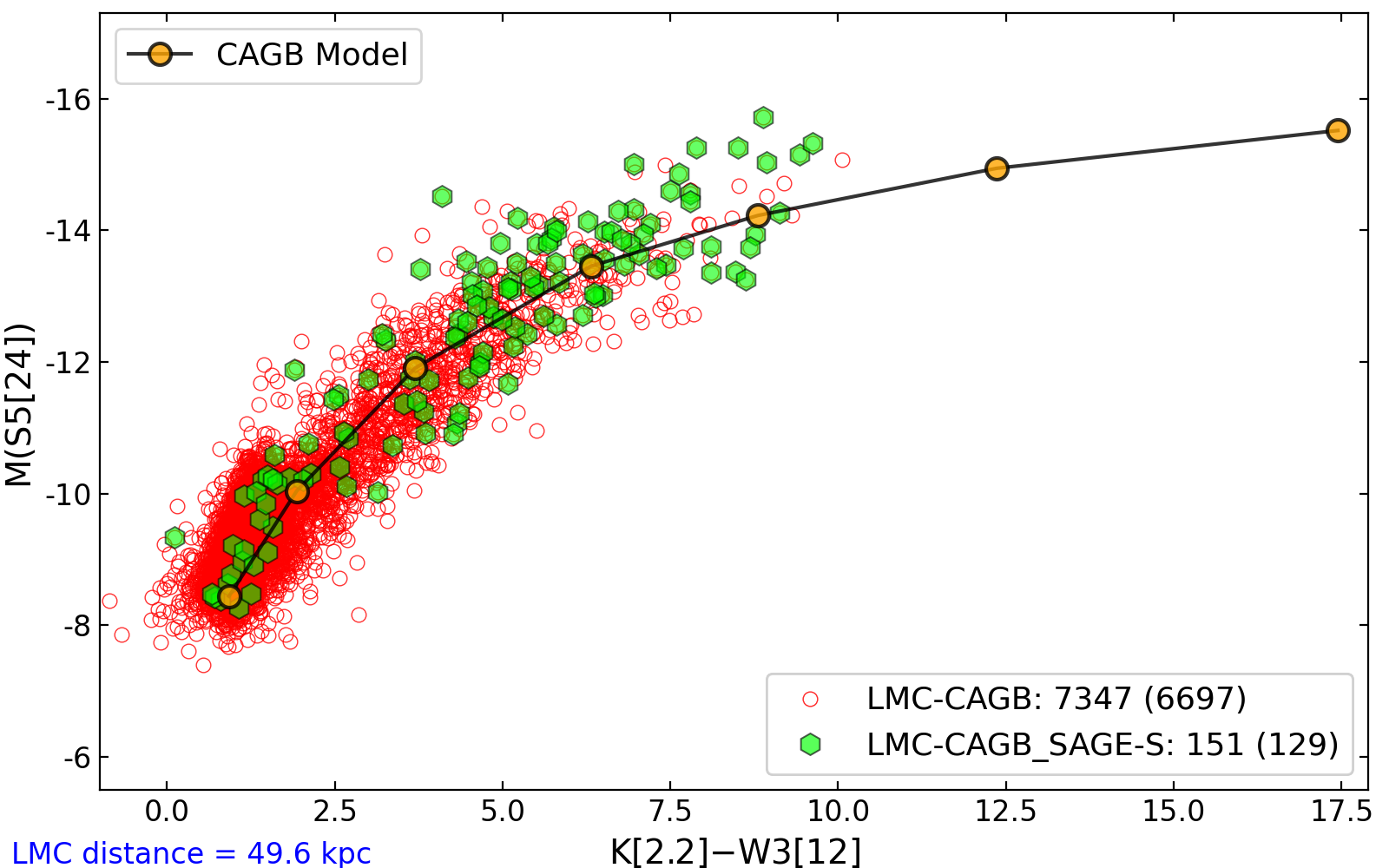}{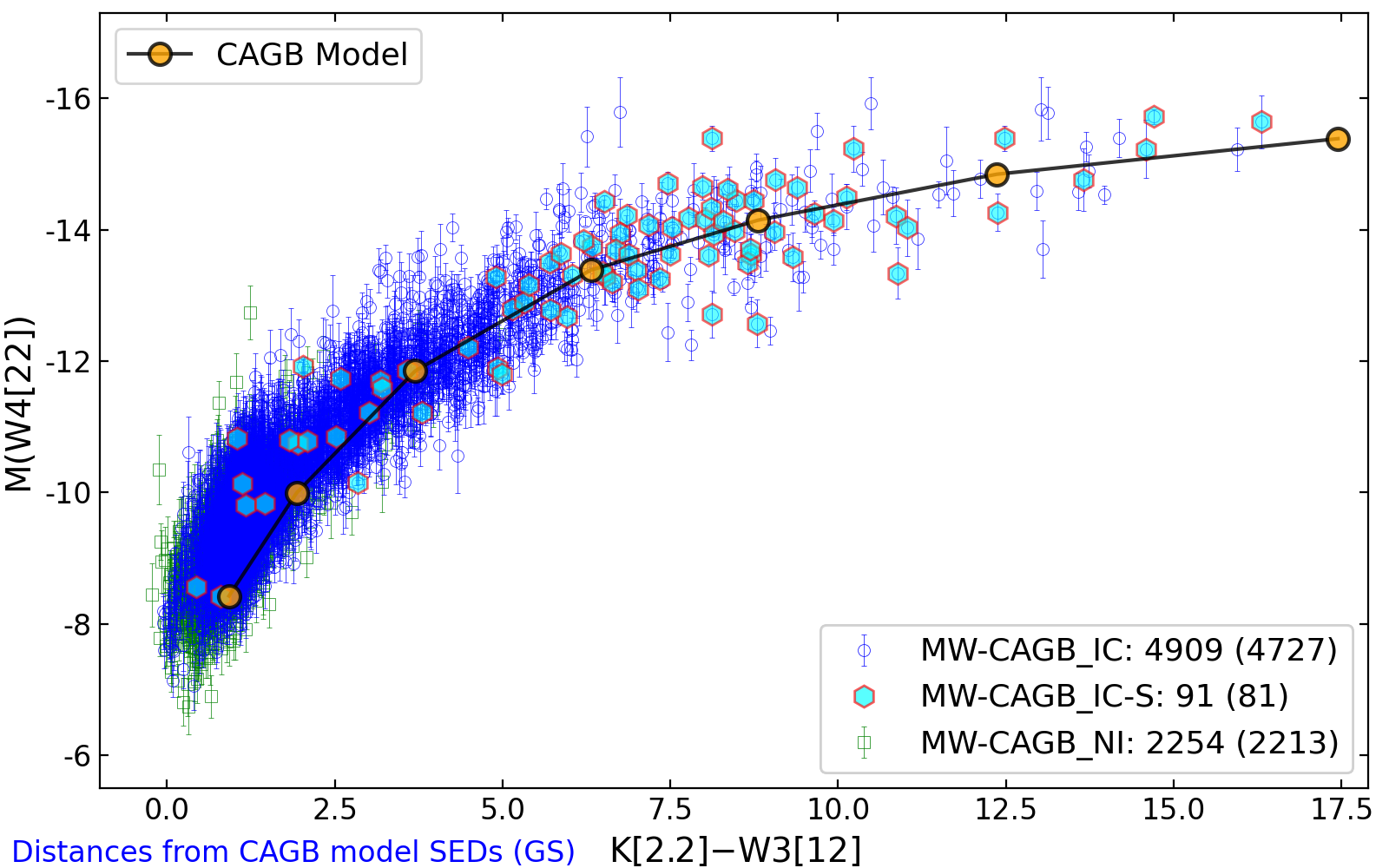}\caption{IR CMDs for CAGB stars in the LMC and the Milky Way.
For CAGB models (AMC, $T_c$ = 1000 K): $\tau_{10}$=0.0001, 0.01, 0.1, 0.5, 1, 2, and 4 from left to right (see Table~\ref{tab:tab3}).
For Galactic CAGB stars, distances are derived from CAGB model SEDs with Gaussian Monte Carlo scatter (GS)
in $\log_{10} d$ ($\sigma$=$\pm$0.0798 dex; obtained from LMC-CAGB stars; see Figure~\ref{f1}).
The total number of objects in each subgroup is given, with the values in parentheses
denoting the number of plotted sources for which data are available.
See Section~\ref{sec:irpro}.}
\label{f9}
\end{figure*}

\section{2CDs and CMDs in IR Bands\label{sec:irpro}}

\citet{srinivasan2011} and \citet{groe2020} modeled CAGB stars in the LMC using
different approaches, presenting various 2CDs and CMDs and comparing observations
with theoretical models. In this study, we extend this work by employing
additional IR wavelength bands for a larger sample of CAGB stars in both the LMC
and the Milky Way, allowing a more detailed comparison between observations and
theoretical models based on more sophisticated schemes.

The upper two panels of Figure~\ref{f8} show WISE–2MASS 2CDs of CAGB stars in the
LMC and the Milky Way, using W3[12]–W4[22] versus K[2.2]$–$W3[12]. In the
upper-left region of the WISE–2MASS 2CD (see the upper panels of
Figure~\ref{f8}), a subset of LMC-CAGB stars deviates significantly from the
theoretical models with $T_c = 1000$ K. In contrast, only a few Galactic CAGB
stars show similar behavior. A plausible explanation is that the W4[22] band
sensitivity is insufficient to reliably detect faint CAGB stars in the LMC.

The Spitzer S5[24] band, which offers higher sensitivity (5$\sigma$ photometric
sensitivity of 110 $\mu$Jy) than the W4[22] band (5.4 mJy), measures the fluxes
of LMC-CAGB stars much more effectively (see \citealt{suh2020}). Indeed, the 2CD
based on the Spitzer S5[24] band (see the lower-left panel of Figure~\ref{f8})
does not exhibit this discrepancy. The lower-right panel of Figure~\ref{f8}
compares fluxes in the W4[22] and S5[24] bands. The W4[22] fluxes show larger
deviations from the S5[24] measurements for fainter sources.

The overall distributions of CAGB stars on the 2CDs in Figure~\ref{f8} are
broadly similar between the two galaxies, except that the LMC sample lacks
objects with very large dust optical depths ($\tau_{10} > 1.5$).

A comparable patten appeared in the WISE–2MASS CMDs (Figures~\ref{f2} and
\ref{f3}), which compare CAGB stars in the LMC and the Milky Way, plotting
M(W3[12]) and M(K[2.2]) against K[2.2]$-$W3[12]. Likewise, Figure~\ref{f9}
presents Spitzer–WISE–2MASS CMDs extending to longer wavelengths (W4[22] and
S5[24]). The left panel shows M(S5[24]) versus K[2.2]$-$W3[12] for LMC-CAGB
stars, while the right panel displays M(W4[22]) versus K[2.2]$-$W3[12] for
Galactic CAGB stars.

On the CMDs for CAGB stars in the LMC and the Milky Way (Figures \ref{f2},
\ref{f3}, and \ref{f9}), the distributions are broadly similar, except that the
LMC sample lacks stars with very large dust optical depths ($\tau_{10} > 1.5$),
which produce redder IR colors and extremely luminous emission at wavelengths
$\gtrsim$ 10 $\mu$m. This may reflect the lower level of high-mass star formation
in the LMC relative to the Milky Way, since CAGB stars with thick dust envelopes
are generally associated with higher masses (or more advanced evolutionary
stages). Multiple studies showed that the LMC has a mean stellar metallicity
significantly below that of the Milky Way (e.g., \citealt{hocde2023}). The lower
metallicity of the LMC likely reduces the efficiency of dust production, further
contributing to the absence of CAGB stars with extremely thick dust envelopes.

Galactic CAGB stars also show brighter absolute magnitudes at the blue ends on
the CMDs, which may reflect the LMC’s lower metallicity. At lower metallicity,
the mass range of LMC-CAGB stars is expected to have a smaller lower limit and a
larger upper limit compared to the Milky Way (see \citealt{karakas2014}).
Consequently, Galactic CAGB stars are likely more massive and thus brighter than
their LMC counterparts at the blue ends of the CMDs.

\section{IR properties of Mira variables\label{sec:pul}}

AGB stars exhibit long-period, large-amplitude pulsations. It is widely accepted
that more evolved (or more massive) AGB stars tend to show stronger pulsation
amplitudes, longer periods, and enhanced mass-loss rates (e.g.,
\citealt{debeck2010}).

Studying the period–magnitude relation (PMR) of Mira variables is more
straightforward in the LMC than in our Galaxy, as their distances are relatively
uniform. Miras in the LMC consistently exhibit a well-defined single-sequence PMR
that is clearer in the MIR bands than at shorter wavelengths (e.g.,
\citealt{sus09}; \citealt{suh2020}; \citealt{iwanek2021}).

To investigate the variability of CAGB stars in the WISE W1[3.4] and W2[4.6]
bands over the past 16 years, one can use multi-epoch photometry from AllWISE
(2009–2010) and the final data release of the Near-Earth Object WISE Reactivation
(NEOWISE-R) mission (\citealt{mainzer2014}). The latter provides observations
from 21 epochs, typically two per year between 2014 and 2023, plus one in 2024.
Such analyses have been carried out for Galactic CAGB stars by \citet{suh2024}
and for LMC-CAGB stars by \citet{suh2025}.

\begin{figure}
\smallplottwo{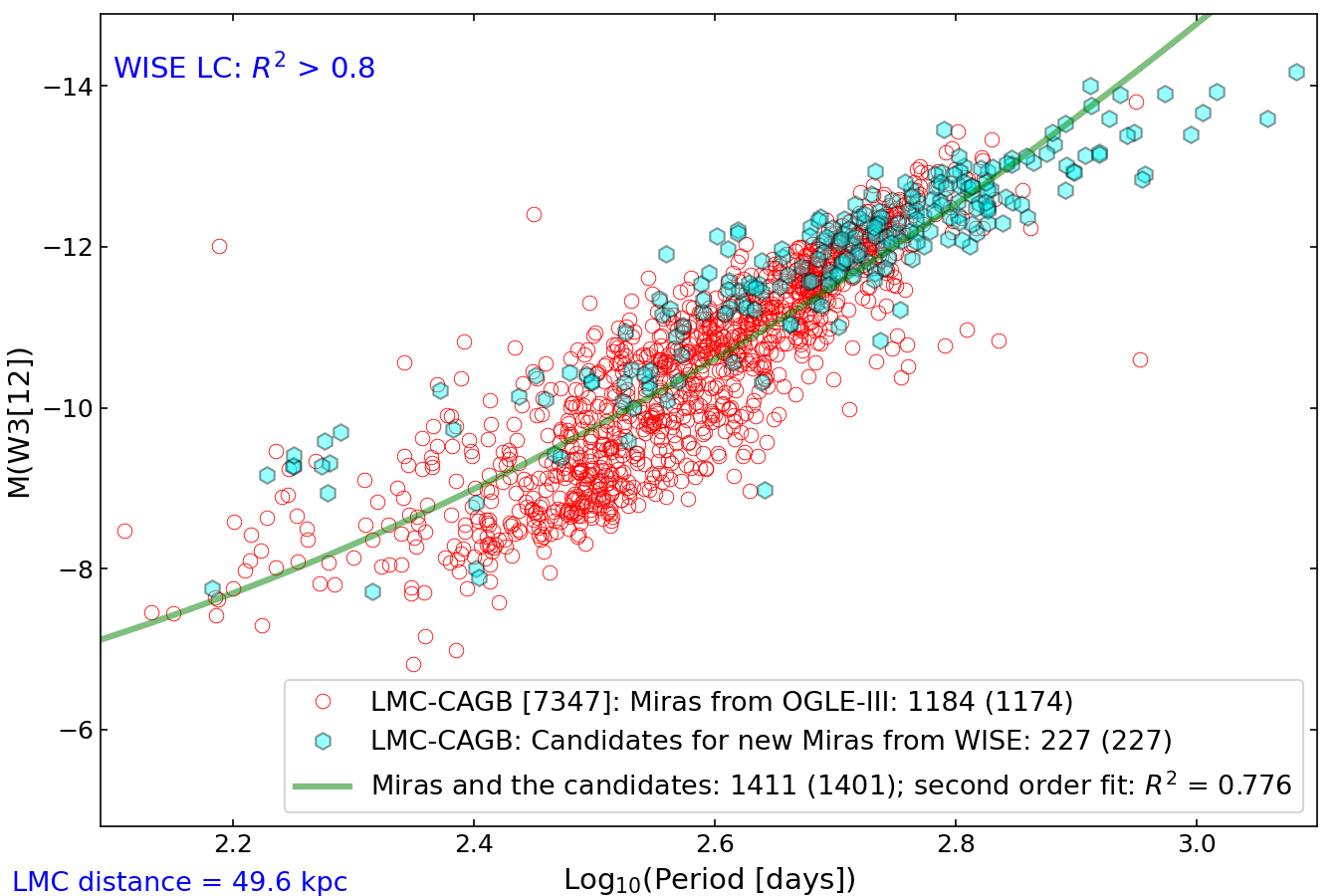}{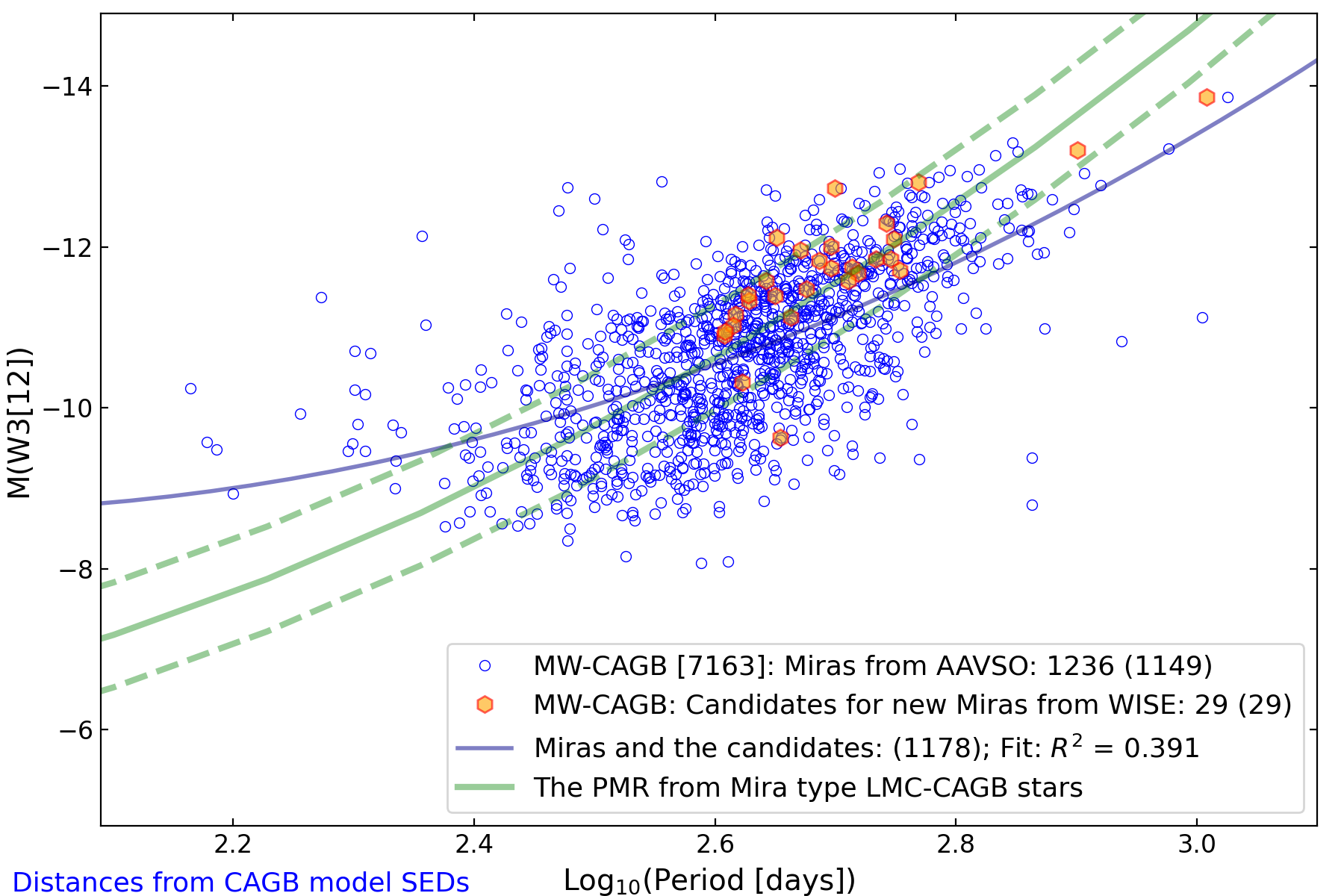}\caption{PMRs for Mira-type CAGB stars.
The top panel shows the PMR for LMC-CAGB objects and the bottom panel shows the PMR for MW-CAGB objects.
The total number of objects in each subgroup is given, with the values in parentheses
denoting the number of plotted sources for which data are available.
See Section~\ref{sec:pul}.}
\label{f10}
\end{figure}

For LMC-CAGB stars, \citet{suh2025} considered both OGLE-III Miras and newly
identified Mira candidates from WISE data, deriving the PMR in the W3[12] band.
After correcting a few misidentified AllWISE counterparts in the sample, we
derived a slightly revised PMR (see the top panel of Figure~\ref{f10}). From a
second-order fit to the 1401 objects, the resulting relation is
\begin{equation}
\mathrm{M(W3[12])} = -3.95 (\log_{10} P)^{2} + 11.7 (\log_{10} P) - 14.3 \pm 0.65, \label{eq:1}
\end{equation}
where $P$ is the pulsation period in days.

\begin{figure}
\centering
\smallplotthree{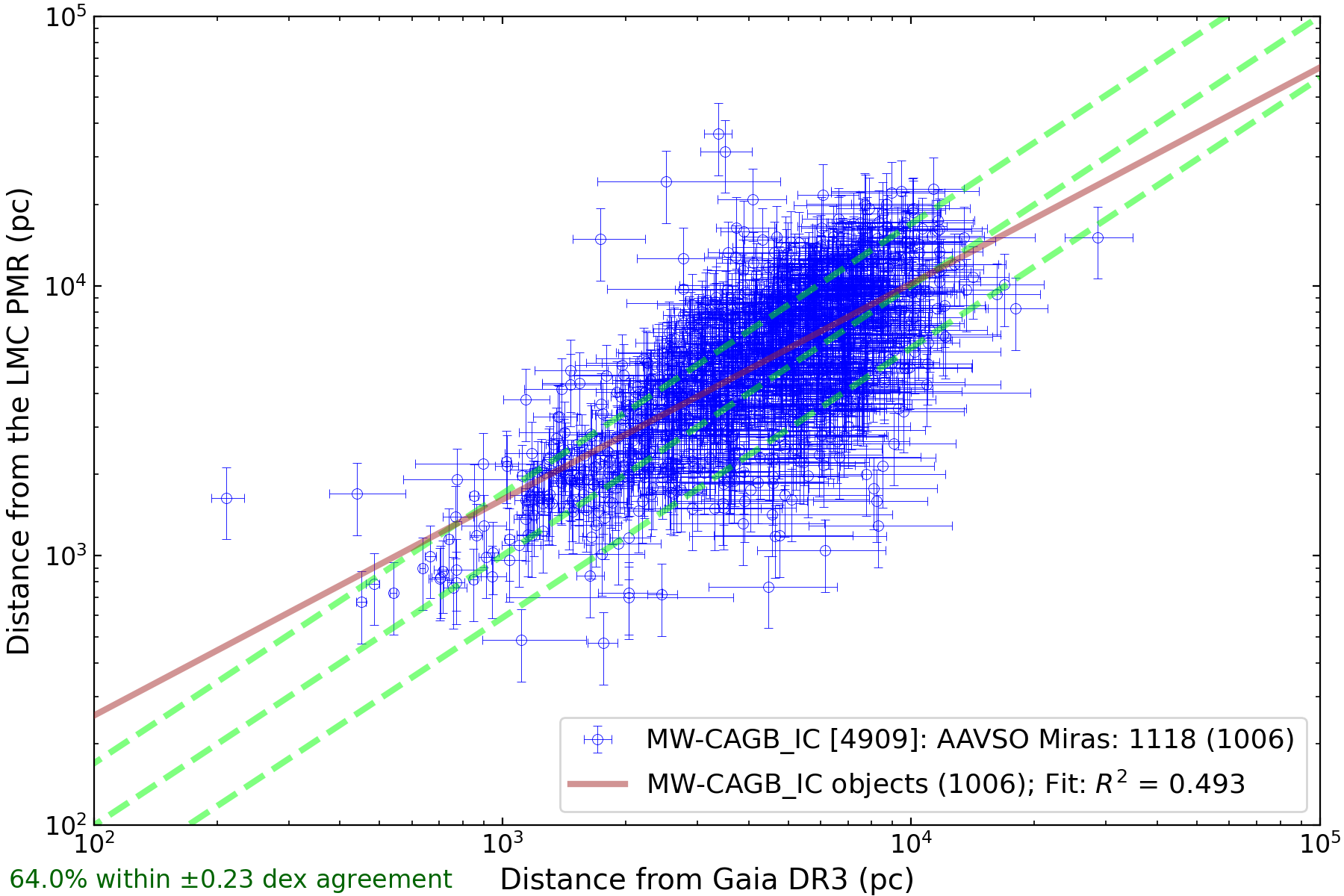}{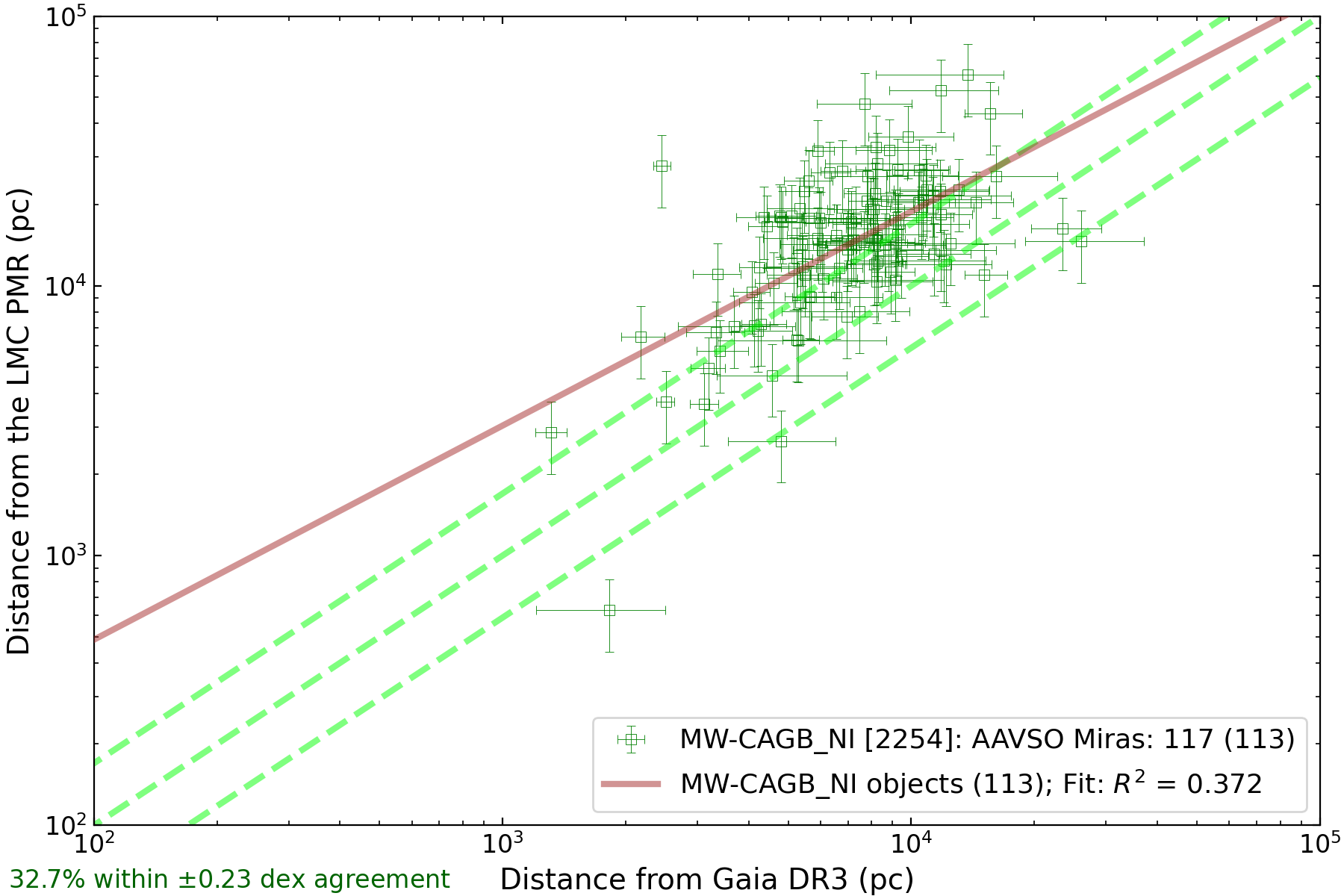}{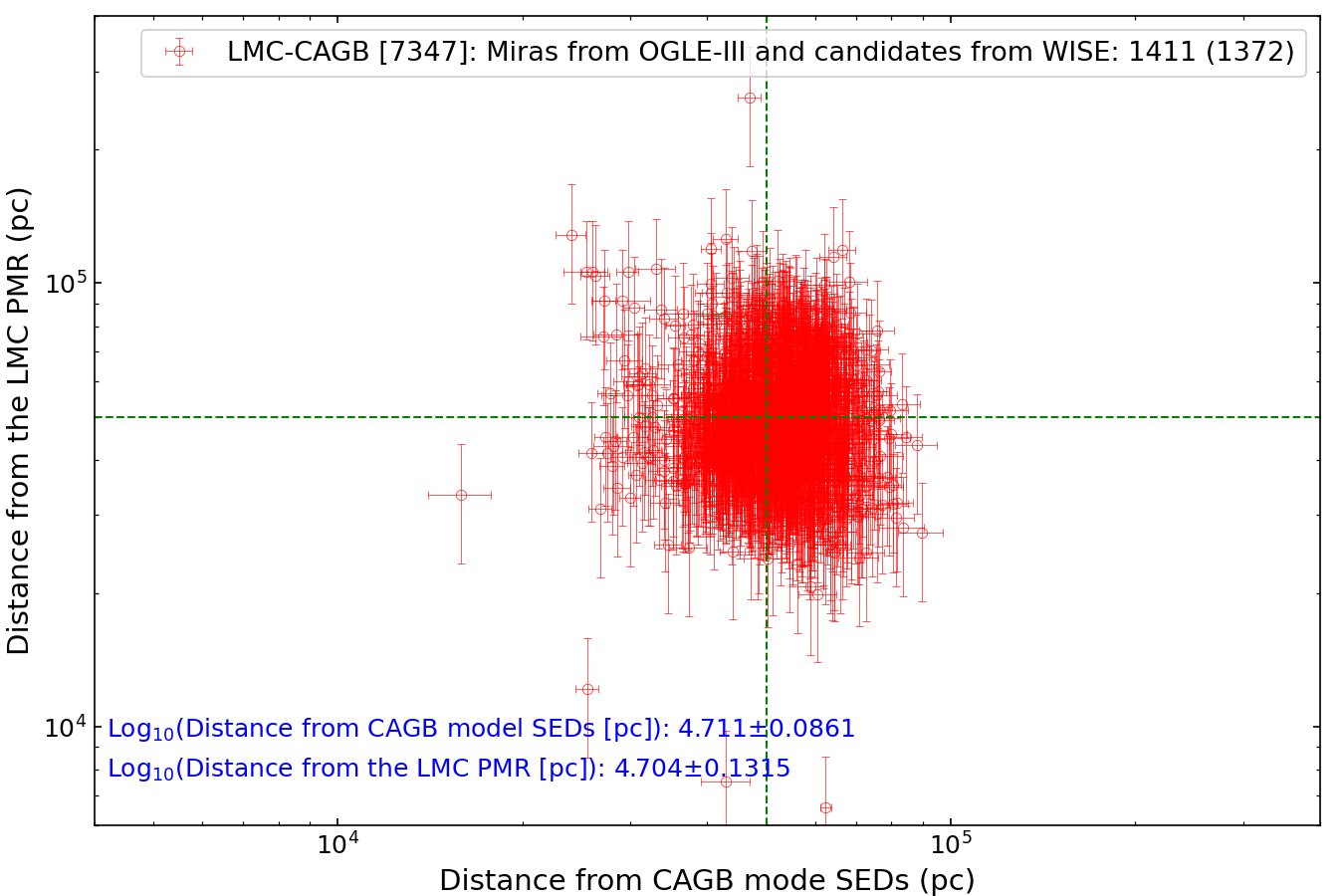}\caption{
The upper two panels compare distances derived from the LMC-based PMR with those
from Gaia DR3 for Mira-type Galactic CAGB stars.
The bottom panel compares distances derived from the LMC-based PMR with those
obtained from CAGB model SEDs for Mira-type LMC-CAGB stars.
The total number of objects in each subgroup is given, with the values in parentheses
denoting the number of plotted sources for which data are available.
See Section~\ref{sec:pul}.}
\label{f11}
\end{figure}

The bottom panel of Figure~\ref{f10} presents the PMR for Galactic CAGB stars,
combining known Mira variables from AAVSO with new Mira candidates identified
from WISE data by \citet{suh2024}, using distances derived from CAGB model SEDs.
For the WISE-based candidates, only stars with periods longer than 400 days were
included. Although the Galactic PMR exhibits greater scatter than that of the
LMC, it broadly follows a similar trend.

For Galactic CAGB stars identified as Miras from AAVSO, the upper two panels of
Figure~\ref{f11} compare the alternative distances derived from the LMC-based PMR
(Equation~\ref{eq:1}) with the Gaia DR3 distances. The vertical error bars for
the PMR distances reflect the uncertainties in the LMC-based PMR. Among the 1006
Mira type MW-CAGB\_IC objects shown in the top panel of Figure~\ref{f11}, 64.0\%
fall within a $\pm$0.23 dex agreement between the two distance estimates. For the
113 Mira type MW-CAGB\_NI objects shown in the middle panel of Figure~\ref{f11},
32.7\% lie within this range of agreement. These results indicate that overall
trends and correlations are consistent with those obtained from the CAGB model
SEDs (see Figure~\ref{f7}).

For Mira-type LMC-CAGB stars, the bottom panel of Figure~\ref{f11} compares
distances derived from the LMC-based PMR (Equation~\ref{eq:1}) with those
obtained from the CAGB model SEDs. Both methods show a similar correlation with
the true distance, although the PMR-based distances exhibit a larger dispersion.
These results further support the reliability of the distance estimates derived
from the CAGB model SEDs.

\section{summary\label{sec:sum}}

We have investigated the properties of CAGB stars in the LMC and the Milky Way,
using samples of 7347 stars in the LMC and 7163 stars in the Milky Way. For all
sample stars, we cross-identified counterparts from Gaia DR3, 2MASS, IRAS, MSX,
ISO, AKARI, WISE, Spitzer, SIMBAD, and the AAVSO. Observed SEDs are compared with
theoretical models to characterize the central stars and their circumstellar dust
envelopes and to estimate distances.

For LMC-CAGB stars, we obtained a set of 37 CAGB models that best reproduced the
shapes of the observed SEDs and minimized the differences between the scaling and
true distances. The resulting distance distribution, with a median of 49.6 kpc
and a small dispersion, confirms the consistency and reliability of the adopted
CAGB models.

For Galactic CAGB stars, distance estimates from Gaia DR3 parallaxes remain
uncertain. To address this, we derived distances by fitting observed SEDs with
the CAGB models validated against LMC-CAGB stars. In addition, for Mira type
Galactic CAGB stars, we determined distances using the period–magnitude relation
(PMR) calibrated from LMC Miras.

Among the three distance estimates for Galactic CAGB stars, model-based SED
distances were found to be both reliable and practical for a large sample of
Galactic CAGB stars. Gaia DR3 parallaxes have inherent limitations for AGB stars,
and PMR-based distances apply only to Mira variables. In contrast, CAGB model SED
distances can be derived whenever observed SEDs of sufficient quality and
wavelength coverage are available.

Overall, CAGB stars in the LMC and the Milky Way exhibit broadly similar
properties in IR CMDs, SEDs, and 2CDs, but an important difference emerges: the
LMC sample lacks stars with very large dust optical depths. Such heavily obscured
CAGB stars are generally thought to represent the most massive or most highly
evolved members of the population. Their relative scarcity in the LMC may
therefore be a consequence of the galaxy’s lower rate of high-mass star formation
compared with the Milky Way.

We have updated the catalog of Galactic CAGB stars from \citet{suh2024} by
including three independent distance estimates for each object, derived in this
work from Gaia DR3 parallaxes, CAGB model SED fitting, and the PMR for LMC Miras.
The dataset for the updated catalog will be available on doi:
\url{https://doi.org/10.5281/zenodo.17262756}.


\begin{acknowledgments}
I thank the anonymous reviewer for constructive comments and suggestions. This
research was supported by Basic Science Research Program through the National
Research Foundation of Korea (NRF) funded by the Ministry of Education
(RS-2022-NR075638). This research has made use of the SIMBAD database and VizieR
catalogue access tool, operated at CDS, Strasbourg, France. This research has
made use of the NASA/ IPAC Infrared Science Archive, which is operated by the Jet
Propulsion Laboratory, California Institute of Technology, under contract with
the National Aeronautics and Space Administration.
\end{acknowledgments}

\end{document}